\documentclass[useAMS,usenatbib]{mnras}

\usepackage{newtxtext,newtxmath,bm}

\usepackage[T1]{fontenc}
\usepackage{ae,aecompl}

\usepackage{graphicx}	
\usepackage{amsmath}	
\usepackage{adjustbox}
\usepackage{rotating}
\usepackage{longtable}
\usepackage{bm}
\usepackage{float}
\bmdefine{\balpha}{\alpha}
\bmdefine{\bbeta}{\beta}
\bmdefine{\bSigma}{\Sigma}
\bmdefine{\btheta}{\theta}
\bmdefine{\bS}{S}
\bmdefine{\bs}{s}
\usepackage{subcaption}
\usepackage{threeparttable}

\newcommand{\red}[1]{\textcolor{red}{#1}}

\newcommand{\Msol}{M_{\odot}}

\newcommand{\SExtractor}{\texttt{SExtractor}}
\newcommand{\Swarp}{\texttt{swarp}}
\newcommand{\Scamp}{\texttt{scamp}}
\newcommand{\THELI}{\texttt{THELI}}

\newcommand{\Elixir}{\texttt{Elixir}}
\newcommand{\automask}{\texttt{automask}}

\title[CODEX Lensing Mass Catalogue]{CODEX Weak Lensing Mass Catalogue and implications on the mass-richness relation}

\author[K. Kiiveri et al.] {K. Kiiveri$^{1, 2}$,   D. Gruen$^{3,4}$,  A. Finoguenov$^{1}$,
T. Erben$^{5}$,  L. van Waerbeke$^{6}$, E. Rykoff$^{3,4}$,\newauthor L. Miller$^{7}$, S.  Hagstotz$^{8}$, R. Dupke$^{9}$,  J. Patrick Henry$^{10}$,  J-P. Kneib$^{11}$,  G. Gozaliasl$^{1,2}$, \newauthor C. C. Kirkpatrick$^{1,2}$, N. Cibirka$^{13,14}$, N. Clerc$^{15}$, M. Costanzi$^{16}$, E. S. Cypriano$^{13}$, \newauthor E. Rozo$^{17}$,  H. Shan$^{18}$, P. Spinelli$^{19}$, J. Valiviita$^{2,20}$,  J. Weller$^{14,21}$\\
$^{1}$Department of Physics, University of Helsinki, PO Box 64, FI-00014 Helsinki, Finland\\
$^{2}$Helsinki Institute of Physics, Gustaf H\" allstr\" omin katu 2, University of Helsinki, Helsinki, Finland \\
$^{3}$Kavli Institute for Particle Astrophysics \& Cosmology, P. O. Box 2450, Stanford University, Stanford, CA 94305, USA \\ 
$^{4}$ SLAC National Accelerator Laboratory, Menlo Park, CA 94025, USA\\
$^{5}$Argelander-Institut f\"ur Astronomie, Auf dem H\"ugel 71, D-53121 Bonn, German\\
$^{6}$Department of Physics and Astronomy, University of British Columbia, 6224 Agricultural road, Vancouver, BC V6T 1Z1, Canada\\
$^{7}$Department of Physics, Oxford University, Keble Road, Oxford OX1 3RH, UK\\
$^{8}$Oskar Klein Centre, Department of Physics, Stockholm University, AlbaNova University Centre, SE 106 91 Stockholm, Sweden\\
$^{9}$Observat\'orio Nacional, Rua Gal. Jos\'e Cristino, 20921-400, Rio de Janeiro, Brasil \\ \;
Department of Astronomy, University of Michigan, 311 West Hall 1085 South University Ave. Ann Arbor, MI 48109-1107, USA\\ \;
Department of Physics and Astronomy, University of Alabama, Box 870324, Tuscaloosa, AL 35487, USA\\ \;
Eureka Scientific Inc., 2452 Delmer St. Suite 100, Oakland, CA 94602, USA\\
$^{10}$Institute for Astronomy, 2680 Woodlawn Drive, Honolulu, HI 96822, USA\\
$^{11}$Laboratoire d'Astrophysique, Ecole Polytechnique F\'ed\'erale de Lausanne (EPFL), Observatoire de Sauverny, CH-1290 Versoix, Switzerland\\ \;
Aix Marseille Universit\'e, CNRS, LAM (Laboratoire d'Astrophysique de Marseille) UMR 7326, 13388, Marseille, France\\
$^{13}$Instituto de Astronomia, Geof\'isica e Ci\^encias Atmosf\'ericas, Universidade de S\~ao Paulo, Brasil\\
$^{14}$Universit\"ats-Sternwarte, Fakult\"at f\"ur Physik, Ludwig-Maximilians-Universit\"at M\"unchen, Scheinerstra\ss e 1, D-81679 M\"unchen, Germany\\
$^{15}$CNRS, IRAP, 9 Av. colonel Roche, BP 44346, F-31028; Universit\'{e} de Toulouse, UPS-OMP, Toulouse, France\\
$^{16}$INAF-Osservatorio Astronomico di Trieste, Via G. B Tiepolo 11, I-34143; IFPU-Institute for Fundamental Physics of the Universe, Via Beirut 2, 34014 Trieste, Italy \\
$^{17}$Department of Physics, University of Arizona, 1118 E. Fourth St, Tucson, AZ 85721, USA\\
$^{18}$Shanghai Astronomical Observatory (SHAO), Nandan Road 80, Shanghai 200030, China \\
$^{19}$Museu de Astronomia e Ci\^encias Afins (MAST), Rua General Bruce 586, 20921-030 Rio de Janeiro, Brasil\\
$^{20}$Department of Physics, P.O.Box 35 (YFL), FI-40014 University of Jyv\"askyl\"a, Finland \\
$^{21}$Excellence Cluster Origins, Boltzmannstra\ss e 2, D-85748 Garching, Germany}

\date{Accepted XXX. Received YYY; in original form ZZZ}

\pubyear{2020}

\begin{document}
\label{firstpage}
\pagerange{\pageref{firstpage}--\pageref{lastpage}}
\maketitle

\begin{abstract}
The COnstrain Dark Energy with X-ray clusters (CODEX) sample contains the largest
flux limited sample of X-ray clusters at $0.35 < z < 0.65$. It was selected from ROSAT data in the 10,000 square degrees of overlap with BOSS, mapping a total number of 2770 high-z galaxy clusters. We present here the full results of the CFHT CODEX program on cluster mass measurement, including a reanalysis of CFHTLS Wide data, with 25 individual lensing-constrained cluster masses. We employ \emph{lensfit} shape measurement and perform a conservative colour-space selection and weighting of background galaxies. Using the combination of shape noise and an analytic covariance for intrinsic variations of cluster profiles at fixed mass due to large scale structure, miscentring, and variations in concentration and ellipticity, we determine the likelihood of the observed shear signal as a function of true mass for each cluster. 
We combine 25 individual cluster mass likelihoods in a Bayesian hierarchical scheme with the inclusion of optical and X-ray selection functions to derive constraints on the slope $\alpha$, normalization $\beta$, and scatter $\sigma_{\ln \lambda | \mu}$ of our richness--mass scaling relation model in log-space: $\left<\ln \lambda | \mu \right> = \alpha \mu + \beta$, with $\mu = \ln (M_{200c}/M_{\mathrm{piv}})$, and $M_{\mathrm{piv}} = 10^{14.81} M_{\odot}$. We find a slope $\alpha = 0.49^{+0.20}_{-0.15}$, normalization $ \exp(\beta) = 84.0^{+9.2}_{-14.8}$ and $\sigma_{\ln \lambda | \mu} = 0.17^{+0.13}_{-0.09}$ using CFHT richness estimates. In comparison to other weak lensing richness-mass relations, we find the normalization of the richness statistically agreeing with the normalization of other scaling relations from a broad redshift range ($0.0<z<0.65$) and with different cluster selection (X-ray, Sunyaev-Zeldovich, and optical).

\end{abstract}




\section{Introduction}
Clusters of galaxies represent the end product of hierarchical structure formation. They play a key role in understanding the cosmological interplay of dark matter and dark energy. Their number density, baryonic content, and their growth are sensitive probes of cosmological parameters, such as the mean dark matter and dark energy density $\Omega_{\rm{m}}$ and $\Omega_{\rm{\Lambda}}$, the dark energy equation of state parameter $w$ and the normalization of the matter power spectrum $\sigma_8$ (see \citealt{allen11} for a recent review).
The idea of using cluster counts to probe cosmology is based on the halo mass function, which predicts their number density as a function of mass, redshift and cosmological parameters (see e.g. \citealt{press74,sheth99,tinker08}). The observational task consists of obtaining an ensemble of galaxy clusters with an observable that correlates with their true mass and a well defined selection function.
In recent years a number of multiwavelength, deep, and wide observations and  surveys have been conducted which allow to detect galaxy clusters with a high signal-to-noise ratio (S/N) out to high redshifts (e.g., $z\sim2.5$). Observations are based on properties of baryonic origin, among them the number count of red galaxies (called richness, see e.g. \citealt{gladders05,koester07,rykoff14}) or the inverse Compton scattering of cosmic microwave photons on the hot intra-cluster gas (the \citealt{sunyaev80} effect, see \citealt{bleem15} and \citealt{planckcollaboration15} for the latest observational results). Another approach is to select a galaxy cluster sample from X-ray observations (see e.g. \citealt{ebeling98,boehringer04,ebeling10, gozaliasl14, gozaliasl19}). However, hydrodynamical simulations have shown that even for excellently measured X-ray observables with small intrinsic scatter at fixed mass and dynamically relaxed clusters at optimal measurement radii ($r \sim r_{2500}$), non-thermal pressure support from residual gas bulk motion and other processes are expected to bias the hydrostatic X-ray mass estimates down by up to 5-10 per cent (see \citealt{nagai07,rasia12}), which represents the currently dominant systematic uncertainty in constraining cosmology from X-ray cluster samples (see \citealt{henry09,vikhlinin09,mantz10,rozo10,benson13,mantz15}).

For this reason, the idea of absolute calibration of the mass scale of large cluster samples by weak gravitational cluster lensing (see e.g. \citealt{hoekstra07,marrone12,gruen14,vonderlinden14a,vonderlinden14b,melchior16,herbonnet19}) has gained traction over the last years. Weak gravitational lensing is sensitive to the entire gravitational matter and is  therefore mostly free of systematic uncertainty that relates to the more complex interaction of baryons.

However, weak lensing mass measurements for individual clusters are inherently quite noisy, as the measured ellipticities of background galaxies do not only depend on the gravitational shear induced by the analysed galaxy cluster but also on the quite broad intrinsic ellipticity distribution, and on the gravitational imprint of all matter along the line of sight, including unrelated projected structure (see e.g. \citealt{hoekstra01,hoekstra03,spinelli12}). On top of this, at fixed true mass the density profiles of clusters intrinsically vary, causing additional scatter in weak lensing mass estimation \citep{becker11,gruen11,gruen15}. For this reason, relatively large samples of galaxy clusters need to be investigated to statistically meet the calibration requirements of cosmology. 

Even with large samples of clusters and sufficiently deep optical data to measure the shapes of numerous background galaxies, several systematic uncertainties limit the power of weak lensing mass calibration. Firstly, shape measurement algorithms commonly recover the amplitude of gravitational shear only with a one to several per-cent level multiplicative bias (e.g. \citealt{mandelbaum15,jarvis16,fenechconti17}, but see the recent advances of \citealt{huff17,sheldon17}). Secondly, the amplitude of the weak lensing signal does not only depend on the cluster mass, but also on the geometric configuration between observer, lens and background objects, more specific on their angular diameter distances among the observer, lens, and source. For interpreting the shear signal, additional photometric data are required to obtain the necessary distance information by photometric redshifts (\citealt{lerchster11,gruen13}), colour cuts or distance estimates by colour-magnitude properties (\citealt{gruen14,cibirka16}). All these methods suffer from systematic uncertainties (see e.g. \citealt{applegate14,gruen16}) that translate to systematic errors in cluster masses. On a related note, cluster member galaxies can enter the photometrically selected background galaxy sample and lower the observed gravitational shear signal (see e.g. \citealt{sheldon04,gruen14,melchior16} for different methods of estimating and correcting the impact of this). Finally, a mismatch between the fitted density profile and the underlying true mean profile of clusters at a given mass (including the miscentring of clusters relative to the assumed positions in the lensing analysis) can cause significant uncertainty in weak lensing cluster mass estimates (see e.g. \citealt{melchior16}).

In this COnstrain Dark Energy with X-ray clusters (CODEX) study we present weak lensing mass analysis for a total of 25 galaxy clusters. The initial CODEX sample of 407 clusters, from which the main lensing sample is obtained, is cut at $0.35 < z < 0.65$ with $\lambda \geq 60$ with X-ray based selection function.
To this end, we also develop new methods to provide a full likelihood of the lensing signal as a function of individual cluster mass, and carefully characterize the systematic uncertainty.

This paper is structured as follows. In \autoref{sec:data and analysis} we present the data and analysis, including data reduction, photometric processing, richness estimation, shape measurement and mass likelihood. In \autoref{sec:bayesmodel} we describe our Hierarchical Bayesian model, which we use to estimate richness-mass relation. In \autoref{sec:application} we apply the Hierarchical Bayesian model to find the richness--mass relation of all 25 clusters in the weak lensing mass catalog. In \autoref{sec:results} we present our results of the Bayesian analysis, and in \autoref{sec:conclusion} we summarize and conclude. In the Appendix, we detail our systematic uncertainties, fields with incomplete colour information, and present weak lensing mass measurements for 32 clusters excluded from the richness--mass calibration. 
 
We adopt a concordance $\Lambda$CDM cosmology and WMAP7 results \citep{Komatsu_2011} with $\Omega_{\rm{m}}=0.27$, $\Omega_{\rm{\Lambda}}=0.73$ and
$H_0 = 70$ km\, s$^{-1}$\, Mpc$^{-1}$. The halo mass of galaxy clusters in this study corresponds to $M_{\mathrm{200c}}$, defined  as  the  mass  within radius $r_{200c}$, the radius in which the mass and concentration definitions is taken to be 200 times the critical density of the Universe ($\rho_{\rm{c}}$).

\section{Data and Analysis}
\label{sec:data and analysis}

\subsection{Cluster catalogue}
\label{sec:cluster catalog}

The CODEX sample was initially selected by a $4\sigma$ photon excess inside a wavelet aperture in the overlap of the ROSAT All-Sky Survey (RASS, \citealt{voges99}) with the Sloan Digital Sky Survey (SDSS). 
We use RASS photon images and search for X-ray sources on scales from 1.5 arcmin to 6 arcmin using wavelets. Any source detected is considered as a cluster candidate and enters the redMaPPer code (see \citealt{rykoff14} and \autoref{sec:redmapper}), which associates an optical counterpart for each source and reports its richness and redshift. For this sample, we consider a high threshold of richness 60 and redshifts above 0.35, which yields the sample of most massive X-ray selected high-z clusters, for which we seek to perform a weak lensing calibration. While other X-ray source catalogues using RASS data exist (e.g. \citealt{boller16}), the advantage of our approach consists of performing detailed modelling of the cluster selection function using our detection pipeline, which takes into account the RASS sensitivity as a function of sky position, Galactic absorption, and cluster detectability as a function of mass and redshift. Availability of such a selection function enables precise modelling of the cluster appearance in the catalog, critically important for the Bayesian modelling of the scaling relations.

At the positions of these overdensities, the redMaPPer algorithm is run to extract estimates of photometric redshift, richness, and a refined position and ROSAT X-ray flux estimate. For more details on the catalog construction, see \citet{clerc16}, \citet{cibirka16} and \citet{finoguenov2019codex}.

The initial sample of 407 clusters is selected by the richness $\lambda_{\rm RM,SDSS}$ and redshift $z_{\rm{RM,SDSS}}$ estimated from the redMaPPer run on SDSS photometric catalogues, cut at $\lambda_{\rm RM,SDSS} \ge 60$ and $0.35 < z_{\rm{RM,SDSS}} < 0.65$. A subsample of the initial sample was chosen as a weak lensing follow-up with CFHT (Canada-France-Hawaii Telescope) designed to calibrate richness--mass relation for this survey. This deeper CFHT survey of 36 clusters, that we call S-I, falls into the CFHT Legacy Survey\footnote{http://www.cfht.hawaii.edu/Science/CFHTLS/} (CFHTLS) footprint, and is selected only by observability. To have an optically clean sample without missing data in CFHT richness or weak lensing mass, we exclude a total of 11 clusters, and define the remaining 25 cluster sample as our main lensing sample. The main lensing sample of 25 clusters is listed in Table \ref{tab:cleanedWL}. The excluded clusters of S-I are described in section \ref{sec:application}, and listed in the Appendix Table \ref{tab:primaryWL}. 

Since weak lensing analysis requires precise knowledge of the cluster redshift,
for 20 clusters without spectroscopic redshifts in S-I, we targeted red-sequence member galaxies for spectroscopy. The clusters observed as a part of the CFHT program, are targeted by several Nordic Optical Telescope (NOT) programs (PI A. Finoguenov, 48-025, 52-026, 53-020, 51-034). Each cluster is observed in multi-object spectroscopy mode, targeting $\sim$20 member galaxies including Brightest Cluster Galaxies (BCGs) and having spectral resolving power of $\sim$500. The typical exposure per mask is 2700 s with a grism that provides wavelength coverage between approximately 400 - 700 nm. The average seeing over the four programmes is near 1 arcsec. Because we are solely interested in the redshift of the Ca H+K lines, only wavelength calibration frames are additionally obtained. Standard IRAF7 packages are used in the data reduction, spectra extraction and wavelength calibration process. The redshifts are determined finally using RVIDLINES to measure the positions of the two calcium lines for a weighted average fit. The acquired spectroscopic cluster redshifts for the weak lensing sample are listed in Table \ref{tab:cleanedWL}, along with X-ray observables, richness estimates, and available photometric data. 
\subsection{Imaging data and data reduction}
\label{sec:imaging data}

This study comprises imaging data covering 34 pointings centred on CODEX clusters observed with the wide field optical camera MegaCam \citep{boulade03} at the CFHT.
For 28 of these pointings full colour information of filters $u$.MP9301, $g$.MP9401, $r$.MP9601, $i$.MP9702, $z$.MP9801 is available. All considered pointings possess $i$-band information. A summary of the imaging data of S-I can be seen in Appendix Table \ref{tab:primary images}.
\\
A detailed description of the data reduction can be found in \citet{cibirka16}. We only give a brief overview here.
\\
We process the CODEX data using the algorithms and processing pipelines (\THELI ) developed within the CFHTLS-Archive Research Survey (CARS, see \citealt{erben09,erben05}; \citealt{schirmer13}) and CFHT Lensing Survey\footnote{http://cfhtlens.org}, (CFHTLenS, see \citealt{erben13,heymans12}).
\\
Starting point is the CODEX data, preprocessed with \Elixir, available at the Canadian Astronomical Data Centre\footnote{http://www4.cadc-ccda.hia-iha.nrc-cnrc.gc.ca/cadc} (CADC). The \Elixir\space preprocessing removes the entire instrumental signature from the raw data and provides all necessary photometric calibration information.
\\
The final data reduction comprises deselection of damaged raw images or images of low quality, astrometric and relative photometric calibration using \Scamp\footnote{http://www.astromatic.net/software/scamp}  \citep{bertin02}, coaddition of the final reduced single frames with \Swarp\footnote{http://www.astromatic.net/software/swarp} and creation of image masks by running the \automask tool\footnote{http://marvinweb.astro.uni- bonn.de/data\_\\products/THELIWWW/automask.html} \citep{dietrich07} to indicate photometrically defective areas (satellite and asteroid tracks, bright, saturated stars and areas which would influence the analysis of faint background sources).

\subsection{Photometric catalogue creation}
\label{sec:photometry}

The photometric redshift calibration, photometric catalogue creation and the photometric redshift estimation are presented  in \citet{brimioulle13}. We only give a brief overview here. 

The estimation of meaningful colours from aperture fluxes requires same or at least similar shape of the point spread function (PSF) in the different filters of one pointing. Therefore in the first step we adjust the PSF by convolving all images of one pointing/filter with a fixed Gaussian kernel, degrading the PSF to the value of the worst band (in general $u$). We select the appropriate kernel in an iterative process, so the observational stellar colours no longer depend on the diameter of the circular aperture they are measured in. We then run \SExtractor\footnote{http://www.astromatic.net/software/sextractor} (see \citealt{bertin96}) in dual-image-mode, selecting the unconvolved $i$-band as detection band and extracting the photometric information from the convolved images. We extract all objects which are at least 2$\sigma$ above the background on at least four contiguous pixels.

Unfortunately the original magnitude zeropoint determination by the \Elixir\space pipeline proved to be inaccurate. The colours of stars and galaxies can vary from field to field due to galactic extinction and because of remaining zero-point calibration errors. Since the CFHTLS-Wide fields are selected to be off the galactic plane, the extinction is rather small and does not change a lot over one square degree tiles: the maximum and minimum extinction in all Wide fields is 0.03 and 0.14, respectively, and the difference between maximum and minimum extinction value per square degree can be up to 0.03 for high extinction fields and 0.01 for fields with low extinction values. We account for one zero-point and extinction correction value per square degree field by shifting the observed stellar colours to those predicted from the Pickles stellar library \citep{pickles98} for the given photometric system. In this way we do not only correct for the inaccurate magnitude zeropoints, but do also correct for galactic extinction and field-to-field zeropoint variations. 

\subsection{redMaPPer}
\label{sec:redmapper}

redMaPPer \citep{rykoff14} is a red-sequence photometric cluster finding procedure that builds an empirical model for the red-sequence colour-magnitude relation of early type cluster galaxies. It is built around the optimized richness estimator developed in \cite{rozo09} and \cite{rykoff2012robust}. redMaPPer detects clusters as overdensities of red galaxies, and measures the probability that each red galaxy is a member of a cluster according to a matched filter approach that models the galaxy distribution as the sum of a cluster and background component. The main design criterion for redMaPPer is to provide a galaxy count based mass proxy with as little intrinsic scatter as possible. To this end, member galaxies are selected at luminosities $L>0.2L_{\star}$, based on their match to the red-sequence model, and with an optimal spatial filter scale \citep[see][]{rykoff16}.

The redMaPPer richness of clusters is the sum of the membership probabilities of all galaxies. The aperture used as a cluster radius to estimate the cluster richness is self-consistently computed with the cluster richness, ensuring that richer clusters have larger cluster radii. This radius is selected to minimize the scatter of richness estimates at a given mass. The cluster richness estimated by redMaPPer has been shown to be strongly correlated with cluster mass by comparing the richness to well-known mass proxies such as X-ray gas mass and Sunyaev–Zel'dovich (SZ) decrements. The main (v5.2) redMaPPer algorithm was presented in \cite{rykoff14}, to which the reader is referred for more details. 

Especially at higher cluster redshift, the shallow SDSS photometry only allows for a relatively uncertain estimate of richness due to incompleteness at a magnitude corresponding to galaxies fainter than the redMaPPer limit of 0.2$L_{\rm \star}$. 
The acquired follow-up CFHT photometry is significantly deeper, and therefore allows for improved estimates of $\lambda$ for the observed CODEX lensing sample. This, however, requires an independent calibration of the red sequence in the used set of filters $g$.MP9401, $r$.MP9601, $i$.MP9702 and $z$.MP9801.
In section \ref{sec:application}, we calibrate the richness--mass relation based on these improved CFHT richness estimates, and use the observed SDSS richnesses only to determine the shape of the sampling function, as is described in \autoref{sec:subsample_selection_function} . 

Due to incomplete observations in $g$ and $z$ for some of the clusters in our sample, we perform this in three separate variants, namely based on $griz$, $gri$ and $riz$ photometry. In the case of CODEX35646, where no $i$.MP0702 band data is available, we generate artificial magnitudes by adding the $i$.MP9702-colour of a red galaxy template at the cluster redshift to the available $i$.MP9701-magnitude of all galaxies in the field. 

For calibrating the red sequence, we use the spectroscopic cluster redshifts (see Table \ref{tab:primaryWL} and Table \ref{tab:subsamples} ), where available. To account for masking to correct galaxy counts for undetected members, we convert the polygon masks applied to the CFHT object catalogues to a \textsc{healpix} mask \citep{gorski05} with $N_{\rm side}=4096$.

\begin{figure}
\includegraphics[width=8.45cm]{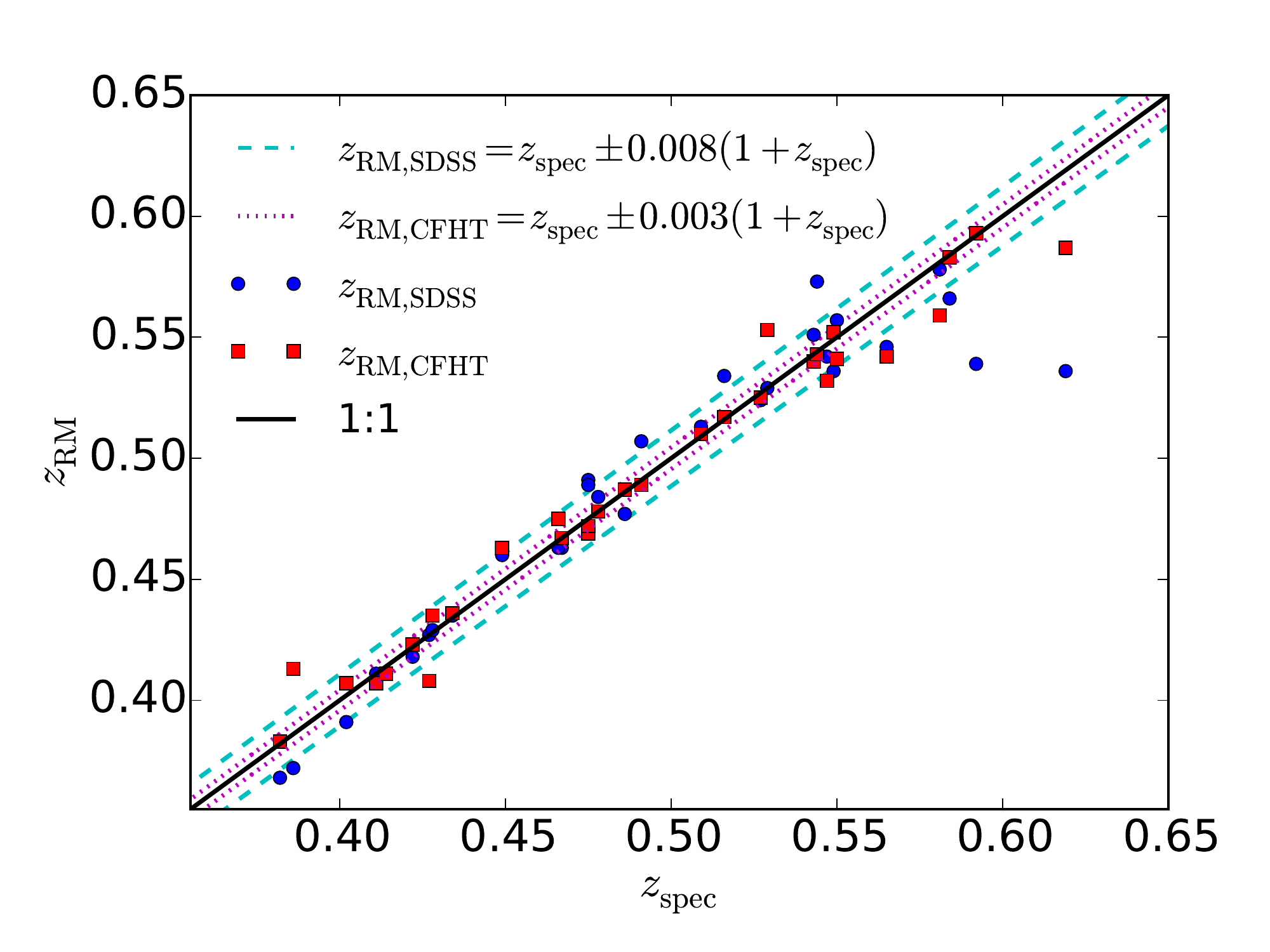}
\caption{Spectroscopic redshifts versus CFHT/SDSS photometric cluster redshift estimates by redMaPPer for all spectroscopically covered clusters. Through a comparison with the spectroscopic redshifts of clusters, we measure photometric redshift precision of $\sigma_{\Delta_{z_{\mathrm{RM,SDSS}}}/(1+z_{\mathrm{spec}})}=0.008$ and $\sigma_{\Delta_{z_{\mathrm{RM,CFHT}}}/(1+z_{\mathrm{spec}})}=0.003$.  }
\label{fig:zspeczSDSS}
\end{figure}

Using the spectroscopic redshifts obtained for this sample we can verify redMaPPer redshift determination. \mbox{Fig. \ref{fig:zspeczSDSS}} shows spectroscopic redshift of the cluster BCGs versus the redMaPPer photometric redshift estimate $z_{\rm{RM}}$. Through this comparison the photometric redshift precision for both samples of SDSS and CFHT are found to correspond to $\sigma_{\Delta_{z_{\mathrm{RM,SDSS}}}/(1+z_{\mathrm{spec}})}=0.008$ and $\sigma_{\Delta_{z_{\mathrm{RM,CFHT}}}/(1+z_{\mathrm{spec}})}=0.003$. 
While the redMaPPer photometric redshift precision of the SDSS-DR8 catalog is $\sigma_{\Delta_{z_{\mathrm{SDSS,DR8}}}/(1+z_{\mathrm{spec}})}=0.006$, as estimated by \cite{rykoff14}.

\subsection{Shape measurement}
\label{sec:shape measurement}

We use the \textsc{lensfit} algorithm (see \citealt{miller13}) to measure galaxy shapes. We chose the $i$-band images for shape extraction as this band has usually smaller FWHM and lower atmospheric differential diffraction than the bluer bands. 

The extracted quantities are the measured ellipticity components $e_1$ and $e_2$ and the weight taking into account shape measurement errors and the expected intrinsic shape distribution as defined in \citet{miller13}. In order to sort out failed measurements and stellar contamination of our background sample we only consider background objects with a \emph{lensfit weight} greater than 0 and a \emph{lensfit fitclass} equal to 0.

For our sample S-I we make use of the latest `self-calibrating' version of the \emph{lensfit} shape measurement (see \citealt{fenechconti17}). Here we only highlight a few important facts about the self-calibration, for a detailed description we refer the reader to its first application in the Kilo-Degree Survey (KiDS, see \citealt{fenechconti17,hildebrandt17}). The main motivation for the self-calibration is given by the noise bias problem plaguing shape measurements techniques (see e.g. \citealt{ melchior12,refregier12,miller13, fenechconti17, Kannawadi_2019}). However, the self-calibration is not perfect as it is shown to contain a residual calibration of the order of 2 per cent. \citet{fenechconti17} discussed how to further reduce this with help of image simulations to the sub-per cent level as required for cosmic shear studies as presented by \citet{hildebrandt17}, but given the residual statistical uncertainties in our cluster lensing studies, we discard this step and use the self-calibrated shapes directly. 
We estimate the uncertainty associated with this step to be around 3-5
percent of the actual shear value.

\subsection{Source selection and redshift estimation}
\label{sec:photoz}
The observable in a weak lensing analysis is the mean tangential component of reduced gravitational shear $g_{\rm t}$ (see equation~\ref{eqn:meanshear}) of an ensemble of sources. At a given projected radius $r$ from the centre of the lens, it is related to the physical surface mass density profile of the latter, $\Sigma(r)$, by
\begin{equation}
g_{\rm t}(r)=\frac{\Delta\Sigma(r)/\Sigma_{\rm crit}}{1-\Sigma(r)/\Sigma_{\rm crit}} + \rm{Noise}\; ,
\label{eqn:reducedgt}
\end{equation}
where we have defined $\Delta\Sigma(r)=\langle\Sigma(r')\rangle_{r'<r}-\Sigma(r)$. In the limit where $\Sigma\ll\Sigma_{\rm crit}$, $g_t$ is equal to the tangential gravitational shear $\gamma_t$,
\begin{equation}
    g_t(r)\approx\gamma_t(r)=\Delta\Sigma(r)/\Sigma_{\rm crit} \; .
    \label{eqn:gammag}
\end{equation}

The critical surface mass density,
\begin{equation}
\Sigma_{\rm crit}=\frac{c^2}{4\pi G D_{\rm d}}\frac{D_{\rm s}}{D_{\rm ds}} \; ,
\label{eqn:sigmac}
\end{equation}
is a function of the angular diameter distances between the observer and lens $D_{\rm d}$, observer and source $D_{\rm s}$, and lens and source $D_{\rm ds}$. The ratio of the latter two is denoted in the following as the shorthand
\begin{equation}
\beta=\frac{D_{\rm ds}}{D_{\rm s}} \; .
\label{eqn:beta}
\end{equation}
This is the part of equation~\ref{eqn:sigmac} that depends on source redshifts, illustrating that the latter need to be known for converting lensing observables to physical surface densities.

Based on five-band photometry, redshifts of individual galaxies cannot be estimated unambiguously. However, since the lensing signal of each cluster is measured as the mean $\langle g_t\rangle$ over a large number of galaxies, for an unbiased interpretation of the signal it is sufficient to know the overall redshift \emph{distributions} of the lensing-weighted source sample only. Here, we do this by defining subregions of the CFHT color-magnitude space with a decision tree algorithm. Each source galaxy can then be assigned to one of these subregions. A reference sample of galaxies with measurements in the same and additional photometric bands can be assigned to the same subregions. The redshift distribution of galaxies in each subregion can be estimated as the histogram of the high-quality photometric redshifts for the reference sample of galaxies assigned to the same subregion. The redshift distribution of the whole sample is a linear combination of the redshift distributions of the contributing subregions.

To this end, we follow the same algorithm as in \citet{cibirka16}, described in more detail in \citet{gruen16}. In a nutshell, we divide five-band colour-magnitude space into boxes (hyper-rectangle subregions) and estimate the redshift distribution in each box from a reference catalog of 9-band optical+near-Infrared photo-$z$.

The reference catalogue of high-quality photo-$z$ is based on a magnitude-limited galaxy sample with 9-band ($u$.MP9301, $g$.MP9401, $r$.MP9601, $i$.MP9701, $i$.MP9702, $z$.MP9801, $J$.WC8101, $H$.WC8201, $Ks$.WC8302)-photometry from the four pointings of the CFHTLS Deep and WIRCam Deep \citep{bielby12} Surveys. The outlier rate of these redshift estimates is $\eta=2.1$ per cent, with a photo-$z$ scatter $\sigma_{\Delta z/(1+z)}=0.026$ for $i<24$ (see \citealt{gruen16}, their Fig. 4). We emphasize that the photometric catalogues in this work and the reference catalogue in \citet{gruen16} have been created in the exact same way. In order to reduce contamination and enhance signal-to-noise-ratio we apply several cuts during the construction of the colour-magnitude decision tree, as in \citet{cibirka16}. This way, we remove parts of colour-magnitude space in which contamination with galaxies at the cluster redshift is possible. In addition, we identify and remove parts of color-magnitude space in which our 9-band photometric redshifts disagree with the COSMOS2015 photo-$z$ of \citet{laigle16}. We also use the latter catalog to identify systematic uncertainties due to potential remaining biases in the high-quality photo-$z$ (see Appendix~\ref{sec:bias_z_distribution}). 

To perform the cuts described above, \emph{before} construction of the decision tree we remove all galaxies from cluster and reference fields whose colour is in the range spanned by galaxies in the reference catalogue best fitted by a red galaxy template in the redshift interval $z_{\rm{d}} \pm 0.04$.

\emph{After} construction of the decision tree we remove
\begin{itemize}
\item{all galaxies in colour-magnitude hyper-rectangles for which $\langle \beta \rangle$ from COSMOS2015 photometric redshifts are below 0.2.}
\item{all galaxies in colour-magnitude hyper-rectangles populated with any galaxies in the reference catalogue for which the redshift estimate is within $z_{\rm{d}} \pm 0.06$. In particular we remove all galaxies with a \emph{cprob}-estimate unequal 0 to prevent contamination of the source sample with cluster members. We estimate the precision of the resulting estimate might still be biased up to a level of 2 per cent.}
\item{all galaxies in colour-magnitude hyper-rectangles where the ratio of $\langle \beta \rangle$-estimates from COSMOS2015 versus our 9-band photometric redshifts deviates by more than 10 per cent from the median ratio over all hyper-rectangles.}
\end{itemize}

The final estimate of the redshift distribution of a color-magnitude box comes from the 9-band photometric redshifts. We estimate the $\beta$ of a source galaxy as the mean $\beta$ of galaxies in the same box which it falls into. We refer the reader to Appendix \ref{sec:bias_z_distribution} for details on  systematic errors in the redshift calibration.

\subsection{Tangential shear and \texorpdfstring{$\Delta\Sigma$}{DeltaSigma} profile}
For a cluster $C$ and any radial bin $R$, we use the weighted mean of tangential ellipticities measured for a set of source galaxies $i$,
\begin{equation}
g_{\rm t}(C,R)=\sum_{i}w_i\epsilon_{\mathrm{t},i} \; ,
\label{eqn:meanshear}
\end{equation}
where $\epsilon_{\mathrm{t},i}$ is the component of the measured shape of galaxy $i$ tangential to the cluster centre and the sum runs over all sources around $C$ in a radial bin $R$, with weights $w_i$ that are normalized to $1=\sum_i{w_i}$.

Equivalently, in the limit of equation~\ref{eqn:gammag}, we can estimate 
\begin{equation}
\Delta\Sigma(C, R)=\sum_i W_i \Delta\Sigma_i = \sum_i W_i \epsilon_{\mathrm{t},i}/\langle \Sigma_{\mathrm{crit},i}^{-1}\rangle \; ,
\label{eqn:meandeltasigma}
\end{equation}
with a different set of weights $W_i$, again with $1=\sum_i{W_i}$. The expectation value of $\Sigma_{\rm crit}^{-1}$ is calculated from equation~\ref{eqn:sigmac} with the value of $\beta$ estimated in \autoref{sec:photoz}.
The statistically optimally weighted mean (i.e., the one with the highest signal-to-noise ratio) is achieved by using weights equivalent to the $\Delta\Sigma$ estimator of \citet{sheldon04}, namely
\begin{equation}
w_i\propto\frac{\beta_i}{\sigma^2_{\rm intr}+\sigma^2_{\rm obs}} \;,
\label{eqn:sourceweight}
\end{equation}
\begin{equation}
W_i\propto\frac{\langle \Sigma_{\mathrm {crit},i}^{-1}\rangle^2}{\sigma^2_{\rm intr}+\sigma^2_{\rm obs}}\propto\frac{\beta_i^2}{\sigma^2_{\rm intr}+\sigma^2_{\rm obs}}\;,
\label{eqn:dssourceweight}
\end{equation}
where $\beta_i$ is the estimate of a galaxy's $\beta$ as described above, $\sigma^2_{\rm intr}$ is the intrinsic variance of an individual component of galaxy ellipticity, and $\sigma^2_{\rm obs}$ is the variance in an individual component of galaxy shape due to observational uncertainty, both variances obtained from \emph{lensfit}.

Equation~\ref{eqn:meanshear} with these weights $w$ yields what we will call, in the following, mean tangential shear, and equation~\ref{eqn:meandeltasigma} with $W$ what we will call mean $\Delta\Sigma$. The above definitions and normalization conditions lead to the relation
\begin{equation}
\Delta\Sigma(C,R)=g_t(C,R)/\langle\Sigma_{\rm crit}^{-1}\rangle \; ,
\end{equation}
where 
\begin{equation}
\langle\Sigma_{\rm crit}^{-1}\rangle = \sum_i w_i \Sigma_{\mathrm {crit},i}^{-1} \; .
\end{equation}
Mean shear and mean $\Delta\Sigma$ are therefore identical, up to normalization by the $w$-weighted mean of $\Sigma_{\mathrm {crit},i}^{-1}$.
We do not show individual shear profiles, as they are rather noisy, but stacked profiles of the same cluster sample, that we have used in this work, can be found in \citet{cibirka16}.
\subsection{Surface density model}
\label{sec:model}
The interpretation of the weak lensing signal in order to derive a mass estimate for the galaxy cluster requires modelling of the surface density profile $\Sigma$. $\Sigma$ is related to the tangential reduced gravitational shear $g_t$ (equation~\ref{eqn:reducedgt}) through the critical surface mass density (equation~\ref{eqn:sigmac}).

In our analysis we assume the galaxy cluster mass profile to follow a universal density profile, also known as NFW profile (see \citealt{navarro96,navarro97}), which is described by
\begin{equation}
\rho(r)=\frac{\delta_{\rm c} \rho_{\rm c}(z) }{(r/r_{\rm s})(1 + r/r_{\rm s})^2} ,
\label{eqn:NFW}
\end{equation}
where $\rho_{\rm c} = \frac{3 H(z)^2}{8 \pi G}$ represents the critical density of the Universe at redshift $z$, $r_s$ refers to the scale radius where the logarithmic profile slope changes from -1 to -3, and $\delta_c$ describes the characteristic over-density of the halo
\begin{equation}
\delta_{\rm c} = \frac{ 200}{3} \frac{c^3}{\rm{\ln}(1+c) - c/(1+c)}\ .
\label{eqn:deltac}
\end{equation}
The characteristic over-density $\delta_c$ itself is a function of the so-called concentration parameter $c=r_{200}/r_s$.

For the explicit parametrizations for NFW shear components $\gamma_{\rm{t}}$, $g_{\rm{t}}$ and density contrast $\Delta\Sigma$ we refer to equations 11-16 of \cite{wright00}. Note that the measured mean $\Delta\Sigma$ of equation~\ref{eqn:meandeltasigma} is equal to the true $\Delta\Sigma$ only in the weak shear limit, $\kappa \ll1,$ where $\kappa \equiv \Sigma(r)/\Sigma_{\mathrm{crit}}$ denotes convergence, i.e. the dimensionless surface-mass density (cf. equation~\ref{eqn:reducedgt}). To compensate the effect of reduced shear, we boost our model by $(1-\kappa)^{-1}$ when comparing it to the data. 

In order to evaluate the weak lensing signal, we calculate the average of $\Delta \Sigma$ in logarithmically equidistantly binned annuli, both for the observational data and the analytic NFW profile that we use as a model. The radial range around the gravitational lens has to be chosen to minimize systematic effects but maximize our statistical power. Removing too much information on small scales results in loss of the region with the highest S/N. However, it is those small scales which are affected the most by off-centring. This subject will be investigated in further detail in Appendix~\ref{sec:profilecalibration} by examining simulated galaxy cluster halo profiles. As a trade-off, we decide to discard all background sources closer to the cluster centre than 500 $h^{-1}$ kpc, reducing a possible mass bias from off-centring to a minimum. 
On the other side large scales come with two effects. Firstly, the integrated NFW mass diverges for infinite scales, i.e. at a certain point the integrated analytic mass will exceed the physical cluster mass and thus bias low. On the other hand large scales start to be affected by higher-order effects as e.g. 2-halo-term, enhancing the observational mass profile, counter-acting at least partially the first effect. However, since these effects are not trivial to model, in our case the safer option is to discard those regions where these complicating effects start increasing, selecting as an outer analysis radius cut a distance of 2500 $h^{-1}$ kpc. In a nutshell, we logarithmically bin our data in 12 radial annuli within 500 and 2500 $h^{-1}$ kpc. Remaining biases by off-centring, large scale effects and other differences between our assumed NFW profile and the actual profile of galaxy clusters will be determined by calibration on recovered masses from simulated cluster halo profiles from \cite{becker11} in Appendix~\ref{sec:profilecalibration} as mentioned before and be taken into account.
Given this choice of scales, we fit mass only, fixing the concentration parameter by the concentration-mass relation of \citet{dutton14} to
\begin{equation}
\rm{\log_{10}} \ c = \it{a} + \it{b} \ \rm{\log_{10}}\ (\it{M}/[10^{12}h^{-1} \rm{M_{\odot}}]) ,
\end{equation}
with
\begin{align*}
b = -0.101 + 0.026z
\end{align*}
and
\begin{align*}
a = 0.520 + (0.905 - 0.520)\ \rm{exp} {(-0.617\it{z}^{\rm{1.21}})}.
\end{align*}
\subsection{Covariance matrix}
The measured profile $\Delta\Sigma_{\rm obs}$ of any cluster of true mass $M$ deviates from the mean profile $\Delta\Sigma(M)$ of clusters of the same mass and redshift. In some annulus $i$, we can write
\begin{equation}
\Delta\Sigma_{\mathrm{obs},i}=\Delta\Sigma_i(M) + \delta_i \; .
\end{equation}
The covariance matrix element $C_{ij}$ required when determining a likelihood of $\Delta\Sigma_{\rm obs}$ as a function of mass is the expectation value
\begin{equation}
C_{ij}=\langle \delta_i \, \delta_j\rangle \; ,
\end{equation}
which contains several components:
\begin{enumerate}
\item \emph{shape noise}, i.e.~the scatter in measured mean shear due to intrinsic shapes and measurement uncertainty of shapes of background galaxies,
\item \emph{uncorrelated large-scale structure}, i.e.~statistical fluctuations of the matter density along the line of sight to the cluster, influencing the light path from the ensemble of background galaxies to the observer,
\item \emph{intrinsic variations of cluster profiles} that would be present even under idealized conditions of infinite background source density and perfectly homogeneous lines of sight.
\end{enumerate}
All these components can be described as independent contributions to the covariance matrix, i.e.
\begin{equation}
C_{ij}(M)=\langle \delta_i \, \delta_j\rangle=C_{ij}^{\rm shape}+C_{ij}^{\rm LSS}+C_{ij}^{\rm intr}(M) \; .
\label{eqn:covcomponents}
\end{equation}
We have made the dependence of the intrinsic variations of the cluster profile on mass $M$ explicit.
The following sections describe these terms in turn. Since the overlap of annuli of pairs of different clusters in our sample is minimal, we assume that there is no cross-correlation of shears measured around different clusters. 
\subsubsection{Shape noise}
The \emph{lensfit} algorithm provides the sum of intrinsic and measurement related variance of the ellipticity of each source $i$, $\sigma_{g,i}^2=\sigma_{\rm intr}^2+\sigma_{\rm obs}^2$. 

Using this to get the shape noise related variance in $\Delta\Sigma_i$,
\begin{equation}
\sigma^2_{\Delta\Sigma, i}=\left(\frac{\sigma_{g,i}}{\langle\Sigma_{\mathrm{crit},i}^{-1}\rangle}\right)^2\propto W_i^{-1}
\end{equation}
the mean $\Delta\Sigma$ with the weights $W_i$ of equation~\ref{eqn:dssourceweight} has a variance
\begin{equation}
C^{\rm shape}_{ii}=\frac{1}{\sum_i \sigma^{-2}_{\Delta\Sigma, i}} \propto \frac{1}{\sum_i W_i} \; .
\end{equation}
Due to the negligible correlation of shape noise between different galaxies, off-diagonal components are set to zero.
\subsubsection{Uncorrelated large-scale structure}
Random structures along the line of sight towards the source galaxies used for measuring the cluster shear profiles cause an additional shear signal of their own. The latter is zero on average, but has a variance (and co-variance between different annuli) that is an integral over the convergence power spectrum and therefore depends both on the matter power spectrum and the weighted distribution of source redshifts.
We analytically account for this contribution to the covariance matrix as \citep[e.g.][]{schneider98,hoekstra03,umetsu11,gruen15}
\begin{equation}
C^{\rm LSS}_{ij}=\int \frac{\ell \mathrm{d}\ell}{2\pi} P_{\kappa}(\ell) \hat{J}_0(\ell\theta_i) \hat{J}_0(\ell\theta_j) \; .
\label{eqn:integral}
\end{equation}
Here, $\hat{J}_0(\ell\theta_i)$ is the area-weighted average of the Bessel function of the first kind $J_0$ over annulus $i$. The convergence power spectrum $P_{\kappa}$ is obtained from the matter power spectrum by the \citet{limber54} approximation as
\begin{eqnarray}
P_{\kappa}(\ell)=\frac{9H_0^2\Omega_m^2}{4c^2}&\int_{0}^{\chi_{\rm max}}\mathrm{d}\chi a^{-2}(\chi)P_{\rm nl}(\ell/\chi,\chi)\nonumber \\ &\int_{\chi}^{\chi_{\rm max}}\mathrm{d}\chi_s p(\chi_s)\left(\frac{\chi_s-\chi}{\chi_s}\right)^2 \; .
\end{eqnarray}
Here $\chi$ denotes comoving distance to a given redshift, and $p(\chi_s)$ is the PDF of comoving distance to sources in the lensing sample, defined as the sum of each individual source PDF (\autoref{sec:photoz}), weighted by the $w$ of equation~\ref{eqn:sourceweight}. For the non-linear matter power spectrum $P_{\rm nl}$ we use the model of \citet{smith03} with the \citet{eisenstein98} transfer function including baryonic effects. Note that since the source sample, weighting, and angular size of annuli is different for each cluster, we calculate a different $C^{\rm LSS}$ for each one of them. 
\subsubsection{Intrinsic variations of cluster profiles}
Even under perfect observing conditions without shape noise, and in the hypothetical case of a line of sight undisturbed by inhomogeneities along the line of sight, the shear profiles of a sample of clusters of identical mass would still vary around their mean.
The reason for this are intrinsic variations in cluster profiles, halo ellipticity and orientation, and subhaloes in their interior and correlated environment. We describe these variations using the semi-analytic model of \citet{gruen15}, which proposes templates for each of these components and determines their amplitudes to match the actual variations of true cluster profiles at fixed mass seen in simulations \citep{becker11}. 
We write
\begin{equation}
C_{ij}^{\rm intr}(M) = C_{ij}^{\rm conc}(M) + C_{ij}^{\rm ell}(M) + C_{ij}^{\rm corr}(M) + C_{ij}^{\rm off}(M) \; ,
\end{equation}
where we assume the best-fit re-scaled model of \citet{gruen15} for the contributions from halo concentration variation $C_{ij}^{\rm conc}$, halo ellipticity and orientation $C_{ij}^{\rm ell}$ and correlated secondary haloes $C_{ij}^{\rm corr}$. For the purpose of this work, the templates in \citet{gruen15} are resampled from convergence to shear measurement and re-scaled to the $\Delta\Sigma$ units of our measurement with the weighted mean $\Sigma_{\rm crit}$ of the source sample.
The final component, $C_{ij}^{\rm off}$ is added to account for variations in off-centring width of haloes. It is calculated as the covariance of shear profiles of haloes of fixed mass, with miscentring offsets drawn according to the prescription of \citet{rykoff16}.
We note that each of these components depends on halo mass, halo redshift, and angular binning scheme. We therefore calculate a different $C_{ij}^{\rm intr}(M)$ for each cluster in our sample. The code producing these covariance matrices is available at \texttt{https://github.com/danielgruen/ccv}.
%

\begin{table*}
\caption{Main weak lensing sample ($\lambda_{\rm{RM,SDSS}}>60$ and $z \ge 0.35$) of 25 clusters}
\label{tab:cleanedWL}
\begin{adjustbox}{angle=90}
\begin{tabular}{c|c|c|c|c|c|c|c|c|c|c|c|c|c}
\hline \hline
 CODEX      &  SPIDERS  & R.A.    &  Dec    &  R.A.   &   Dec    & Filters &   z   &    $\rm{z_{RM}}$  &  $\rm{\lambda_{RM}}$    &    $\rm{z_{RM}}$  & $\rm{\lambda_{RM}}$    &  $\rm{\log M_{200, WL}}$ & $\rm{L_X}$ \\
 ID & ID & opt & opt & X-ray & X-ray & CFHT & spec  & SDSS & SDSS & CFHT & CFHT & $\ \rm{M_{\odot}}$ & $\rm[h_{70}^{-2}\ 10^{44}erg/s]$\\
\hline \hline
 16566   & 1\_2639  &  08:42:31  &    47:49:19  &  08:42:28  &    47:50:03  &  ugriz  &    0.382  &  0.368  &  $108 \pm    7$    &  0.383  &   $120  \pm  3$ &   $ 14.61 _{ -0.29 } ^{ +0.20 } \pm 0.02 $  &  $  3.1 \pm  1.2 $  \\
 24865   & 1\_5729  &  08:22:42  &    41:27:30  &  08:22:45  &    41:28:09  &  ugriz  &    0.486  &  0.477  &  $138 \pm   23$    &  0.487  &   $ 91  \pm  3$ &   $ 14.91 _{ -0.27 } ^{ +0.19 } \pm 0.03 $  &  $  4.9 \pm  1.7 $  \\
 24872   & 1\_5735  &  08:26:06  &    40:17:31  &  08:25:59  &    40:15:19 &  ugriz  &    0.402  &  0.391  &  $149 \pm   10$    &  0.407  &   $116  \pm  4$ &   $ 14.76 _{ -0.35 } ^{ +0.23 } \pm 0.02 $  &  $  5.4 \pm  1.3 $   \\
 24877   & 1\_5740  &  08:24:27  &    40:06:19  &  08:24:40  &    40:06:53 &  ugriz  &    0.592  &  0.539  &  $ 63 \pm   59$    &  0.593  &   $ 71  \pm  4$ &   $ 15.28 _{ -0.24 } ^{ +0.18 } \pm 0.03 $  &  $  4.9 \pm  2.0 $  \\
 24981   & 1\_5830  &  08:56:13  &    37:56:16  &  08:56:14  &    37:55:52 &  ugriz  &    0.411  &  0.411  &  $123 \pm   12$    &  0.407  &   $107  \pm  3$ &   $ 14.68 _{ -0.34 } ^{ +0.23 } \pm 0.02 $  &  $  7.6 \pm  1.9 $  \\
 25424   & 1\_6220  &  11:30:56  &    38:25:10  &  11:31:01  &    38:24:42 &  ugriz  &    0.509  &  0.513  &  $ 65 \pm   17$    &  0.510  &   $ 69  \pm  3$ &   $ 14.51 _{ -0.35 } ^{ +0.23 } \pm 0.02 $  &  $  5.5 \pm  2.1 $  \\
 25953   & 1\_6687  &  14:03:44  &    38:27:04  &  14:03:42  &    38:27:38 &  ugriz  &    0.478  &  0.484  &  $131 \pm   19$    &  0.478  &   $ 88  \pm  3$ &   $ 14.70 _{ -0.29 } ^{ +0.20 } \pm 0.03 $  &  $  4.6 \pm  1.3 $  \\
 27940   & 1\_7312  &  00:20:09  &    34:51:18  &   00:20:10  &    34:53:36 & ugriz  &    0.449  &  0.46   &  $116 \pm   24$    &  0.463  &   $ 89  \pm  3$ &   $ 14.82 _{ -0.31 } ^{ +0.21 } \pm 0.03 $  &  $  9.2 \pm  2.1 $   \\
 27974   & 2\_6669  &  00:08:51  &    32:12:24  &  00:08:55  &    32:11:12  &  ugriz  &    0.475  &  0.491  &  $100 \pm   25$    &  0.469  &   $ 75  \pm  3$ &   $ 14.74 _{ -0.26 } ^{ +0.19 } \pm 0.03 $  &  $  8.1 \pm  3.6 $  \\
 29283   & 1\_7697  &  08:04:35  &    33:05:08  &  08:04:36  &    33:05:27 &  ugriz  &    0.549  &  0.536  &  $129 \pm   30$    &  0.552  &   $107  \pm  3$ &   $ 15.02 _{ -0.30 } ^{ +0.21 } \pm 0.03 $  &  $  7.0 \pm  2.3 $  \\
 29284   & 1\_7698  &  08:03:30  &    33:01:47  &  08:03:30  &    33:02:06 &  ugriz  &    0.550  &  0.557  &  $122 \pm   33$    &  0.541  &   $ 68  \pm  3$ &   $ 14.50 _{ -0.61 } ^{ +0.31 } \pm 0.03 $  &  $  4.9 \pm  1.9 $  \\
 35361   & 1\_11298 &  14:56:11  &    30:21:04  &  14:56:13  &    30:21:12  &  ugriz  &    0.414  &  0.411  &  $103 \pm    9$    &  0.411  &   $ 98  \pm  3$ &   $ 14.76 _{ -0.23 } ^{ +0.18 } \pm 0.03 $  &  $  6.0 \pm  1.3 $  \\
 35399   & 1\_11334 &  15:03:03  &    27:54:58  &  15:03:10  &    27:55:00  &  ugriz  &    0.516  &  0.534  &  $153 \pm   31$    &  0.517  &   $ 81  \pm  3$ &   $ 14.87 _{ -0.26 } ^{ +0.19 } \pm 0.03 $  &  $  4.8 \pm  1.8 $  \\
 41843   & 1\_14643 &  23:40:45  &    20:52:04  &  23:40:45  &    20:53:02 &  ugriz  &    0.434  &  0.435  &  $119 \pm   13$    &  0.436  &   $ 75  \pm  3$ &   $ 14.52 _{ -0.48 } ^{ +0.28 } \pm 0.02 $   &  $  3.7 \pm  1.4 $  \\
 41911   & 1\_14706 &  00:23:01  &    14:46:57 &  00:23:01  &    14:46:31  &  ugriz  &    0.386  &  0.372  &  $104 \pm    7$    &  0.413  &   $ 81  \pm  3$ &   $ 14.84 _{ -0.25 } ^{ +0.19 } \pm 0.03 $  &  $  4.0 \pm  1.4 $   \\
 43403   & 1\_15084 &  08:10:18  &    18:15:18 &  08:10:20  &    18:15:13   &  ugriz  &    0.422  &  0.418  &  $130 \pm   10$    &  0.423  &   $ 94  \pm  3$ &   $ 14.94 _{ -0.23 } ^{ +0.17 } \pm 0.03 $  &  $  4.7 \pm  1.7 $  \\
 46649   & 1\_17215 &  01:35:17  &    08:47:50 &  01:35:17  &    08:48:14   &  ugriz  &    0.619  &  0.536  &  $ 85 \pm   31$    &  0.587  &   $128  \pm  5$ &   $ 15.13 _{ -0.23 } ^{ +0.18 } \pm 0.03 $   &  $ 14.2 \pm  4.6 $  \\
 47981   & 1\_17406 &  08:40:03  &    08:37:54 &  08:40:02  &    08:38:04  &  ugriz  &    0.543  &  0.551  &  $136 \pm   33$    &  0.540  &   $ 69  \pm  3$ &   $ 14.83 _{ -0.53 } ^{ +0.30 } \pm 0.02 $  &  $  6.9 \pm  2.6 $  \\
 50492   & 1\_18933 &  23:16:43  &    12:46:55 &  23:16:46  &    12:47:12   &  ugriz  &    0.527  &  0.524  &  $163 \pm   30$    &  0.525  &   $105  \pm  3$ &   $ 15.24 _{ -0.22 } ^{ +0.17 } \pm 0.03 $  &  $  7.4 \pm  2.2 $  \\
 50514   & 1\_18954 &  23:32:14  &    10:36:35 &  23:32:14  &    10:35:32  &  ugriz  &    0.466  &  0.463  &  $ 82 \pm   13$    &  0.475  &   $ 73  \pm  3$ &   $ 14.50 _{ -0.53 } ^{ +0.29 } \pm 0.03 $   &  $  3.6 \pm  1.3 $   \\
 52480   & 1\_19778 &  09:34:39  &    05:41:45 &  09:34:37  &    05:40:53 &  ugriz  &    0.565  &  0.546  &  $106 \pm   54$    &  0.542  &   $ 83  \pm  3$ &   $ 15.09 _{ -0.29 } ^{ +0.21 } \pm 0.02 $  &  $  7.6 \pm  2.4 $   \\
 54795   & 1\_21153 &  23:02:16  &    08:00:30 &  23:02:17  &    08:02:14  &  ugriz  &    0.428  &  0.429  &  $125 \pm   35$    &  0.435  &   $ 73  \pm  3$ &   $ 14.57 _{ -0.46 } ^{ +0.27 } \pm 0.03 $  &  $  5.9 \pm  1.6 $  \\
 55181   & 1\_21510 &  00:45:12  &   -01:52:32  &  00:45:10  &   -01:51:49 &  ugriz  &    0.547  &  0.542  &  $149 \pm   43$    &  0.532  &   $ 97  \pm  4$ &   $ 14.67 _{ -0.38 } ^{ +0.24 } \pm 0.03 $  &  $  5.9 \pm  2.3 $  \\
 59915   & 1\_23940 &  01:25:05  &   -05:31:05 &  01:25:01  &   -05:31:53  &  ugriz  &    0.475  &  0.489  &  $143 \pm   25$    &  0.472  &   $ 98  \pm  3$ &   $ 15.17 _{ -0.18 } ^{ +0.15 } \pm 0.03 $  &  $  3.9 \pm  1.4 $  \\
 64232   & 1\_24833 &  00:42:33  &   -11:01:58 &  00:42:32  &   -11:04:07  &  ugriz  &    0.529  &  0.529  &  $112 \pm   37$    &  0.553  &   $ 66  \pm  3$ &   $ 14.53 _{ -0.72 } ^{ +0.34 } \pm 0.02 $   &  $  4.8 \pm  1.8 $  \\
\hline \hline
\end{tabular}
\end{adjustbox}
\end{table*}
\subsection{Mass likelihood}
\label{sec:likelihood}
The lensing likelihood for an individual cluster is proportional to the probability of observing the present mean $\Delta\Sigma$ given a true cluster mass $M=M_{\rm{200c}}$. Assuming multivariate Gaussian errors in the observed signal, it can be written as
\begin{equation}
p(\Delta\Sigma|M)\propto\frac{1}{\sqrt{\det C(M)}}\times\exp\left(-\frac{1}{2}\bm{E}(M)^{\rm T} C^{-1}(M)\bm{E}(M)\right) \; ,
\label{eqn:likelihood}
\end{equation}
where $\bm{E}$ is the vector of residuals between data and model evaluated at mass $M$, 
\begin{equation}
E_i(M)=\Delta\Sigma^{\rm obs}_i-\Delta\Sigma^{\rm model}_i(M) \; ,
\end{equation}
and $C$ is the covariance matrix (cf. equation~\ref{eqn:covcomponents}). The mass dependence of the covariance, due entirely to $C_{\rm intr}$, causes a complication relative to a simple minimum-$\chi^2$ analysis: the normalization of the Gaussian PDF depends on mass that needs to be accounted for by the $\det^{-1/2}C(M)$ term in equation~\ref{eqn:likelihood}. If the covariance is modelled perfectly, including the mass dependence, the above is the correct likelihood (see e.g.~\citealp{kodwani19}). If, however, the mass dependence of the covariance is modeled with some statistical or systematic uncertainty, the $\det^{-1/2}C(M)$ term can cause a bias in the best-fit masses.

For this reason, we use a two-step scheme:
\begin{enumerate}
\item determine the best-fit mass using a covariance that consists of shape noise and LSS contributions only, i.e. has no mass dependence
\item evaluate $C^{\rm intr}$ at the best fit mass of step (i), add this to the covariance without mass dependence and repeat the likelihood analysis with this updated, full, yet mass-independent covariance.
\end{enumerate}

\section{Hierarchical Bayesian model}
\label{sec:bayesmodel}

Below we describe the hierarchical Bayesian model, which we use to determine the posterior distribution of the parameters of interest. The following section follows a similar framework as in \cite{Nagarajan_2018} and \cite{Mulroy2019}, except, instead of one selection function, we introduce two separate selection functions, the CODEX selection function and the sampling function, for our lensing subsample. 

The true underlying halo mass of the cluster $i$ in log-space $\mu_i = \ln(M_i)$ is related to all other observables through a scaling model $P(\bs_i, \mu_i | \theta)$, where $\bs_i = \ln(\bS_i)$ is the vector of true quantities in log-space and $\btheta$ represents a vector of parameters of interest. At given redshift, the joint probability distribution that there exist a cluster of mass $\mu_i$ can be written as 
\begin{equation}
P(\bs_i,\mu_i | \btheta, z_i) = P(\bs_i | \mu_i, \btheta)P(\mu_i | z_i)P(z),
\end{equation}
where we model the conditional distribution for the mass at given redshift $z_i$, $P(\mu_i |z_i)$, as the halo mass function (HMF) $\frac{dn}{d \ln m}(\mu_i | z_i)$ and $P(z)$ is the comoving differential volume element $dV/dz(z)$. In practice, $P(\mu_i |z_i)$ is evaluated as a \citet{tinker08} mass function using fixed $\Lambda$CDM cosmology, where $\Omega_m = 0.27$, $\Omega_{\Lambda} = 0.73$, $\Omega_b = 0.049$, $H_0 = 70$ km\, s$^{-1}$\, Mpc$^{-1}$, $\sigma_8 = 0.82$, $n_s = 0.962$, for a density contrast of $200 \times \rho_{c}$. 

The underlying true values of the observables in log-space $\bs_i$ are assumed to come from a multivariate Gaussian distribution:
\begin{equation}
P(\bs_i | \mu_i, \btheta) \propto \det (\bSigma_i^{-1/2}) \exp \bigg[-\frac{1}{2}(\bs_i - \langle \bs_i \rangle)^T \bSigma_i^{-1} (\bs_i - \langle \bs_i \rangle)\bigg],
\end{equation}
where the mean of the probability distribution of observables is modelled as a linear function in log-space $ \langle \bs_i \rangle = \balpha \mu_i + \bbeta $. The model parameters are defined as $\btheta = \{  {\balpha}, \bbeta, \bSigma_i  \}$, where $\balpha$ is the vector of slopes, $\bbeta$ is vector of intercepts and $\bSigma$ is the intrinsic covariance matrix of the cluster observables at fixed mass. The diagonal elements of the intrinsic covariance matrix, $\sigma_{\ln s_i | \mu}$, represent the intrinsic scatter for a cluster observable $s_i$ at fixed mass. The off-diagonal elements are the covariance terms between different cluster observables at fixed mass. 

However, we cannot directly access cluster observables, but only have estimates through observations, which contain observational uncertainties. We denote the observed logarithmic quantities with tilde: $\tilde{\bs}_i, \tilde{\mu}_i, \tilde{z}_i$, and the vector of all observables as $\tilde{\bf{o}} \in \{\tilde{\bs}_i, \tilde{\mu}_i, \tilde{z}_i\}$. 
To connect them to their respective underlying true observables $\bf{o} \in \{ s_{i}$, $\mu_i$, $z_i \}$, we assume the full lensing likelihood from equation \ref{eqn:likelihood} for the mass, which we denote here $P(\tilde{\mu}_i$|$\mu_i)$, and, for other parameters, a multivariate Gaussian distribution $P(\tilde{\bs}_i, \tilde{z}_i | \bs_i, z_i) $,
which acts as our measurement error model:
\begin{equation}
P(\tilde{\bs}_i, \tilde{z}_i | \bs_i, z_i) \propto \det (\tilde{\bSigma}_i^{-1/2}) \exp \bigg[-\frac{1}{2}(\tilde{\bs}_i - \bs_i )^T \tilde{\bSigma}_i^{-1} (\tilde{\bs_i} - \bs_i )\bigg].
\label{eq:measurement_error_model}
\end{equation}

The diagonal elements of the covariance matrix in equation \ref{eq:measurement_error_model} represent the relative statistical errors in the observables for cluster $i$ and the off-diagonal elements the covariance between the relative errors of different observables. In practice, instead of using the evaluated richness measurement errors from the redMaPPer algorithm, we assume a Poisson noise model, described further in equations \ref{eq:sdss} and \ref{eq:cfht}. For simplicity, for a single cluster, we expect independent measurement errors between different observables.

For the total population, the probability of measuring the observed cluster property $\tilde{\bs}_i$ for a single cluster $i$ at fixed observed mass $\tilde{\mu}_i$ and observed redshift $\tilde{z}_i$, can be expressed as
 \begin{equation}
 \label{eq:prob_single_cluster_population}
  \begin{split}
    P(\tilde{\bs}_i, \tilde{\mu}_i, \tilde{z}_i | \btheta ) = &\int d\bs_i \int d\mu_i \int dz_i P(\tilde{\bs}_i, \tilde{z}_i | \bs_i, z_i) \\
    \ \cdot &P(\tilde{\mu}_i | \mu_i)P(\bs_i | \mu_i, \btheta)P(\mu_i | z_i)P(z).
  \end{split}
 \end{equation}
Note that in equation \ref{eq:prob_single_cluster_population}, we have to marginalize over all the unobserved cluster properties, i.e., underlying halo mass, true observables and true redshift.

In reality, one cannot directly observe the full population of clusters, but a subsample of it based on some easily observable cluster property, such as luminosity or richness of the cluster. In order to rectify the bias coming from the observed censored population, one has to include the selection process in the model. If the selection is done several times with different observables, e.g., taking a subsample from a sample that represents the population, one should introduce all different selection processes into the modelling.

In order to introduce a selection effect into the Bayesian modelling, we define a boolean variable for the selection $I$, which we will use as a conditional variable to specify whether a cluster is detected or not. 
Let's first consider a single selection variable $\tilde{\lambda}$. Assume we have made a cut at $\tilde{\lambda}$, and we observe all the clusters above this limit. Then $P(I=1 | \tilde{\lambda} \ge \mathrm{cut}) = 1$ for all observed clusters, and $P(I=0 | \tilde{\lambda} < \mathrm{cut}) = 0$, for all unobserved clusters.

However, if we don't detect all the clusters above the cut, just a subsample of clusters, but know how many clusters we miss, we can calculate the fraction of clusters from the subsample that belong to the sample at certain richness $f(\tilde{\lambda}_{i,\mathrm{sub}})=\tilde{N}_{\mathrm{sub}}/\tilde{N}_{\mathrm{sample}}(\tilde{\lambda}_{i, \mathrm{sub}})$, and treat this fraction as our subsample detection probability, for which $P(I=1 | \tilde{\lambda}_{i,\mathrm{sub}}) = f(\tilde{\lambda}_{i,\mathrm{sub}}) \leq 1$. We note that $f$ returns to the heaviside step function, if we observe all the clusters above the cut $\tilde{\lambda}$. Below, we generalize the selection probability $P(I | \tilde{\bf{o}}_i, \btheta)$ by considering any selection function to depend on multiple selection variables $\tilde{\bf{o}}_i$, and the vector of parameters of interest $\btheta$.

Using the Bayes' theorem, the probability of measuring the observed cluster properties $\tilde{\bf{o}}_i$, given fixed vector of parameters $\btheta$ and that the cluster passed the selection is
\begin{equation}
\label{eq:prob_w_sel}
    P(\tilde{\bf{o}}_i | I, \btheta) = \frac{P(I | \tilde{\bf{o}}_i, \btheta)P(\tilde{\bf{o}}_i | \btheta ) }{P(I | \btheta)},
\end{equation}
where $P(I | \tilde{\bf{o}}_i , \btheta)$ quantifies the probability of detecting a single cluster, and $P(I | \btheta)$, is the overall probability for all the clusters to be selected, which can be evaluated by marginalizing over the observed cluster properties from the numerator in equation \ref{eq:prob_w_sel}:
\begin{equation}
    P(I | \btheta) = \int d\tilde{\bf{o}}_i P(I | \tilde{\bf{o}}_i, \btheta)P(\tilde{\bf{o}}_i | \btheta).
\end{equation}

In the case where the selection depends on \textit{both} observed and true quantities, equation \ref{eq:prob_w_sel} becomes, according to Bayes theorem:
\begin{equation}
    \label{eq:prob_obs_unobs}
    P(\tilde{\bf{o}}_i, | I_{\mathrm{tot}}, \btheta)
    =
    P(\tilde{\bf{o}}_i, | I_{\mathrm{obs}}, I_{\mathrm{true}}, \btheta) = \frac{P(I_{\mathrm{obs}} | \tilde{\bf{o}}_i, \btheta) P(I_{\mathrm{true}}, \tilde{\bf{o}}_i | \btheta ) }{P(I_{\mathrm{obs}}, I_{\mathrm{true}} | \btheta)} ,
\end{equation}
where we have introduced a second selection parameter $I_{\mathrm{true}}$, that denotes the selection based on true quantities. The first term is the same selection function $P(I | \tilde{\bf{o}}_i, \btheta)$
as in equation \ref{eq:prob_w_sel}, and the second term in the numerator can be expressed as 
\begin{equation}
\label{eq:sel_unobs}
\begin{split}
    P(I_{\mathrm{true}}, \tilde{\bf{o}}_i | \btheta ) = 
&\int d\bs_i \int d\mu_i P(I_{\mathrm{true}} | \bs_i, \mu_i ) \\  \ \cdot &P(\tilde{\bs}_i, \tilde{\mu}_i | \bs_i, \mu_i)P(\bs_i, \mu_i | \btheta).
\end{split}
\end{equation}

Equation \ref{eq:prob_single_cluster_population} is assumed to work only if no censoring is involved, but equation \ref{eq:sel_unobs} assumes that the observed set belongs to a larger population, and the selection $P(I_{\mathrm{true}} |\bs_i, \mu_i)$ can be modelled with simulations, where the true observables are known. In section \ref{sec:xray_selection}, we introduce the CODEX X-ray selection, $P(I_X | \bs_i, \mu_i)$, which is defined as a function of true observables.

The normalization of the likelihood function in equation \ref{eq:prob_obs_unobs} can also be expressed as an integral over all observables:
\begin{equation}
    P(I_{\mathrm{obs}}, I_{\mathrm{true}} | \btheta) = \int d\tilde{\bf{o}}_i P(I_{\mathrm{obs}} | \tilde{\bf{o}}_i, \btheta) P(I_{\mathrm{true}}, \tilde{\bf{o}}_i | \btheta).
\end{equation}

Finally, the full likelihood function for the subsample, with the inclusion of the selection effects, becomes a product of the single cluster likelihood functions from equation \ref{eq:prob_obs_unobs}:
\begin{equation}
    \mathcal{L}( \tilde{\bf{o}}_{N} | \btheta ) = \prod_{i=1}^N 
    P(\tilde{\bf{o}}_i | I_{\mathrm{tot}}, \btheta ),
\end{equation}
where subscript N denotes the full vector of observed measurements from all the clusters. The full posterior distribution, which describes the probability distribution of parameters of interest, given the observed mass, redshift and set of observables is then 
 \begin{equation}
     P(\btheta | \tilde{\bs}_{N}, \tilde{\mu}_{N}, \tilde{z}_N) \propto \pi(\btheta)\mathcal{L}( \tilde{\bs}_{N}, \tilde{\mu}_{N}, \tilde{z}_N | \btheta ),
 \end{equation}
where $\pi(\btheta)$ describes the prior knowledge of the parameters.

\section{Application to the CODEX weak lensing sample}
\label{sec:application}

We apply the above described Bayesian method to the lensing sample S-I, and exclude eleven clusters: CODEX ID 53436 and 53495 as they are missing both CFHT richness and weak lensing information; 37098 as it is missing weak lensing information; 13390, 29811 and 56934 as they are missing CFHT richness information; CODEX ID 13062 (griz) and 35646 (griz) as we only employed our method to clusters measured with five filters (ugriz); CODEX ID 12451, 18127 and 36818 as their CFHT richness are below the 10\% CODEX survey completeness limit, which is further described in section \ref{sec:optical_selection}. 

We aim to constrain both the intrinsic scatter in richness and the scaling relation parameters describing the richness-mass relation, see equation \ref{eq:M-l}. For that we fit a model of richness-mass relation to CFHT richness estimates and weak lensing mass likelihood (see Table \ref{tab:primaryWL} for CFHT richness estimates).
We don't fit for the SDSS richness-mass relation as the SDSS richness estimates have mean relative uncertainty of $\sim20 \%$, in contrast to CFHT richness mean relative uncertainty of $\sim4 \%$. However, since the lensing sample of 25 clusters, i.e., a subsample of the initial CODEX sample, is based purely on observability, such that not all clusters above the $\tilde{\lambda}_{\mathrm{SDSS}} = 60$ cut are observed, we use the fraction of SDSS richnesses $P(I = 1 | \ln \tilde{\lambda}_{\mathrm{SDSS}})$ as our subsample selection function, and treat the SDSS richness in our likelihood function as one of the selection variables, which we will marginalize over. As for CFHT and SDSS richnesses, we assume both are coming from the similar log-normal richness distribution, i.e., $P(\ln \tilde{\lambda} | \ln \lambda) = \mathcal{N}(\ln \tilde{\lambda} ; \ln \lambda,  \sigma_{\ln \lambda  })$, but with somewhat larger scatter for the SDSS richness, which is described below.  

The relation between underlying true richness and true mass of the cluster is assumed to be a Gaussian distribution in logarithmic space, with the mean of this relation given by the logarithm of a power-law:
\begin{equation}
\langle \ln\lambda_i | \mu_i \rangle = \alpha \mu_i +  \beta,
\label{eq:M-l}
\end{equation}
where we have defined $\mu_i \equiv \ln(M_i/M_{\mathrm{piv}})$ with pivot mass set to $M_{\mathrm{piv}} = 10^{14.81} \Msol$, i.e., the median mass of the lensing subsample. The model parameters of interest, $\alpha$ and $\beta$, describe the scaling relation slope and intercept, respectively. This parametrization follows \citet{saro2015}. We write the full scatter in $\tilde{\lambda}_{\mathrm{SDSS}}$ as the sum in quadrature of a Poisson and an intrinsic variance terms. Thus, the total variance in observed SDSS richness at a fixed true mass $\mu_i$ can be written as \citep{spiders2018}:
\begin{equation}
\label{eq:sdss}
\sigma^2_\mathrm{tot, SDSS}(\ln \lambda_i| \mu_i) = \frac{\eta(z_i)}{\exp{\left \langle \ln\lambda_i | \mu_i \right \rangle}} \\
+\sigma^2_{\ln \lambda|\mu, \mathrm{intr}} \ ,
\end{equation}
where $\sigma^2_{\ln \lambda|\mu, \mathrm{intr}}$ is the third free parameter of our model.
As described in \citet{spiders2018}, a redshift dependent correction factor $\eta(z)$ is estimated for high redshift clusters to remedy the effect that the SDSS photometric data is not deep enough to correctly measure the richness after a certain magnitude limit is reached. As the CFHT photometric richnesses come from a sufficiently deep survey, we can set the survey depth correction factor to unity, so that the total variance in CFHT richness can be modelled as:
\begin{equation}
\label{eq:cfht}
\sigma^2_\mathrm{tot, CFHT}(\ln \lambda_i| \mu_i) = \frac{1}{\exp{\left \langle \ln\lambda_i | \mu_i \right \rangle}} + \sigma^2_{\ln \lambda|\mu, \mathrm{intr}} \ .
\end{equation}

We also test the Poisson term in terms of true richness, in contrast to mean richness, and the difference between these two error estimation methods are negligible. 

For the observed mass estimation, we use the single cluster mass likelihood function $P(\tilde{\mu} | \mu)$, from equation \ref{eqn:likelihood}. We introduce a fourth scalar parameter, $l_{\mathrm{sys}}$ with standard normal distributed prior, to draw how different the noiseless logarithmic lensing masses are from the true logarithmic masses due to imperfect calibration of lensing shapes, redshifts, and the cluster density profiles.  

We assume that the observed spectroscopic redshift is close to the true redshift of the cluster, i.e., we model the term $P(\tilde{z} |z)$ as a delta function. 

In the case the sample is only limited by observed richness $\tilde{\lambda}_i$, with the calibration of the richness-mass scaling relation based on weak lensing data, the probability distribution can be written according to equation \ref{eq:prob_w_sel}. The initial CODEX sample contains both optical and X-ray selection. The X-ray selection requires the inclusion of the CODEX selection function, replacing equation \ref{eq:prob_w_sel} with equation \ref{eq:prob_obs_unobs}. 

\subsection{Optical selection functions} 
\label{sec:optical_selection}
We consider two separate optical selection functions below that account for optical cleaning and incompleteness of the survey.
We describe by $P(I_{\mathrm{clean}} |  \tilde{\lambda}, \tilde{z})$ the optical cleaning applied to the catalog. In practice, this is a redshift dependent cut in observed richness used to minimize false X-ray sources while keeping as many true systems as possible. For the CODEX survey, this redshift cut is chosen by the $10 \%$ sensitivity limit. We adopt the 10\% CODEX sensitivity limit
\begin{equation}
    P(I_{\mathrm{clean}} | \tilde{\lambda}, \tilde{z}) =  \left\{
\begin{array}{ll}
   1,&  \mathrm{if} \tilde{\lambda} > 22\left(\frac{\tilde{z}}{0.15}\right)^{0.8} \\
   0,& \mathrm{otherwise}.
   \end{array} 
\right.
\end{equation}
from \citet{finoguenov2019codex} to CFHT richnesses to only account for clusters which have richness completeness over 10\%. This cut excludes three clusters from S-I (CODEX ID 12451, 18127, and 36818).

We also consider the $50 \%$ SDSS richness completeness boundary:
\begin{equation}
\begin{split}
    \ln \lambda_{50 \%}(z) = \ln\left(17.2 +  \exp\left(\frac{z}{0.32}\right)^2\right)
\end{split}
\end{equation}
i.e., clusters with SDSS richness above these limits have at least 50\% completeness, respectively. 
We include the 50\% SDSS richness completion as an optical selection function 
\begin{equation}
    P(I_{\mathrm{opt}} | \ln \lambda) = 1-\frac{1}{2}\mathrm{erfc}\left(\frac{\ln \lambda - \ln \lambda_{50\%}}{\sqrt{2}\sigma}\right)
\end{equation}  
in the likelihood function with a scatter of $\sigma=0.2$, as described in \citet{finoguenov2019codex}. 
This term accounts for incompleteness due to limited photometric depth of the SDSS survey causing a fraction of clusters to go unobserved. 

\subsection{X-ray selection function}
\label{sec:xray_selection}
Details of the CODEX selection function are given in \citet{finoguenov2019codex}. The CODEX selection function $P(I_X | \mu, \mathrm{z}, \nu)$ provides an effective survey area at a given mass, redshift, and deviation from the mean richness at fixed mass $\nu \equiv \frac{\ln \lambda_i - \left<\ln \lambda | \mu_i \right>}{\sigma_{\ln\lambda}^\mathrm{intr}}$, which accounts for the covariance between scatter in richness and X-ray luminosity. The limits for $\nu$ is fixed between $\pm 4$. In the modelling the CODEX selection function, the $L_x$-mass scaling relations are fixed to those by the XMM-XXL survey (\citealt{lieu16, giles}), but the richness-mass relation is not modelled explicitly in the selection function, only the covariance between richness and luminosity. 
For the selection function modelling, the covariance coefficient is fixed to $\rho_{\mathrm{L}_\mathrm{X} - \lambda} = -0.3$, which is based on results from \citet{farahi}. In this work, the CODEX selection function is evaluated at fixed cosmology with $\Omega_m = 0.27$. 
The formulation of selection function allows us to propagate these effects into the full selection function. 

As the CODEX selection function depends on $\nu(\lambda, \left<\ln \lambda \right>)$, and the mean richness in $\nu$ depends on scaling relation parameters, we can simplify the likelihood function by evaluating it in $\nu$-space instead of in $\lambda$-space. In $\nu$-space, equation \ref{eq:sel_unobs} can be rewritten as
\begin{equation}
\begin{split}
    P(I_X, \ln \tilde{\lambda}, \tilde{\mu}, \tilde{z} | \btheta) = &\int d\nu \int d\mu \int dz P(I_X | \mu, \nu, z) P( \tilde{\mu} | \mu)P(\tilde{z} | z) \\
    &\ \cdot P(\ln \tilde{\lambda} | \nu, \theta, \mu)P(\nu)P(\mu | z)P(z), \\
\end{split}
\end{equation}
which is the probability of observing a full sample with the inclusion of CODEX selection. However, we are dealing with a subsample, which gets selected with the sampling function, described below.

\subsection{Subsample selection function}
\label{sec:subsample_selection_function}
\begin{figure}
\includegraphics[width=8cm]{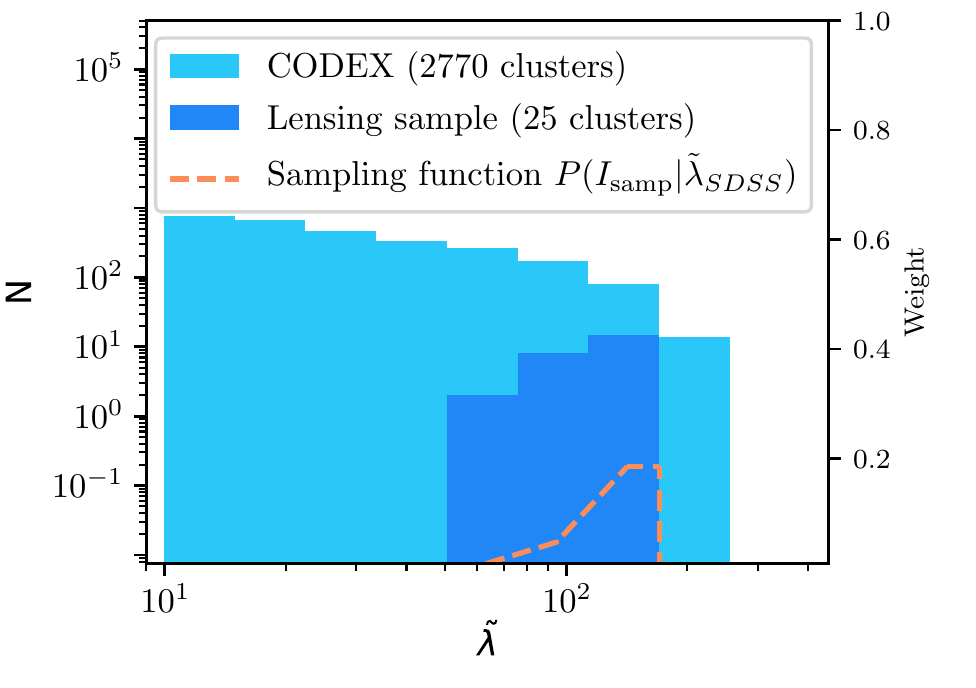}
\caption{SDSS richness distributions of CODEX sample and lensing sample, from which the sampling function (weight as a function of observed richness) is derived.}
\label{fig:optical_sel}
\end{figure}

For evaluating the sampling function, based on SDSS richness, we use the initial CODEX sample (407 clusters, three light blue bins behind the three dark blue bins in Fig. \ref{fig:optical_sel}) and its subsample (25 clusters, three dark blue bins in Fig. \ref{fig:optical_sel}). 
We bin both the initial sample and the subsample, the lensing sample, into equal bin widths and evaluate the ratio of the height of the bins. We then fit a linear piecewise function between the mean of the bins, which becomes our sampling function that depends on observed SDSS richness, depicted by the orange curve in Fig. \ref{fig:optical_sel}. 

The sampling function has the following form:
\begin{equation}
    P(I_\mathrm{samp} |§Kii \tilde{\lambda}_{\mathrm{SDSS}}) =  \left\{
\begin{array}{ll}
      0 & \tilde{\lambda} < 60 \\
      \frac{1}{1000}(\tilde{\lambda} - 60) + \frac{7}{1000} &  60 \leq \tilde{\lambda} < 91 \\
      \frac{33}{1000}(\tilde{\lambda} - 91) + \frac{38}{1000} &  91 \leq \tilde{\lambda} < 136 \\
      \frac{186}{1000} & 136 \leq \tilde{\lambda} \leq 163, \\
\end{array} 
\right.
\end{equation}
where $\tilde{\lambda} \equiv \tilde{\lambda}_{\mathrm{SDSS}}$.

As the clusters in the 407 cluster initial sample has cut at $\tilde{\lambda}_{\mathrm{SDSS}} \ge 60$, the sampling function defines a null probability for clusters below this cut.
Since the lensing sample, a subsample of the initial sample, is selected based only by observability, some of the clusters in the initial sample above the richness cut are unobserved, the sampling function differs from a typical heaviside step function.

The sampling function depends only on SDSS richness, which we can consider as an effective richness. We introduce an additional Gaussian distribution $P(\ln \tilde{\lambda}_{SDSS} | \ln \lambda)$ to account for the connection between SDSS richness and true richness and marginalize the likelihood function over the SDSS richness.

\subsection{Full data likelihood function}
Included for completeness is the full likelihood function in $\nu$-space that we use to constrain the parameters of interest $\theta = \{\alpha, \beta, \sigma_{\ln \lambda}^{\mathrm{intr}}\}$:
\begin{equation}
\begin{split}
    \mathcal{L} = \prod_{i=1}^{N}\phi(I_X, I_{\mathrm{samp}}, I_{\mathrm{opt}} | \theta)^{-1} &\int d\nu_i \int d \mu_i \int d\ln \tilde{\lambda}_{i,SDSS} \\
    \ \cdot &P(I_\mathrm{samp} | \ln \tilde{\lambda}_{i,SDSS}) \\
    \ \cdot&P(I_X | \mu_i, \tilde{z}_i, \nu_i) \\ 
    \ \cdot&P(I_{\mathrm{opt}} | \nu_i, \mu_i, \theta  ) \\
    \ \cdot&P(\ln \tilde{\lambda}_{i,SDSS} | \nu_i, \mu_i, \theta, \tilde{z}_i) \\
    \ \cdot&P(\ln \tilde{\lambda}_{i,CFHT} | \nu_i, \mu_i, \theta) \\ 
    \ \cdot&P(\tilde{\mu}_i | \mu_i ) \\ 
    \ \cdot&P(\nu_i) \\
    \ \cdot&P(\mu_i,\tilde{z}_i),  \\
    \end{split}
\end{equation}
where the normalization of the likelihood is :
\begin{equation}
\begin{split}
\phi(I_X, I_{\mathrm{samp}}, I_{\mathrm{opt}} | \theta) =& \int d\nu \int d\mu \int d\ln\tilde{\lambda}_{SDSS} \int d\tilde{z} \\
\ \cdot&P(I_{\mathrm{samp}} | \ln \tilde{\lambda}_{SDSS}) \\ 
\ \cdot&P(I_X | \mu, \tilde{z}, \nu) \\
\ \cdot&P(I_{\mathrm{opt}} | \nu, \mu, \theta  ) \\
\ \cdot&P(\ln \tilde{\lambda}_{SDSS} | \nu, \mu, \theta, \tilde{z}) \\
\ \cdot&P(\nu)  \\
\ \cdot&P(\mu,\tilde{z}).
\end{split}
\end{equation}
The subscript $i$ is omitted in the normalization as it is identical for all clusters. We note, that the full likelihood function incorporates three of the four selection effects: X-ray selection $P(I_X | \mu_i, \tilde{z}_i, \nu_i) $, to account for covariance between X-ray cluster properties with richness, optical selection $P(I_{\mathrm{opt}} | \nu_i, \mu_i, \theta  )$, to account for the incompleteness of the SDSS richness, and the sampling function $P(I_{\mathrm{samp}} | \ln \tilde{\lambda}_{SDSS})$, to account for the fact that we analyse a subsample of the initial CODEX sample. We don't include the fourth selection function, the optical cleaning function $P(I_{\mathrm{clean}} |  \tilde{\lambda}, \tilde{z})$ in the data likelihood, as it is only used to make the redshift dependent cut, removing cluster ID 12451, 18127, and 36818 from the S-I sample. 

\section{Results and Discussion}
\label{sec:results}
\begin{figure}
\includegraphics[width=8cm]{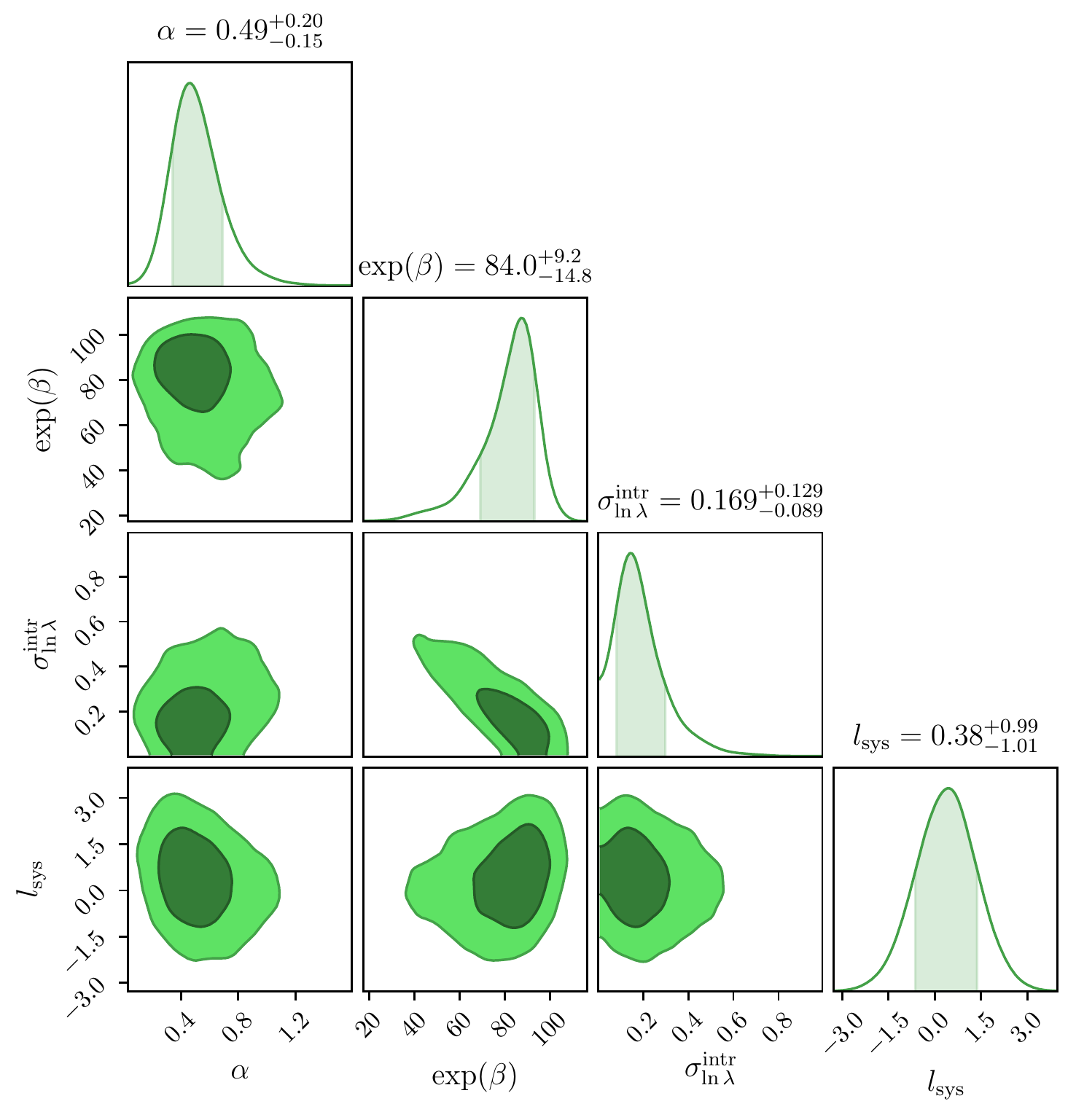}
\caption{Result from the MCMC fitting, with the one and two dimensional projections of the posterior distributions for the CFHT samples. Contours indicate the statistical 1$\sigma$ ($68 \%$) , and 2$\sigma$ ($95 \%$)  credible regions.}
\label{fig:cfht-sdss-corner}
\end{figure}

\begin{table*}
    \caption{Summary of measured parameters, their initial values, priors and posteriors. The initial parameter values for each of the 24 random walkers in the MCMC run are randomly drawn around a circle with the center value listed in the Initial column and with radius $10^{-2}$. This way all walkers start to scan the parameter space at slightly different initial position.
    }
  \begin{center}
    \begin{tabular}{ccccc} \hline\hline
      \centering
      Parameter & Initial & Prior & Posterior\\
\hline \hline
$\alpha$ & 0.98 & ${\rm flat}(0, 1.6)$ & $0.49^{+0.20}_{-0.15}$ & \\
$\beta$ & 3.68 & ${\rm flat}(0, 6)$ & $4.42^{+0.13}_{-0.20}$ & \\
$\sigma_{\ln \lambda}^{\mathrm{intr}}$ & 0.22 & ${\rm flat}(0, 1)$ & $0.17^{+0.13}_{-0.09}$ \\
$l_{\mathrm{sys}}$ & 0.0 & ${\mathcal{N}}[0, 1]$ & $0.38_{-1.01}^{+0.99}$ & \\

      \hline
\end{tabular}
  \end{center}
  \begin{tablenotes}
\item[1] $\alpha$ is the mass slope of the richness--mass relation $\langle \ln \lambda | \mu \rangle = \alpha\mu + \beta$. 
\item[2] $\beta$ is intercept (normalization) of the richness--mass relation. 
\item[3] $\sigma_{\ln \lambda}^{\mathrm{intr}}$ is the intrinsic scatter in richness, which quantifies how much true richness at given mass scatters from the mean.
\item[4] $l_{\mathrm{sys}}$ is a scalar lensing systematic parameter. It is used to draw how different the noiseless log lensing masses are from the log true masses due to imperfect calibration of lensing shapes, redshifts, and the cluster density profiles. 
\end{tablenotes}
\label{tab:results}
\end{table*}
\begin{table*}
\caption{Scaling relation parameter comparison to literature. The credible intervals refers to $1\sigma$ ($68 \%$)  statistical uncertainties.}
\label{tab:scal_rel_comparison}
\begin{tabular}{c|c|c|c|}
\hline \hline
Bayesian analysis results     &  Intercept & Slope & Scatter  \\
     &  $\lambda_0 = \exp{(\beta)}$ & $\alpha$& $\sigma_{\ln \lambda}^{\mathrm{intr}}$  \\
\hline
CODEX lensing sample  & $84.0^{+9.2}_{-14.8} $ & $0.49^{+0.20}_{-0.15}$ & $0.17^{+0.13}_{-0.09}$ \\
\hline \hline
Previously published results   &  $\lambda_0(10^{14.81}M_{\odot}, z=0.5)$ & $M_{200c}^{\alpha} $ & $\sigma_{\ln \lambda}^{\mathrm{intr}}$ \\
\hline
LoCuSS prediction \textbf{\citep{Mulroy2019}} & $93.66 \pm 7.43$ & $0.74 \pm 0.06$ & $0.24 \pm 0.05$  \\
SPIDERS prediction \textbf{\citep{spiders2018}}& $65.10 \pm 7.21$ & $0.98 \pm 0.07$ & $0.22^{+0.08}_{-0.09}$ \\
SPTpol prediction \textbf{\citep{Bleem_2020}}& $79.15 \pm 8.30$ & $1.02 \pm 0.08$ & $0.23 \pm 0.16$ \\
DES Y1 prediction \textbf{\citep{desy1}}& $70.66 \pm 2.55$ & $0.73 \pm 0.03$ & $-$ \\
\hline \hline
\end{tabular}
\end{table*}

\begin{figure*}
  \centering
  \includegraphics[width=\textwidth]{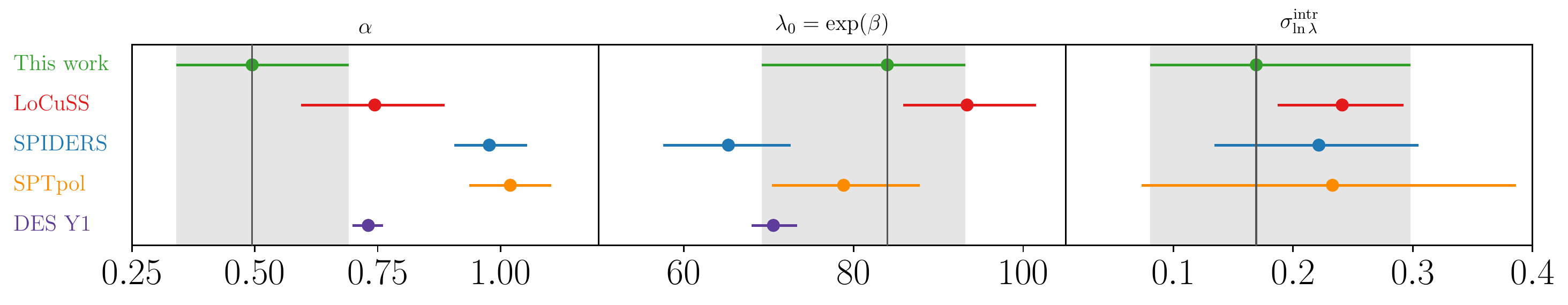}
  \caption{Comparison between the predicted richness and other results from the literature. The predicted richnesses are evaluated at $M_{200c} = 10^{14.81} M_{\odot}$ and $z = 0.5$. Gray bands denote the statistical $1\sigma$ $(68 \%)$ uncertainty of this work. For the DES Y1 analysis, the intrinsic scatter and its $1\sigma$ uncertainty is not shown, as it is not constrained in their work.
 }
\label{fig:params}
\end{figure*}

\begin{figure}
\includegraphics[width=8cm]{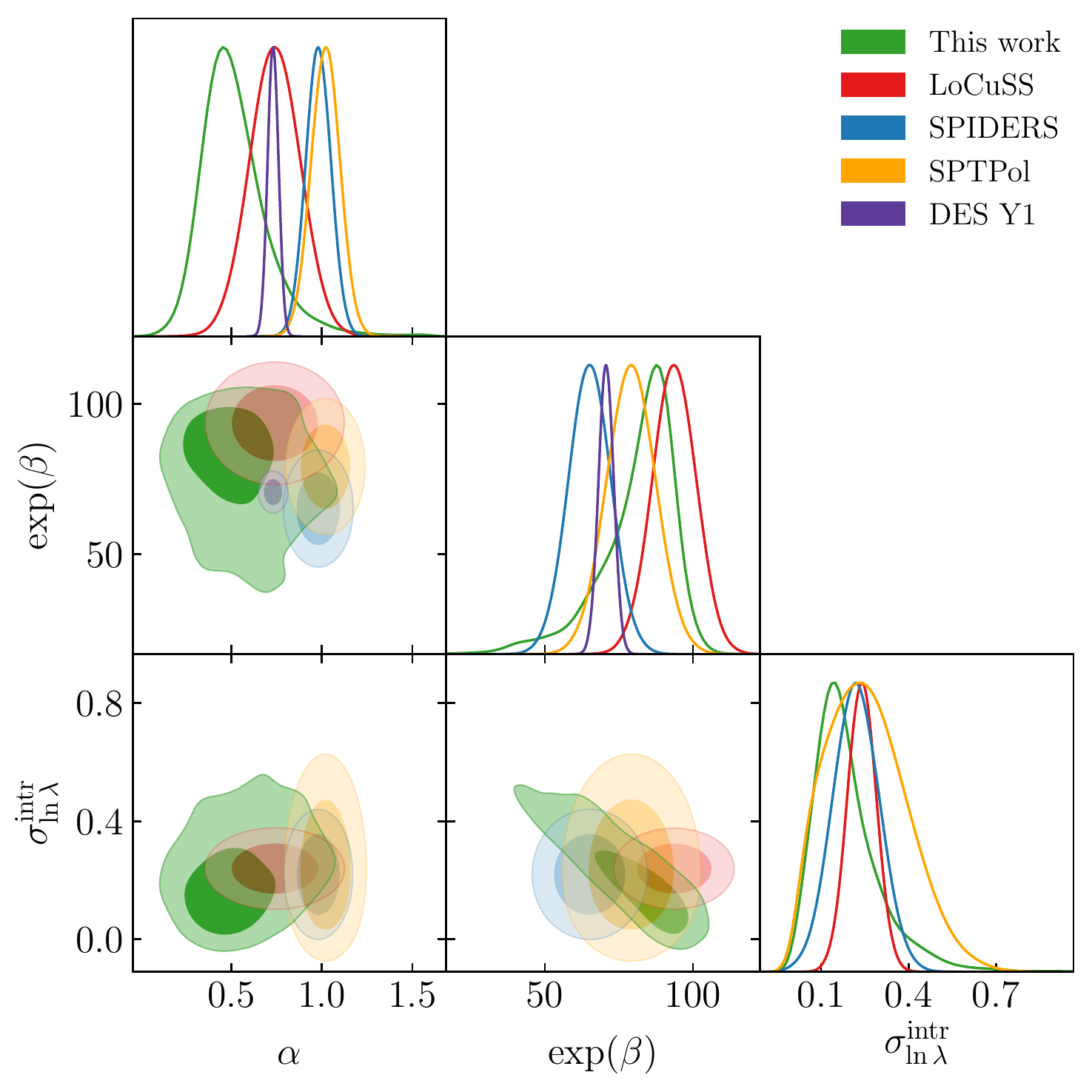}
\caption{Identical MCMC fitting results as in Fig. \ref{fig:cfht-sdss-corner}, but with the inclusion of the scaling relation results from the literature, rescaled at $M_{200c} = 10^{14.81} M_{\odot}$ and $z=0.5$. Contours indicate the statistical 1$\sigma$ ($68 \%$) , and 2$\sigma$ ($95 \%$)  credible regions.}
\label{fig:cfht-corner-comparison}
\end{figure}

\begin{figure}
  \centering
  \includegraphics[width=8cm]{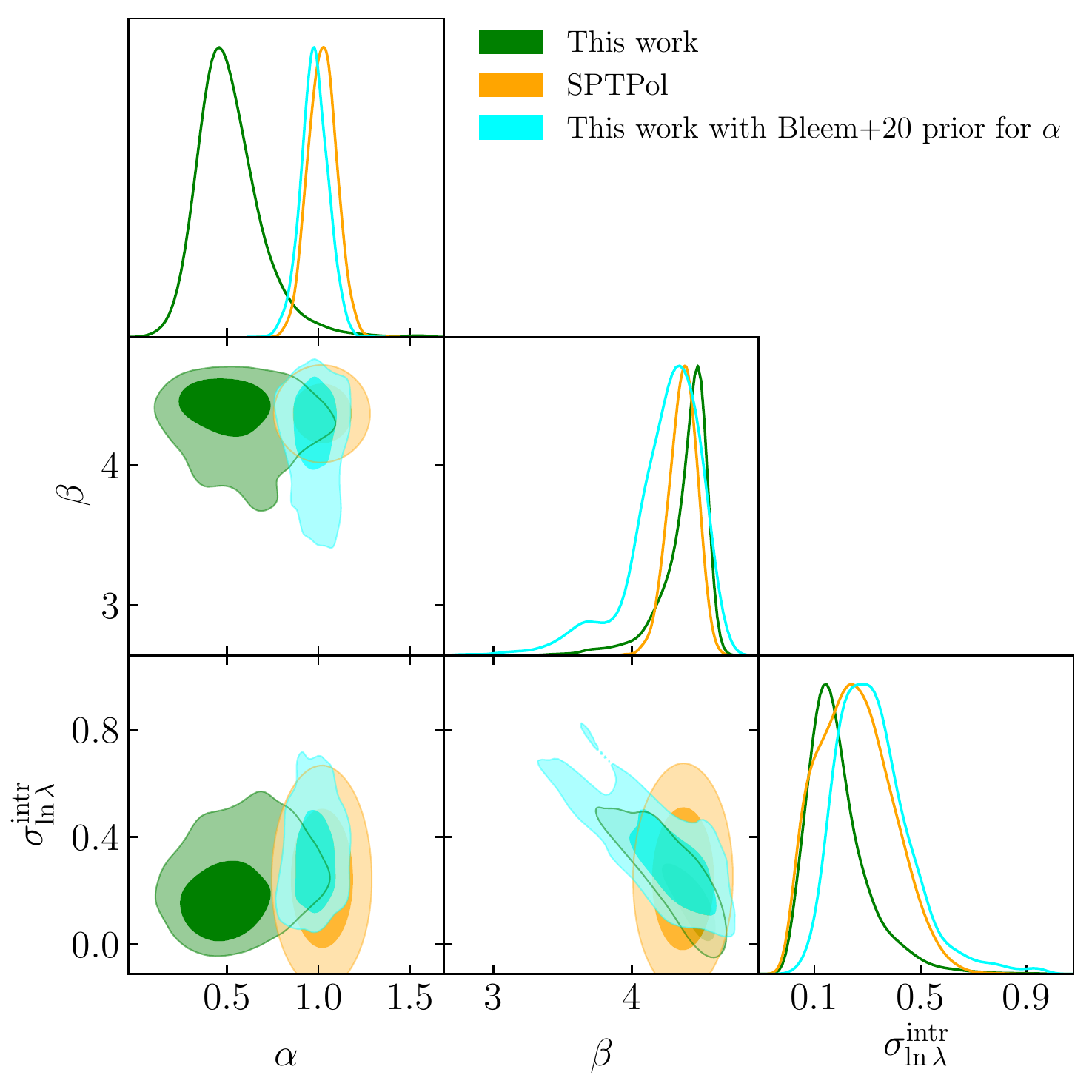}
  \caption{Comparison of the predicted \citet{Bleem_2020} parameter distributions (in orange) with respect to this work, but assuming a similar slope as in \citet{Bleem_2020} (in cyan). For the slope, instead of using a flat prior, we use a Gaussian prior with the mean and scatter set to SPTpol prediction listed in Table \ref{tab:scal_rel_comparison}. Contours indicate the statistical 1$\sigma$ ($68 \%$) , and 2$\sigma$ ($95 \%$)  credible regions.}
\label{fig:cfht-sptpol}
\end{figure}

We sample the likelihood of the parameters using the \emph{EMCEE} package \citep{foremanmackey13}, which is a Markov Chain Monte Carlo (MCMC) algorithm. We run 24 walkers with 2.000 steps each, excluding the first 400 steps of each chain to remove the burn-in region. We checked the chain convergence by running a successful Gelman-Rubin and Geweke statistic for it using the ChainConsumer package \citep{chainconsumer}.
The summary of both initial and prior parameter values used for the MCMC and their posterior values and $1\sigma$ statistical uncertainties are listed in Table \ref{tab:results}. The initial values for these scaling relations are set to the results of the SPIDERS cluster work \citep{spiders2018}. Originally, we set the upper limit of $\alpha$ prior to 3, but above 1.6, this upper limit introduced two additional disconnected regions of relatively good likelihood. The two regions had mean values of $\alpha = 2.4, \beta = 4.4$ and $\sigma_{\ln \lambda}^{\mathrm{intr}}=0.25$, and $\alpha = 2.1, \beta = 4.2$ and $\sigma_{\ln \lambda}^{\mathrm{intr}}=1.00$. The scaling relations of these two regions have nonphysically low true and mean richness at low masses ($< 3 \times 10^{14} \Msol$).
Therefore, we rerun the MCMC algorithm with
the upper limit of $\alpha$ prior set to 1.6, which removed the two nonphysical regions. We report the maximum likelihood of the posterior distribution as our best-fit values, and the uncertainties correspond to the interval containing $68\%$ of the points.

Fig. \ref{fig:cfht-sdss-corner} shows the results of the MCMC fitting. For the normalization $\lambda_0$ of the richness--mass relation, in logarithmic form $\left< \ln \lambda | \mu_{200c} \right> = \ln \lambda_0 + \alpha \mu_{200c}$, we found $\lambda_0 =\exp{\beta}= 84.0^{+9.2}_{-14.8}$, and for the slope $\alpha = 0.49^{+0.20}_{-0.15}$ at pivot mass $M_{\mathrm{200c, piv}} = 10^{14.81} M_{\odot}$. Our result for the intrinsic scatter in richness at fixed mass is $\sigma_{\ln{\lambda}|\mu}^{\mathrm{intr}} = 0.17^{+0.13}_{-0.09}$. 

We compare our richness--mass relation to previous work from \citet{Mulroy2019}, \citet{spiders2018}, \citet{desy1}, and \citet{Bleem_2020}. We give a brief summary of each of their results below.

In \citet{Mulroy2019}, a simultaneous analysis on several galaxy cluster scaling relations between weak lensing mass and multiple cluster observables is done, including 
richness--mass relation in logarithmic space $\left< \ln \lambda | \mu_{500c} \right> = \beta + \alpha\mu_{500c}
$ using a sample of 41 X-ray luminous clusters from the Local Cluster Substructure Survey (LoCuSS), spanning the redshift range of $0.15 < z < 0.3$ and mass range of $2.1 \times 10^{14} \Msol < M_{500c, WL} < 1.6 \times 10^{15} \Msol$, with $z_{\mathrm{piv}} = 0.22$, and $M_{500c, \mathrm{piv}} = 7.14 \times 10^{14} \Msol$. Their method for estimating the data likelihood function has the same basis as this work, thus we expect the least disagreement between their results and ours. 

\begin{figure}
\includegraphics[width=8cm]{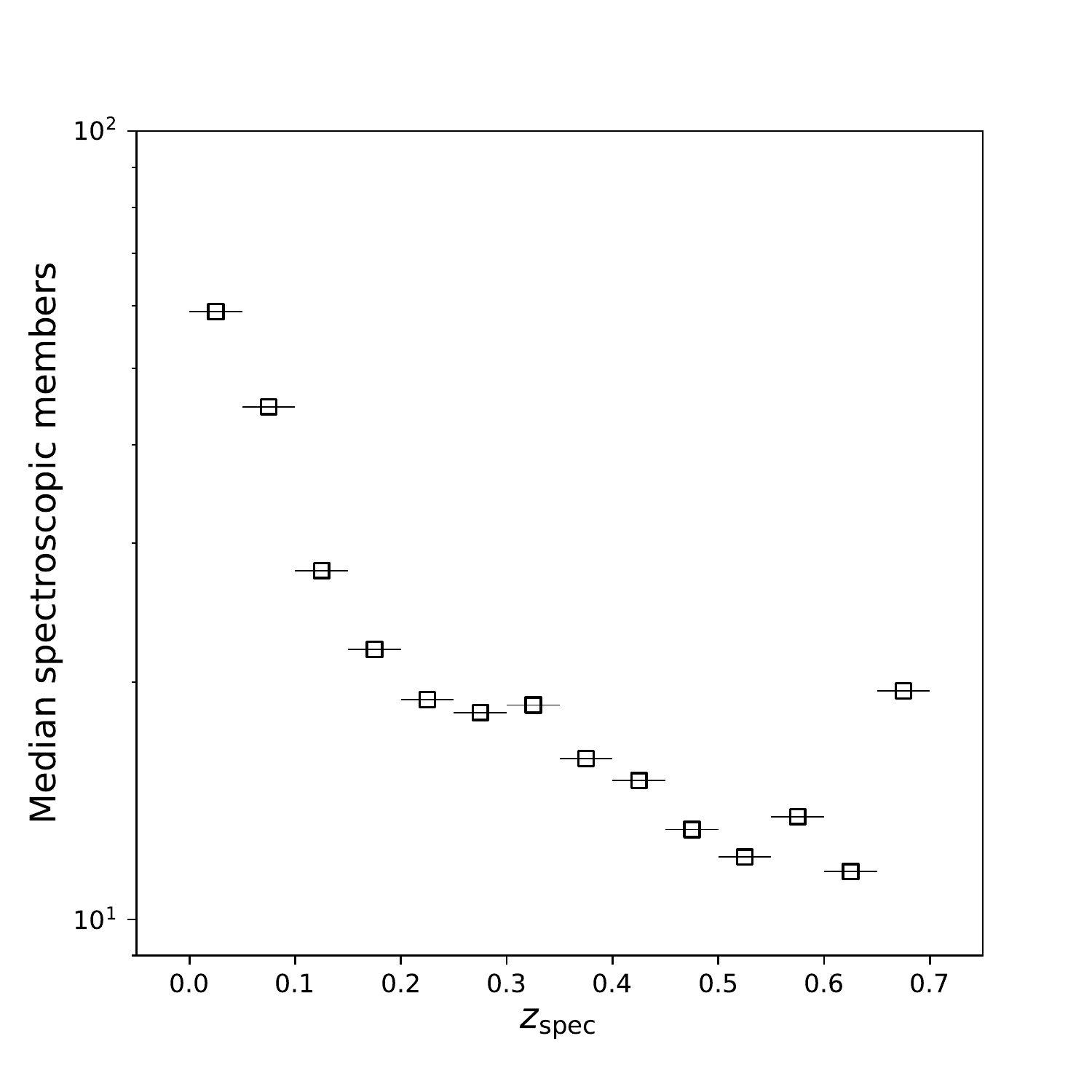}
\caption{Median of the spectroscopic members as a function of spectroscopic redshift of the SPIDERS sample, which CODEX sample is part of. The redshift bin is set to $\Delta z = 0.05$, and the selection cuts are set those of \citet{spiders2018}
($\lambda \geq 60$ and $N_{\mathrm{mem}} \geq 10$).}
\label{fig:median_spec_zbin}
\end{figure}

\citet{spiders2018} derive the richness--mass--redshift relation $\left< \lambda | \mu_{200c}, z \right> = A\mu_{200c}^{\alpha}(\frac{1+z}{1+z_{\mathrm{piv}}})^{\gamma}$ using a sample of 428 X-ray luminous clusters from the SPIDERS survey, spanning the redshift range $0.03 \leq  z \leq 0.66$ and dynamical mass range $1.6 \times 10^{14} \Msol < M_{200c, \mathrm{dyn}}< 1.6 \times 10^{15} \Msol $ with $z_{\mathrm{piv}}=0.18$ and $M_{200c, \mathrm{piv}} = 3 \times 10^{14} \Msol$. We compare our richness-mass results to their baseline analysis that accounted for the CODEX selection function.
Since the CODEX survey is part of the SPIDERS programme, they share a similar CODEX selection function as we do. Between $0.4<z<0.65$ our CODEX cluster sample overlap with \citet{spiders2018} with the cluster mass, richness, and redshift range. However, clusters with $z > 0.4$ in both \citet{spiders2018} and our work have the median number of spectroscopic redshift members $\leq 20$, as can be seen from Fig. \ref{fig:median_spec_zbin}, below, thus the quality of dynamical mass estimates is very different at $z<0.2$, where there are many more than 20 members (median is up to 60 members at $z<0.1$).

\citet{desy1} derive mass--richness--redshift relation $\left< M_{200m} | \lambda, z \right> = M_0 (\lambda/40)^{F}((1+z)/1.35)^G$, and they constrained the normalization of their scaling relation at the 5.0 per cent level, finding $M_0 = (3.081 \pm 0.075) \times 10^{14} M_{\odot} $ at $\lambda = 40$ and $z=0.35$. They find the richness slope at $F = 1.356 \pm 0.051$ and the redshift scaling index $G = -0.3 \pm 0.30$. 
They use redMaPPer galaxy cluster identifier in the Dark Energy Survey Year 1 data using weak gravitational lensing, and $4 \times 3$ bins of richness $\lambda$ and redshift $z$ for $\lambda \ge  20$ and $0.2 \leq z \leq 0.65$. The analysis of \citet{desy1} is the most statistically constraining result from the literature that we consider. However, they consider purely optically selected clusters, which are known to be prone to contamination of low-mass systems.  

\citet{Bleem_2020} derive richness--mass--redshift relation $\langle \ln \lambda | M_{500c}\rangle = \ln A + B \ln(M_{500c}/ 3\times 10^{14} M_{\odot} h^{-1}) + C\ln(E(z)/E(z=0.6))$, 
and found $A = 76.9 \pm 8.2$, $B = 1.020 \pm 0.08$, $C = 0.29 \pm 0.27$. They report finding a $28 \%$ shallower slope $F = 1/B$ than \citet{desy1} with the difference significant at the $4\sigma$ level.
This 2770 $\mathrm{deg}^2$ survey is conducted using the polarization sensitive receiver in the South Pole Telescope (SPTpol) using the identified Sunyaev-Zel'dovich (SZ) signal of 652 clusters to estimate the cluster masses. The richnesses of the clusters are estimated using the redMaPPer algorithm and matched with DES Y3 RM catalog to calibrate the richness--mass relation, taking the SPT selection into account. This sample is closest to ours in terms of sample definition, as both X-ray and SZ signal require the presence of hot intracluster medium (ICM), which cleans the contamination of optical samples.

In a recently published CODEX weak lensing analysis by \citet{Phriksee}, a mass-richness comparison was made to \citet{spiders2018}, with 279 clusters in the optical richness range at $20 \leq \lambda \leq$ 110, and $0.1 \leq z \leq 0.2$. They found an excellent agreement with both dynamical mass estimates and weak lensing mass estimates at $z \leq 0.15$.

We use the {\tt colossus} python package \citep{Diemer2018ab} to convert the $M_{500c}$, and $M_{200m}$ to $M_{200c}$ when necessary, and evaluate the slope and intercept at $M_{200c, \mathrm{piv}} = 10^{14.81} M_{\odot}$, in order to compare our constraints with other results. Since \citet{spiders2018}, \citet{desy1}, and \citet{Bleem_2020} included the $z$ evolution of their scaling relation, we estimate their relation at $z=0.5$, the mean $z$ of our 25 cluster subsample, to make our results comparable. For \citet{Mulroy2019}, we rescale the scaling relation parameters by assuming $\lambda_0(z) = \exp{\beta}(z) = const$. For the \citet{desy1} results, we use the \citet{leauthaud_2009} to invert the mass-richness relation, and evaluate the relation at $z=0.5$, $M_{\mathrm{200c, piv}} = 10^{14.81} M_{\odot}$. The inversion requires a bias term, which depends on the $\sigma_{\ln \lambda}^{\mathrm{intr}}$, for which we use our intrinsic scatter value of $\sigma_{\ln \lambda}^{\mathrm{intr}}=0.17^{+0.13}_{-0.09}$, as \citet{desy1} did not constrain it.  
In Table \ref{tab:scal_rel_comparison}, we show the predicted richness--mass mean parameter values and their $1\sigma$ statistical uncertainties from the LoCuSS, SPIDERS, SPTpol, and DES Y1 work, all evaluated at $z=0.5$ and $M_{200c, \mathrm{piv}} = 10^{14.81} M_{\odot}$. In Fig. \ref{fig:params}, we compare the slope and predicted richness $\lambda_0 = \langle \lambda | M = 10^{14.81} M_{\odot}, z=0.5, \rangle = \exp(\beta)$ from our work (gray bands) to the ones in the literature. 
 
Fig. \ref{fig:cfht-corner-comparison} shows the predicted mean relations from Table \ref{tab:scal_rel_comparison} overplotted to our MCMC fitting results from Fig. \ref{fig:cfht-sdss-corner}. We note that all the predicted mean results fall within $2\sigma$ region of our posterior distributions, where the largest deviation in both slope and intercept is with \citet{spiders2018} and \citet{Bleem_2020}. 

Since our slope is only accurate up to $2\sigma$ for both \citet{spiders2018} and \citet{Bleem_2020}, with both centered around unity, and the latter having shallower constraints for the slope, to see how different prior of the slope affects our parameter estimation, we redo our Bayesian analysis with the same 25 clusters as before, but using a Gaussian prior for the slope, set to the mean and the scatter from SPTpol prediction of Table \ref{tab:scal_rel_comparison}. In Fig. \ref{fig:cfht-sptpol}, we show the posterior distributions of the Gaussian prior for the slope in cyan, and compare the parameter distributions against the predicted SPTpol parameter distributions, shown in orange. When using a Gaussian prior for the slope, we found the posterior slope $\alpha = 0.98 \pm 0.09$, normalization $\lambda_0 = \exp(\beta) = 74.4^{+21.4}_{-18.2}$, and intrinsic scatter in richness $\sigma_{\ln \lambda}^{\mathrm{intr}} = 0.28^{+0.16}_{-0.14}$. We create the SPTpol parameter distributions by using a multivariate Gaussian with mean and elements of the diagonal scatter matrix set at the mean and the square of the $1\sigma$ uncertainties of the SPTpol predictions from Table \ref{tab:scal_rel_comparison}. We note that a tight parameter constraint on the slope loosens both the normalization, and the intrinsic scatter to wider range, forcing the mean of the normalization parameters towards smaller values, but intrinsic scatter towards the predicted SPTpol results. 
Since the number of clusters is small in our subsample, the prior shape has a larger impact on the final marginalized posterior distributions. We have a preference for choosing a flat prior for the slope, as our data points are within narrow mass range with large uncertainty on the mass, and small uncertainty on the richness. 

\begin{figure}
\begin{subfigure}{8.45cm}
\centering\includegraphics[width=7cm]{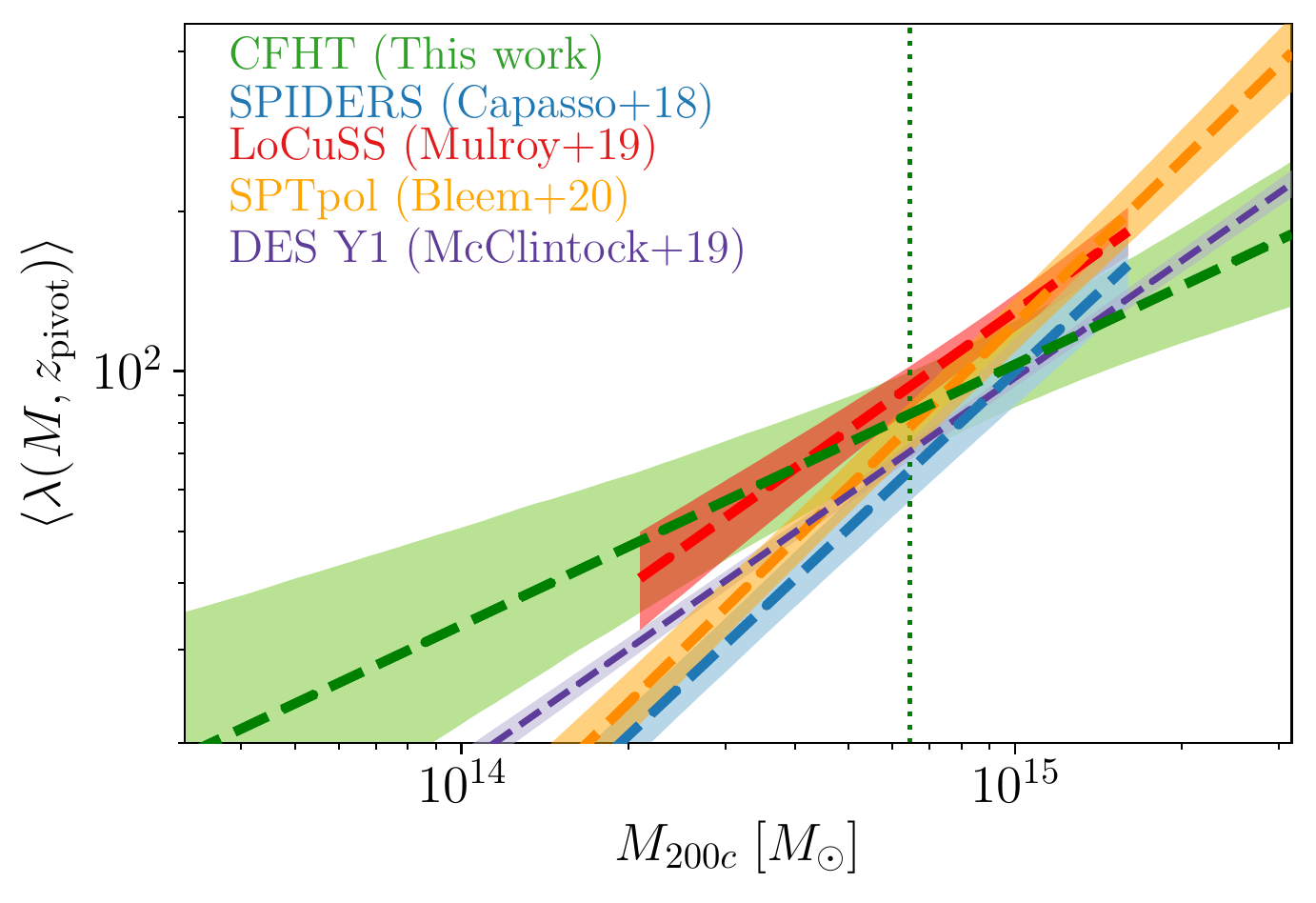}
\end{subfigure}
  \begin{subfigure}{8.45cm}
    \centering\includegraphics[width=7cm]{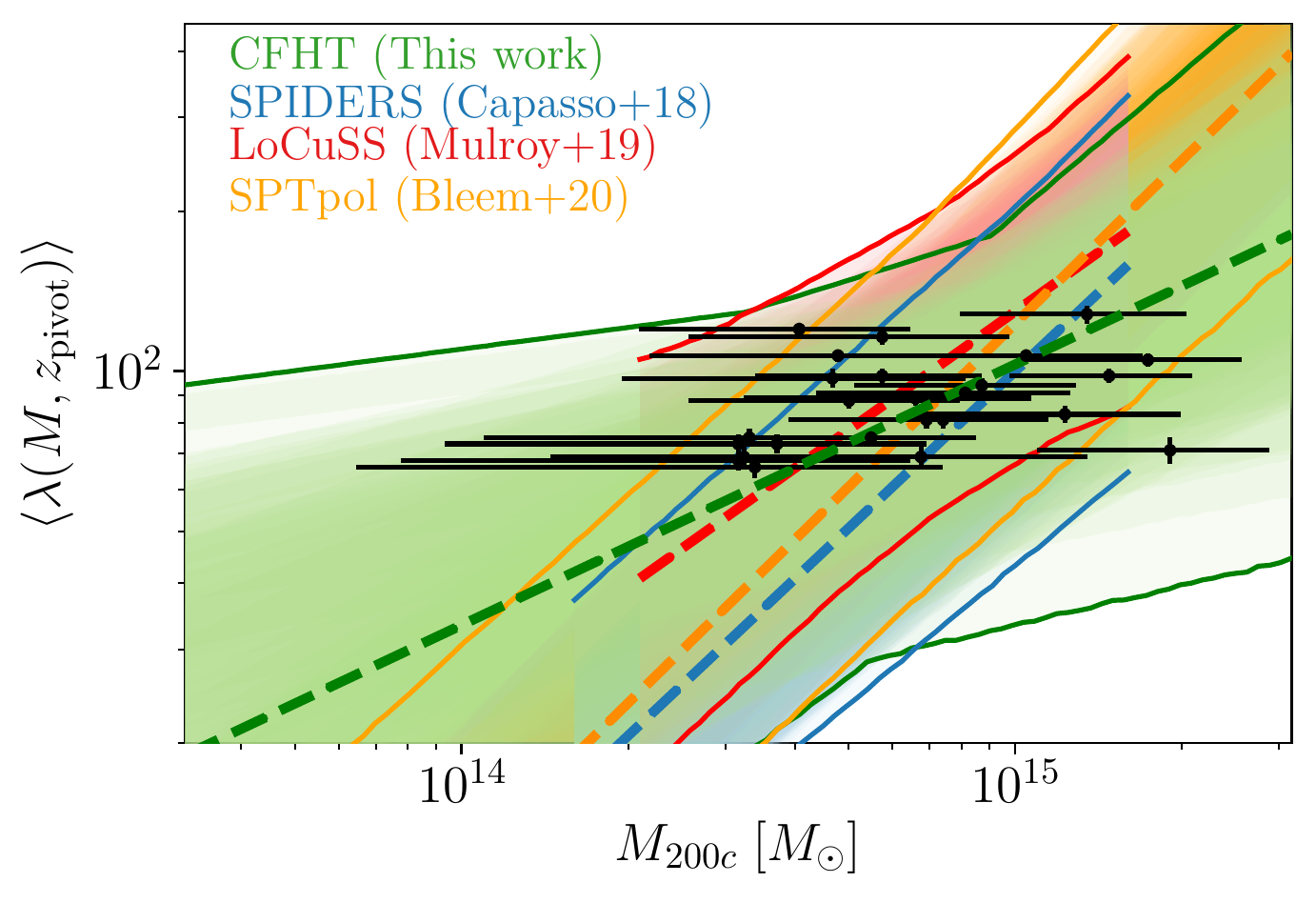}
  \end{subfigure}
\caption{\textbf{Upper panel}: Mean relation comparison with the predicted results from the literature. The confidence regions (light blue, light green, light red, light orange, and light violet envelopes) represent the 1$\sigma$ uncertainty of the slope and intercept of the mean relations (blue, green, red, orange and violet dashed lines, respectively). The predicted relations from DES Y1, SPTpol, and SPIDERS have been scaled to $z_{\mathrm{pivot}}=0.5$, and the DES Y1 relation is inverted according to \citet{leauthaud_2009}. The vertical green line is the pivot mass of this work. We limit each predicted relation to their respective mass and richness range.} 

\textbf{Lower panel}: Since in the data likelihood function, we account for the intrinsic scatter in richness, it is meaningful to include its effect to the overall parameter uncertainty budget. The error envelopes takes into account the $1\sigma$ uncertainties of the slope, intercept and the intrinsic scatter in richness. The uncertainties in data points represent $1\sigma$ statistical error in mass and observed richness.  
\label{fig:best_fit}
\end{figure}

In Fig. \ref{fig:best_fit}, we show the richness--mass relations from Table \ref{tab:scal_rel_comparison}. In the upper panel, we only consider the statistical $1\sigma \space (68\%)$ uncertainty around the mean relations, whereas in the lower panel, we consider the $1\sigma \space (68\%)$ interval, where new richness observations may fall at fixed mass. We do this by introducing the $\sigma_{\ln \lambda}^{\mathrm{intr}}$ and its $1\sigma$ uncertainty to all surveys, except for DES Y1, which lacked intrinsic scatter information.
The $1\sigma$ confidence regions in Fig. \ref{fig:best_fit} are done the following way:
\begin{enumerate}
    \item Draw 5000 new scaling relation parameter samples ($\alpha$, $\beta$, and $\sigma_{\ln \lambda}^{\mathrm{intr}}$) from a multivariate Gaussian distribution with mean and diagonal scatter matrix set to results from Table \ref{tab:scal_rel_comparison},
    \item Use new values of $\alpha$ and $\beta$ to generate 5000 new mean richnesses at each mass point,
    \item For the upper panel, calculate the $1\sigma$ statistics of these 5000 mean richness values and plot them,
    \item For the lower panel, sample 1000 new richness values for each of the 5000 mean richness values from a log-normal distribution with mean and scatter set to values sampled from the multivariate Gaussian in step (i),
    \item Calculate the $1\sigma$ uncertainty from the 1000 new richness values for each of the 5000 mean richnesses and plot those uncertainties to the lower panel.
\end{enumerate}

The error envelopes in the lower panel include the $1\sigma$ uncertainties of the slope, the intercept and the intrinsic scatter in richness. Typically in the literature, only the mean with $1\sigma$ uncertainties are shown as the scaling relation, like in the upper panel of Fig. \ref{fig:best_fit}, but this method only accounts for uncertainty in the slope and intercept, and does not consider that the mean relation may deviate from the fixed data points by the intrinsic scatter. In the lower panel of Fig. \ref{fig:best_fit}, we also take account the effect of intrinsic scatter in richness and its $1\sigma$ uncertainty in the scaling relations. The latter method takes into account both the uncertainty of the mean relation due to intrinsic scatter, along with the uncertainty on the parameters. We note that the data points in Fig. \ref{fig:best_fit} refer to observed values from Table \ref{tab:cleanedWL}, not to their true values. We show these here to point out the narrow mass range of the observed data with large statistical uncertainty in weak lensing mass and small uncertainty in the observed richness.

From Fig. \ref{fig:params}, the richness normalization $\lambda_0$, at $z=0.5$ and $M_{200c}=10^{14.81} \Msol$, from our work overlaps within $1\sigma$ uncertainty with all four different survey richness normalizations that we consider. 
The main difference in the normalization is between LoCuSS, which had measured clusters at $0.15<z<0.3$, and the rest of the surveys, but given that LoCuSS richness relation is estimated without redshift dependent evolution in richness, so this might mean that there is an evolution of cluster richness at a given mass, as discussed in \citep{spiders2018}. 

Relatively flat slopes found in this and in LoCuSS work could be attributed to a combination of probing small mass range, and that intrinsic scatter in richness could increase with decreasing mass $\sigma_{\ln \lambda}^{\mathrm{intr}}(m) \propto 1/m$.
Although, our mass slope is only $1\sigma$ away from the slope found by \citet{desy1}, a steeper slope of $\alpha = 1.0^{+0.22}_{-0.22}$ was robustly established in low-z CODEX studies \citep{Phriksee}, and was attributed to CODEX X-ray clusters being less prone to possible contamination by projected low mass groups of galaxies along line-of-sight than purely optically selected clusters, such as \citet{desy1}.

Also, from Fig. \ref{fig:params}, we see that our result on the intrinsic scatter in richness overlaps within $1\sigma$ with other results found from the literature, however with smaller mean at $\sigma_{\ln \lambda}^{\mathrm{intr}} = 0.17^{+0.13}_{-0.09}$. When the same analysis is done with a Gaussian prior on the slope, $\alpha \sim \mathcal{N}(1.02, 0.08)$ (see Fig. \ref{fig:cfht-sptpol}), we find the intrinsic scatter at $\sigma_{\ln \lambda}^{\mathrm{intr}} = 0.28^{+0.16}_{-0.14}$, indicating the importance of the prior choice, when a small sample size is considered.

Our comparison to the results of the dynamical mass modelling, presented in \citet{spiders2018}, indicate marginally lower mass for a given richness at richness values around $80$. Considering other weak lensing calibrations, performed on X-ray clusters, we quote from \cite{Phriksee} that at $z<0.15$ the weak lensing calibration of CODEX clusters of \cite{Phriksee} agrees well with \citet{spiders2018}, while we find from Fig. \ref{fig:params} that LoCuSS \citep{Mulroy2019} results ($0.15<z<0.3$) are in significant tension with \citet{spiders2018}. These results, if confirmed, could be used to constrain the models of modified gravity \citep{arnold,Sakstein2016,wilcox2016, mitchell, Tamosiunas_2019}. Improvements in spectroscopic follow-up of high-z clusters is however, very critical. As \citet{zhang2017} showed, a low number of spectroscopic redshifts per cluster and fiber-collisions of SPIDERS tiling can have strong effect on bias and scatter of dynamical mass estimates.

\section{Conclusions}
\label{sec:conclusion}

We present the results of Bayesian weak lensing mass calibration analysis of CODEX cluster sample of 25 clusters for high redshift ($0.35 < z < 0.62 $), with redMaPPer richness $\ge 60$, and with a detailed consideration of systematic uncertainties.
The weak lensing data is obtained by pointed CFHT observations of CODEX clusters, to which we add a reanalysis of the public CFHTLS data. We obtain the cluster masses by running a likelihood analysis including a covariance matrix to account for contributions by large scale structure and intrinsic properties. We refine the original richness estimates based on SDSS photometry by rerunning redMaPPer on CFHT photometry and obtain richness-mass relation $\langle \ln \lambda | \mu \rangle = \alpha \mu + \beta$, with $\mu = \ln (M_{200c}/10^{14.81} M_{\odot})$, and compare this relation to the one obtained by \citet{Mulroy2019} ($z\sim0.2$),  and z=0.5 predictions of \citet{spiders2018}, \citet{desy1}, and \citet{Bleem_2020}. We
measure richness-mass relation with slope of $\alpha=0.49^{+0.20}_{-0.15}$ and intercept of $\lambda_0 = \exp(\beta) = 84.0^{9.2}_{-14.8}$, using a data likelihood function that incorporate the overall error budget of the weak lensing mass calibration analysis, along with optical, X-ray, survey incompleteness and subsample selection effects.   

We find our results on the slope, intercept, and intrinsic scatter in richness overlap with the weak lensing analysis of low-z ($0.15<z<0.3$) LoCuSS clusters by \citet{Mulroy2019} within $1\sigma$ uncertainty over the entire LoCuSS mass range.

At masses of $10^{14.81}M_\odot$, our 68\% credible region for the mean cluster richness overlaps with that of \citet{Mulroy2019}, \citet{desy1}, and \citet{Bleem_2020}, and at around the 16th percentile, slightly overlaps the 84th percentile of the \citet{spiders2018}. The $1\sigma$ statistical uncertainty in richness is at the level of difference in the results based on different cluster selection and different mass measurements.
Even though we consider a multitude of selection effects with a narrow mass range and a small sample size, we find relatively flat slope. Thus, future improvements should not be directed solely towards increasing the sample size, but also on understanding the selection effects and improvements in the mass measurements. 
The importance of our work consists in extending the weak lensing calibration of massive X-ray clusters to $z \leq 0.6$, where previously, large disagreements on weak lensing calibrations were reported \citep{Smith_2015}.
\section*{Acknowledgements}
We thank an anonymous referee for thorough review of the manuscript, Raffaella Capasso and Jacob Ider Chitham for discussion of the results.
This work is based on observations obtained with MegaPrime/MegaCam, a joint project of CFHT and CEA/IRFU, at the Canada-France-Hawaii Telescope (CFHT) which is operated by the National Research Council (NRC) of Canada, the Institut National des Science de l'Univers of the Centre National de la Recherche Scientifique (CNRS) of France, and the University of Hawaii.
\\
We use data from the Canada-France-Hawaii Lensing Survey \citep{heymans12}, hereafter referred to as CFHTLenS. The CFHTLenS survey analysis combined weak lensing data processing with THELI \citep{erben13} and shear measurement with lensfit \citep{miller13}. A full systematic error analysis of the shear measurements in combination with the photometric redshifts is presented in \citet{heymans12}.
\\
Based on observations made with the Nordic Optical Telescope, operated by the Nordic Optical Telescope Scientific Association at the Observatorio del Roque de los Muchachos, La Palma, Spain, of the Instituto de Astrofisica de Canarias.
\\
We acknowledge Fabrice Brimioulle for his substantial work on an early version of this manuscript, and we understand his decision not to be listed on the paper, since he is no longer working in astronomy. We thank Matthew R.~Becker and Andrey Kravtsov for making their cluster simulations available. 
\\
KK and JV acknowledge financial support from the Finnish Cultural Foundation, KK the Magnus Ehrnrooth foundation, and the Academy of Finland grant 295113. This work was supported by the Department of Energy, Laboratory Directed Research and Development program at SLAC National Accelerator Laboratory, under contract DE-AC02-76SF00515 and as part of the Panofsky Fellowship awarded to DG. NC acknowledges financial support from the Brazilian agencies CNPQ and CAPES (process \#2684/2015-2 PDSE). NC also acknowledges support from the Max-Planck-Institute for Extraterrestrial Physics and the Excellence Cluster Universe. ESC acknowledges financial support from Brazilian agencies CNPQ and FAPESP (process \#2014/13723-3). LM acknowledges STFC grant ST/N000919/1. AF \& CK acknowledge the Finnish Academy award, decision 266918. HYS acknowledges the support from the Shanghai Committee of Science and Technology grant No. 19ZR1466600.
\\
We acknowledge R. Bender for the use of his photometric redshift pipeline in this work. NC acknowledge J. Weller for the hospitality.

This work made use of the astronomical data analysis software \texttt{TOPCAT} \citep{Taylor2005aa}. Data analysis has been carried out with University of Helsinki computing clusters Alcyone and Kale. We acknowledge the use of the research infrastructures Euclid Science Data Center Finland (SDC-FI, urn:nbn:fi:research-infras-2016072529) and the Finnish Grid and Cloud Computing Infrastructure (FGCI, urn:nbn:fi:research-infras-2016072533), and the Academy of Finland infrastructure grant 292882. The author acknowledges the usage of the following python packages, in alphabetical order: \texttt{astropy} \citep{Astropy-Collaboration2013aa, Astropy-Collaboration2018aa}, \texttt{chainConsumer} \citep{chainconsumer}, 
\texttt{emcee} \citep{emcee3}, \texttt{matplotlib} \citep{matplotlib}, \texttt{numpy} \citep{numpy1, numpy2}, and \texttt{scipy} \citep{scipy}.
\section*{Data availability}
The raw data underlying this article are available in CFHT server, at \url{https://www.cadc-ccda.hia-iha.nrc-cnrc.gc.ca/en/cfht/}
%
%

%
\addcontentsline{toc}{chapter}{Bibliography}
\bibliographystyle{mnras}
\bibliography{MassCat}
%
%
%
\appendix

\section{Systematic uncertainties}
\label{sec:systematics}
 Our lensing signal is affected by two sources of systematics: the errors from shape measurements and distance estimates computed using the colour-magnitude decision tree.
The systematic uncertainties enter in our lensing model as a factor that multiplies the theoretical density profile, changing its amplitude to assimilate the errors. This factor follows a Gaussian prior with the mean shifted by the bias from both shear and photometric redshift measurements, $1 - \delta_{\rm{cm}} - \delta_{\rm{sm}}$, and width corresponding to the quadratic sum of the variances $\sigma_{\rm{cm}}$ and $\sigma_{\rm{sm}}$. In the following sections we describe how we derive these contributions from shape and distance measurements.
\subsection{Shear bias}
As mentioned in \autoref{sec:shape measurement}, we expect the residual uncertainty in the \emph{lensfit} shape measurement to be in the order of 2 per cent (see \citealt{fenechconti17}) and assume the same uncertainty in case of CFHTLenS \emph{lensfit} shapes after applying the corrections shown in equations \ref{eqn:mlensfit} and \ref{eqn:clensfit}. We account for this uncertainty by introducing a shear calibration factor with mean $\delta_{\rm{sm}}=0$ and Gaussian width $\sigma_{\rm{sm}}=0.02$ in our modelling.
\subsection{Bias of source redshift distribution}
\label{sec:bias_z_distribution}
The colour-magnitude decision tree method contributes to the final error budget through two sources of systematic uncertainties: cosmic variance and errors in the reference catalogue of photometric redshifts. We assess the contribution from photo-$z$ errors by comparing the values of $\beta$ from the CFHTLS D2 field ($\beta_{\rm{D2}}$) and COSMOS2015 ($\beta_{\rm{C2015}}$) catalogues. The difference in $\beta$ from this matched catalogue is free of cosmic variance because we use different template fits over the same galaxies. The mean shift of each individual cluster \emph{i} is computed as 
\begin{equation}
\delta_{\rm{cm},i} = \frac{1}{2} \frac{\left \langle \beta_{\rm{C2015}} \right \rangle - \left \langle \beta_{\rm{D2}} \right \rangle}{\left \langle \beta_{\rm{D2}} \right \rangle}
\end{equation}
with the variance, assuming a Gaussian of the same variance as a top-hat distribution between $\beta_{\rm{D2}}$ and $\beta_{\rm{C2015}}$, given by
\begin{equation}
\sigma_{\rm{cm},i} = \frac{1}{\sqrt{3}} \lvert \delta_{\rm{cm, i}} \rvert
\end{equation}
To derive the cosmic variance contribution $\sigma_{\rm{cv}}$ of each cluster we use a jackknife estimate over the four pointings of CFHTLS Deep:
\begin{equation}
\sigma_{\rm{cv}, i} = \sqrt{\frac{3}{4}\sum_{j=1}^4 [ (\langle \beta_i \rangle_{\neg j}-{\langle \beta_i \rangle)^2 ] }/{\langle \beta_i \rangle}} \; ,
\end{equation}
with $\langle\beta_i\rangle_{\neg j}$ being the lensing-weighted mean $\beta$ excluding CFHTLS Deep pointing $j$ and $\langle\beta_i\rangle$ the average over all pointings. 
The final shift of each individual cluster takes into account the lensing weight \emph{w} and is given by
\begin{equation}
\label{mumg}
\delta_{\rm{cm,i}} = w_i \delta_{\rm{cm},i}
\end{equation}
with variance
\begin{equation}
\sigma_{\rm{cm,i}} = w_i \sigma_i
\end{equation}
where $\sigma_i$ incorporates the contribution from cosmic variance and photo-z errors:
\begin{equation}
\label{sigsig}
\sigma_{i} = \sqrt{\sigma_{\rm{cv}, i}^2 + \sigma_{\rm{cm}, i}^2} \ .
\end{equation}

In selecting our sources, we removed regions of colour-magnitude space that are contaminated by galaxies at the cluster redshift. Due to redshift uncertainties we estimate this to be at a level of about 2 per cent (see section 2.6). For this reason, we expect dilution of our source sample with cluster members to be minor. To account for residual cluster member dilution, we assume a value of $\delta_{\rm{cmd}}=0$ with an uncertainty of $\sigma_{\rm{cmd}}=0.02$.

The final shear calibration term is derived by
\begin{equation}
\label{eq:S}
S_{\rm{m}}=1-\delta_{\rm{cm}}-\delta_{\rm{sm}}-\delta_{\rm{cmd}}
\end{equation}
with the uncertainty given by
\begin{equation}
\label{eq:sigS}
\sigma_{\rm{S}}=\sqrt{\sigma_{\rm{cm}}^2+\sigma_{\rm{cv}}^2+\sigma_{\rm{sm}}^2+\sigma_{\rm{cmd}}^2}\ .
\end{equation}
We correct the measured lensing signal dividing by $S_{\rm{m}}$. Given the large statistical errors originating from shape noise the shear calibration error hardly carries weight. Nonetheless we take the uncertainty into account by remeasuring weak lensing masses with shear calibration values of $S_{\rm{m}} \pm \sigma_{\rm{S}}$ and adding the deviation from the actual best-fit value quadratically into our systematic error budget. The shear calibration values for the individual CODEX clusters can be seen in Table \ref{tab:pz shear calibration}. 

\begin{table*}
\caption{Weak lensing shear calibration values from p(z) for all three CODEX subsamples of galaxy clusters. We set $\sigma_{\mathrm{cmd}} = \sigma_{\mathrm{sm}} = 0.02.$ The final shear and total uncertainty are given in equations \ref{eq:S}, and \ref{eq:sigS}, respectively.}
\label{tab:pz shear calibration}
\begin{tabular}{c|c|c|c|c|c|c|}
\hline \hline
Subsample & CODEX ID & $\delta_{\rm{cm}}$ & $\sigma_{\rm{cm}}$ & $\sigma_{\rm{cv}}$  & $S_{\rm{m}}$ & $\sigma_{\rm{S}}$ \\
\hline \hline
\textbf{S-I}&   12451  &  0.023  &  0.013  &  0.005  &  0.977  &  0.032  \\
  & 13062  &  0.012  &  0.007  &  0.004  &  0.988  &  0.029  \\
   &13390  &  0.025  &  0.014  &  0.005  &  0.975  &  0.032  \\
   &16566  &  0.006  &  0.004  &  0.003  &  0.994  &  0.029  \\
   &18127  &  0.014  &  0.008  &  0.004  &  0.986  &  0.030  \\
   &24865  &  0.013  &  0.008  &  0.004  &  0.987  &  0.029  \\
   &24872  &  0.006  &  0.004  &  0.003  &  0.994  &  0.029  \\
   &24877  &  0.023  &  0.013  &  0.006  &  0.977  &  0.032  \\
   &24981  &  0.004  &  0.003  &  0.003  &  0.996  &  0.029  \\
   &25424  &  0.015  &  0.009  &  0.004  &  0.985  &  0.030  \\
   &25953  &  0.013  &  0.008  &  0.003  &  0.987  &  0.030  \\
   &27940  &  0.007  &  0.004  &  0.003  &  0.993  &  0.029  \\
   &27974  &  0.011  &  0.007  &  0.003  &  0.989  &  0.029  \\
   &29283  &  0.017  &  0.010  &  0.006  &  0.983  &  0.031  \\
   &29284  &  0.017  &  0.010  &  0.005  &  0.983  &  0.030  \\
   &29811  &  0.014  &  0.008  &  0.003  &  0.986  &  0.030  \\
   &35361  &  0.007  &  0.004  &  0.003  &  0.993  &  0.029  \\
   &35399  &  0.016  &  0.009  &  0.003  &  0.984  &  0.030  \\
   &35646  &  0.010  &  0.006  &  0.003  &  0.990  &  0.029  \\
   &36818  &  0.022  &  0.013  &  0.005  &  0.978  &  0.031  \\
   &37098  &  0.017  &  0.010  &  0.005  &  0.983  &  0.030  \\
   &41843  &  0.008  &  0.004  &  0.003  &  0.992  &  0.029  \\
   &41911  &  0.007  &  0.004  &  0.003  &  0.993  &  0.029  \\
   &43403  &  0.008  &  0.005  &  0.003  &  0.992  &  0.029  \\
   &46649  &  0.024  &  0.014  &  0.005  &  0.976  &  0.032  \\
   &47981  &  0.018  &  0.010  &  0.004  &  0.982  &  0.030  \\
   &50492  &  0.016  &  0.009  &  0.004  &  0.984  &  0.030  \\
   &50514  &  0.011  &  0.007  &  0.004  &  0.989  &  0.029  \\
   &52480  &  0.020  &  0.012  &  0.006  &  0.980  &  0.031  \\
   &53436  &  0.018  &  0.010  &  0.004  &  0.982  &  0.030  \\
   &53495  &  0.011  &  0.006  &  0.003  &  0.989  &  0.029  \\
   &54795  &  0.009  &  0.005  &  0.003  &  0.991  &  0.029  \\
   &55181  &  0.017  &  0.010  &  0.005  &  0.983  &  0.030  \\
   &56934  &  0.009  &  0.005  &  0.003  &  0.991  &  0.029  \\
   &59915  &  0.011  &  0.007  &  0.003  &  0.989  &  0.029  \\
   &64232  &  0.017  &  0.010  &  0.004  &  0.983  &  0.030  \\

\hline 
\textbf{S-II}  & 13311  &   0.007  &  0.004  &  0.003  &  0.993  &  0.029  \\
  & 13315  &   0.024  &  0.014  &  0.006  &  0.976  &  0.032  \\
  & 13380  &  -0.001  &  0.001  &  0.002  &  1.001  &  0.028  \\
  & 13390  &   0.025  &  0.014  &  0.005  &  0.975  &  0.032  \\
  & 13391  &  -0.001  &  0.001  &  0.002  &  1.001  &  0.028  \\
  & 13400  &   0.013  &  0.007  &  0.003  &  0.987  &  0.029  \\
  & 17449  &   0.009  &  0.005  &  0.003  &  0.991  &  0.029  \\
  & 17453  &   0.002  &  0.001  &  0.003  &  0.998  &  0.028  \\
  & 54652  &   0.017  &  0.010  &  0.005  &  0.983  &  0.030  \\
  & 56934  &   0.009  &  0.005  &  0.003  &  0.991  &  0.029  \\
  & 57017  &   0.002  &  0.001  &  0.002  &  0.998  &  0.028  \\
  & 60076  &  -0.002  &  0.001  &  0.002  &  1.002  &  0.028  \\
  & 60131  &   0.001  &  0.000  &  0.002  &  0.999  &  0.028  \\
  & 60155  &  -0.001  &  0.001  &  0.003  &  1.001  &  0.028  \\
  & 64565  &   0.003  &  0.002  &  0.003  &  0.997  &  0.028  \\
  & 64636  &   0.008  &  0.004  &  0.003  &  0.992  &  0.029  \\
  & 210288  &  -0.002  &  0.001  &  0.002  &  1.002  &  0.028  \\
  & 210306  &  -0.002  &  0.001  &  0.002  &  1.002  &  0.028  \\

\hline 
 \textbf{S-III}  &24925  &   0.002  &  0.001  &  0.003    &  0.998  &  0.028  \\
   &27955  &   0.002  &  0.001  &  0.003   &  0.998  &  0.028  \\
  & 46647  &   0.001  &  0.001  &  0.002   &  0.999  &  0.028  \\
  & 54796  &  -0.001  &  0.001  &  0.002    &  1.001  &  0.028  \\
\hline \hline
\end{tabular}
\end{table*}
\subsection{Surface density profile}
\label{sec:profilecalibration}
Our model for the $\Delta\Sigma$ profile of a cluster of given mass is not perfect: on small scales, the off-centring of redMaPPer-identified BCG candidates smears out the profile; on large scales, truncation reduces the surface density of the main halo, while correlated secondary haloes add to it; and there are additional differences between the mean density profiles of haloes and the NFW prescription, as measured from detailed $N$-body simulations.
We calibrate these effects using simulated clusters of galaxies. To this end, we convert the shear maps extracted by \citet{becker11} from the simulation labelled L1000W in \citet{tinker08}. We use two snapshots with dark matter particles of mass $6.98\times 10^{10} h^{-1}\Msol$ in a box of comoving size 1 $h^{-1}$ Gpc, one at $z_d=0.245$ for all haloes with $M_{\rm{200c}}\geq10^{14} h^{-1} \Msol$ and one at $z_d=0.50$ for haloes with $M_{\rm{200c}}\geq3\times10^{14} h^{-1} \Msol$. We convert the gravitational shear maps centred on these haloes and simulated for sources at $z_s=1$ to observable reduced shear profiles.
We run the mass likelihood described in \autoref{sec:likelihood}, using a covariance matrix including the mean shape noise of our cluster sample, LSS contributions calculated for the lensing-weighted stacked source $p(z)$ of our cluster sample, and intrinsic profile variations at the respective cluster mass and redshift. The surface density model differs from the one described in \autoref{sec:likelihood} in that we use the mass--concentration relation of \citet{duffy08} that better matches the cosmological parameters and resulting halo profiles of the L1000W simulation. 
\begin{figure*}
    \centering\includegraphics[width=8.45cm]{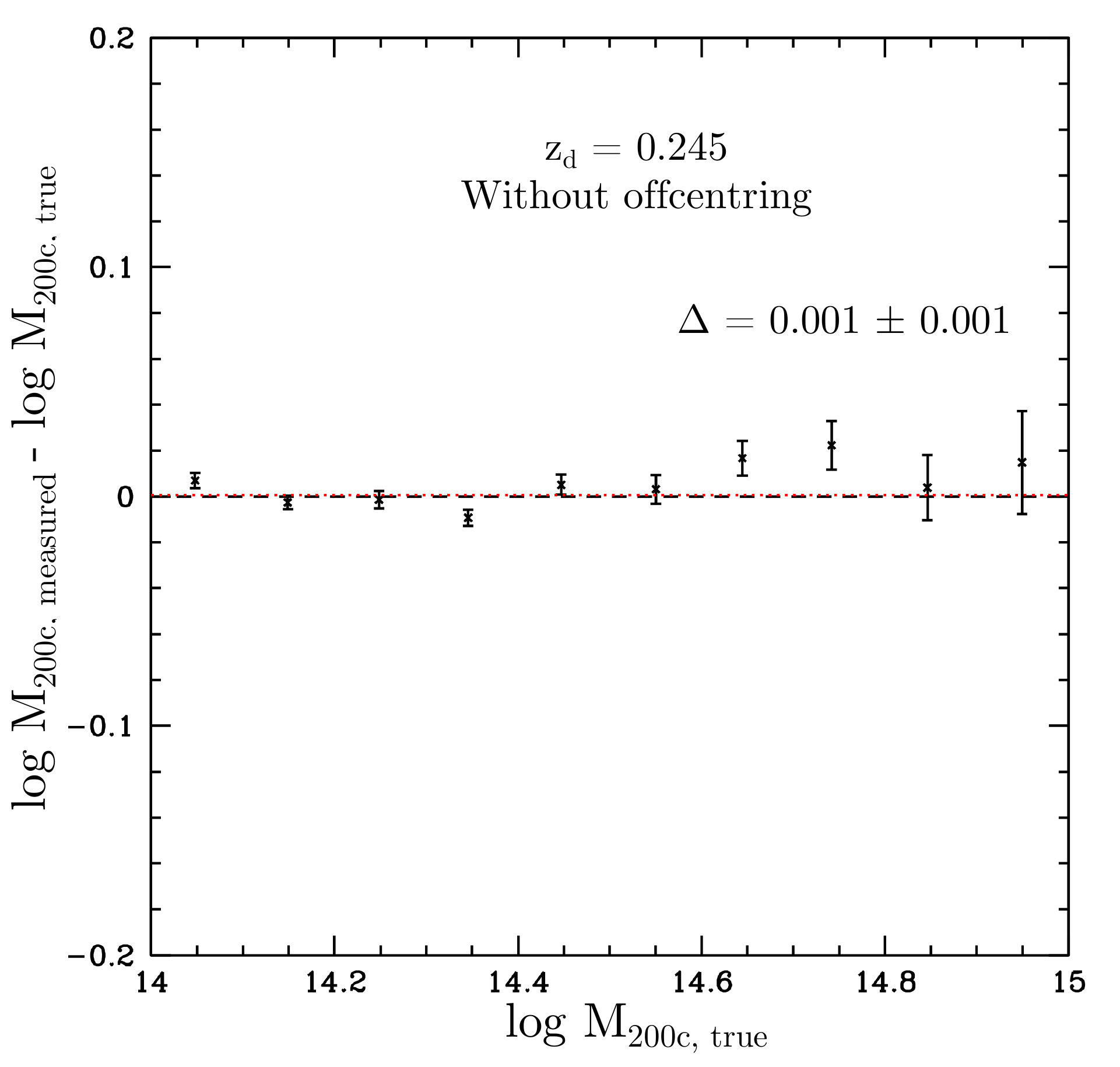}
    \centering\includegraphics[width=8.45cm]{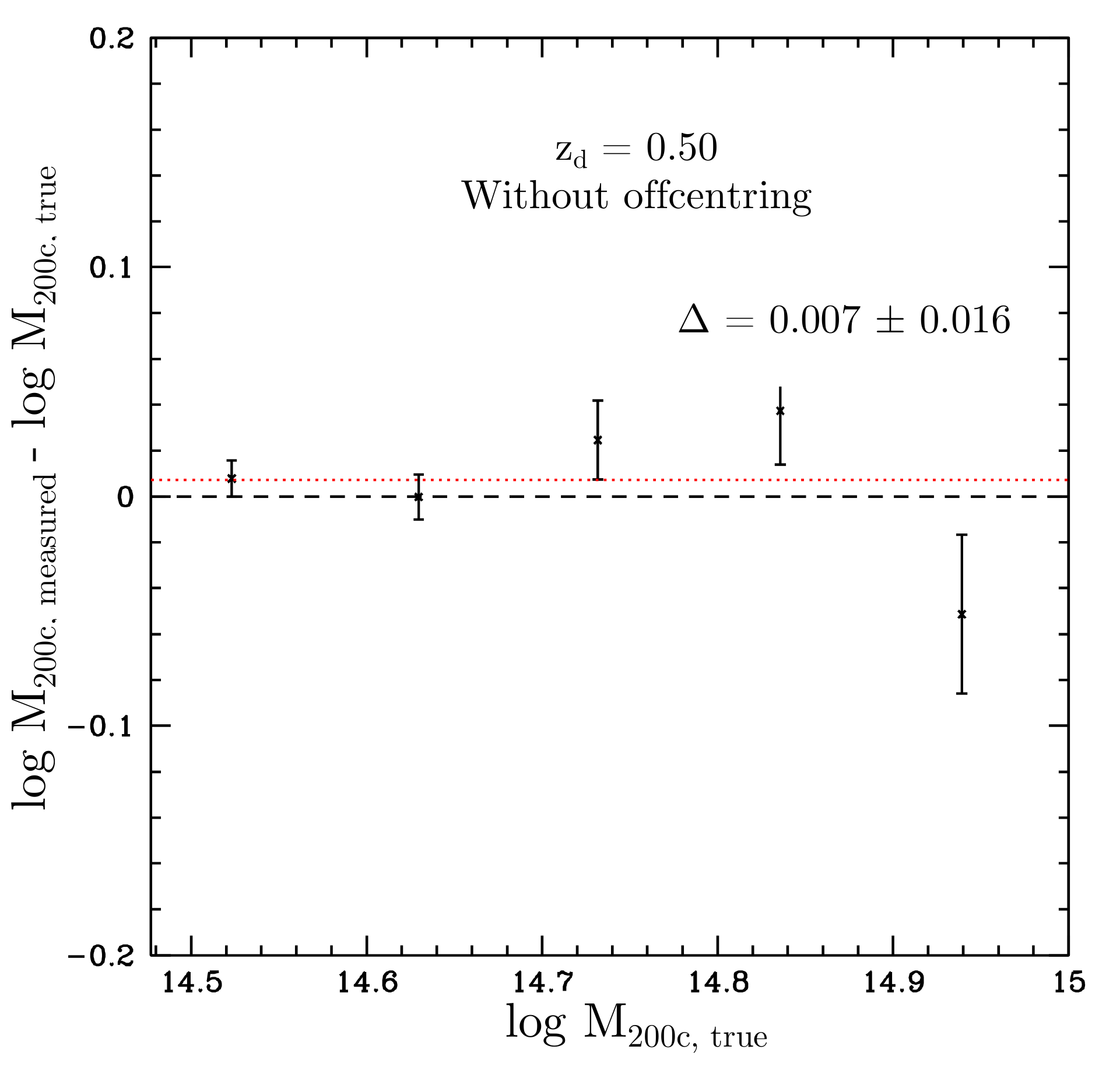}
    \centering\includegraphics[width=8.45cm]{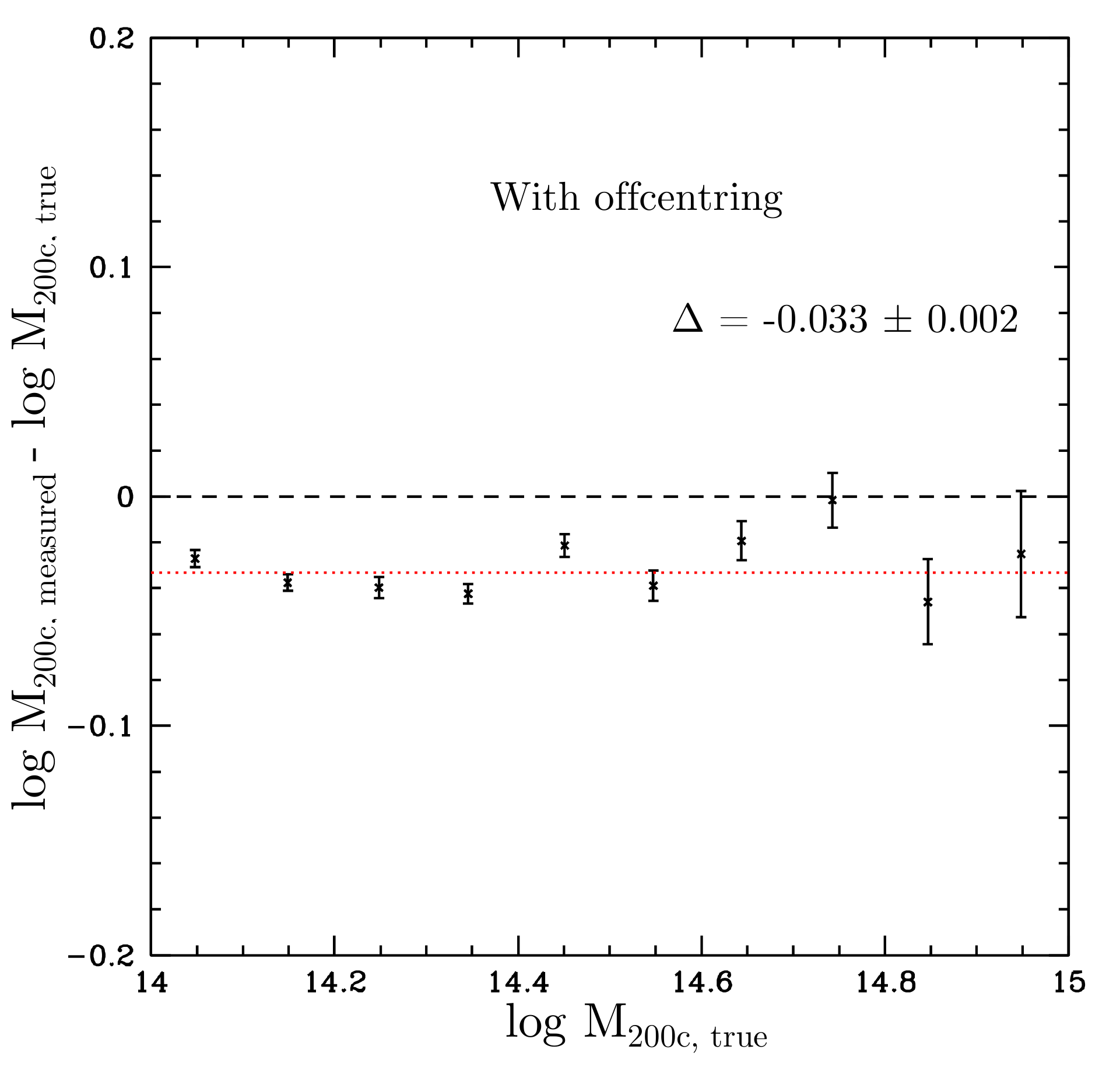}
\caption{True mass vs. measured mass for perfectly well centred simulated galaxy clusters for the simulation at $z_d=0.245$ (upper left panel) and at $z_d=0.50$ (upper right panel). The true masses are recovered to a very good level in the weak lensing analysis. In the lower panel, the true mass vs. measured mass for the off-centred simulated galaxy clusters (all redshifts combined). The measured cluster masses are in average about 0.03 lower in logarithmic scale than the true masses.}

\label{fig:deltam-no-off-centring}
\end{figure*}
Fig. \red{\ref{fig:deltam-no-off-centring}} shows the mean recovered mass in bins of true mass without off-centring (in the two upper panels). We find no significant bias and no significant evolution of bias with cluster mass or redshift. The mean bias of all clusters in the two snapshots is$\Delta\log M_{\rm{200c}}=0.001\pm 0.002$. We include these corrections in our analysis and their uncertainty in our systematic error budget.
\subsubsection{Concentration--mass relation}
We verify the robustness  of the measured cluster masses with respect to the chosen concentration-mass relation. For that purpose we repeat the mass measurements on the simulated cluster sample, modifying the applied concentration-mass relation. In a first run we increase the original concentration value by 10 per cent, in a second run we increase it by 33 per cent with respect to the original value. The retrieved average logarithmic mass is lower by 0.002 in the first case and by 0.007 in the second case. If we further increase the concentration by 50 per cent with respect to the original relation we measure a logarithmic mass lower by 0.010.
\\
\citet{cibirka16} found a mean concentration value for the stacked CODEX sample of $c=3.7^{+0.7}_{-0.6}$ which roughly corresponds to an uncertainty of 20 per cent. If we scale the concentration-mass relation by 20 per cent, once up and once down, we obtain logarithmic masses which are lower by 0.004 in the first case and higher by 0.002 in the second case. Taking these analyses into account we conclude that in the chosen range of between 500 and 2500 $h^{-1}$ kpc the results are quite robust against modest modifications of the concentration-mass relation. Anyhow we will add the scatter ($\sigma_{\log M}=0.003$) based on the 20 per cent modification into our systematic error budget.
\subsubsection{Off-centring}
\label{sec:off-centring}
In our lensing analysis, we will define cluster centres as the most likely centre candidates identified by the redMaPPer algorithm. There are several failure modes of this assumption: sometimes it is not unambiguous from the photometric data which of the cluster galaxies is the central one, other times the true central galaxy might be lost to masking or, e.g. for ongoing mergers, there might not be a single central galaxy at all.
These effects lead to a distribution of centre offsets that, on average, lower the cluster shear profiles on small scales. A lensing analysis could correct for this either by using an appropriately smoothed model \citep[cf., e.g.][for redMaPPer lensing analyses using this approach]{simet16,melchior16} or by correcting for the average mass bias incurred in the fitting process to off-centred haloes.
We do the latter, by shifting the centres of a fraction $(1-p_{\rm cen})$ of the simulated cluster fields by a (RA, dec) offset drawn from two independent Gaussians with a standard deviation of $340 h^{-1}$kpc, the best-fit parameters of \citet{cibirka16}.
We find that the off-centring causes a mass bias of $\Delta\log M_{\rm{200c}}=-0.033\pm0.002$ (see Fig. \ref{fig:deltam-no-off-centring}, lower panel), which we include in our analysis as a calibration factor. The dominant uncertainty in this offset does not result from the size of our simulated cluster sample, but from the width of the off-centring priors. From the constraints on off-centring derived in \citet{rykoff16} from X-ray and SZ estimates of redMaPPer cluster centres, we approximate the uncertainty on the effect of off-centring as 50 per-cent of the fiducial amplitude, i.e.~a systematic bias and uncertainty on mass of $\Delta \log M_{\rm{200c}}=-0.033$ and $\sigma_{\log M_{\rm{200c}}}=0.017$.
\subsection{Systematic error budget}
We summarize our budget of systematic errors here. The following effects contribute to systematic uncertainty of our weak lensing cluster mass measurements:
\begin{itemize}
\item multiplicative error in shape measurements
\item multiplicative error in our photometric estimate of $\beta$ 
\item dilution of the source sample with cluster members
\item mismatch of the fitted density profile to the truth
\item uncertainty of mass-concentration relation
\item uncertain prior on off-centring 
\end{itemize}
Contributions from the effects named above are described in detail in the previous sections. An overview is given in \autoref{tab:systematics}. Multiplicative uncertainties in the measured profile amplitude are scaled up by $4/3$ \citep[][their eqn. 53]{melchior16} to yield multiplicative errors in mass.
The sign of $\Delta\log\Delta\Sigma$ and $\Delta\log M_{\rm{200c}}$ is defined such that if $<0$, the respective effect lowers the reconstructed $\Delta\Sigma$ and mass. 
We correct for the \emph{mean} value of these biases in all massed we quote: the $\Delta\Sigma$ profiles we analyse with our likelihoods are corrected to account for the estimated mean value of the bias in $\beta$, and the recovered masses are re-scaled to correct for the biases expected from model profile mismatch and off-centring.
For each cluster, we calculate a systematic uncertainty in mass as the squared sum of all above effects, where only the $p(z)$ bias differs from cluster to cluster.
\begin{table*}
\caption{Systematic uncertainties in weak lensing mass likelihood.}
\label{tab:systematics}
\begin{tabular}{c|c|c|c|c|c|c|c}
\hline \hline
Effect      &  $\Delta\log\Delta\Sigma$ & $\sigma_{\log \Delta\Sigma}$ & $\Delta\log M_{\rm{200c}}$  & $\sigma_{\log M_{\rm{200c}}}$ & Section/Appendix & Notes \\
\hline \hline
Shear bias  & 0.000 & 0.009 & 0.000 & 0.011 & A1 & Assuming 2\% \\
$p(z)$ bias &   -  &   -  &  - & - & 2.6 & See Table \ref{tab:systematicspz123} \\
Cluster member dilution & 0.000 & 0.009 & 0.000 & 0.011 & 2.6 & Assuming 2\% \\
\hline
Model profile mismatch & - & - & 0.001 & 0.002 & A3 & - &\\
$C(M)$ & - & - & 0.000 & 0.003 & 2.9 & - &\\
Off-centring & - & - & -0.033 & 0.017 & A3.2& - & \\
\hline \hline
\end{tabular}
\end{table*}

\begin{table*}
\caption{Systematic uncertainties in weak lensing mass likelihood for individual clusters for all three subsamples.
\newline
Left column shows lower and upper uncertainty interval combined from p(z), shear bias and cluster member dilution, right column shows total systematic error budget. 
}
\label{tab:systematicspz123}
\begin{tabular}{c|c|c|c}
\hline \hline
Subsample & CODEX ID &  $\sigma_{\ln M_{200c},\sigma_{\rm{S}},\rm{syst}}$ & $\sigma_{\ln M_{200c},\rm{total,syst}}$ \\
\hline \hline
\textbf{S-I} &12451	&  0.021  &  0.027 \\
&13062	&  -      &  -     \\
&13390	&  0.021  &  0.027 \\
&16566	&  0.015  &  0.023 \\
&18127	&  0.027  &  0.032 \\
&24865	&  0.021  &  0.027 \\
&24872	&  0.017  &  0.024 \\
&24877	&  0.023  &  0.029 \\
&24981	&  0.017  &  0.024 \\
&25424	&  0.017  &  0.024 \\
&25953	&  0.019  &  0.026 \\
&27940	&  0.019  &  0.026 \\
&27974	&  0.019  &  0.026 \\
&29283	&  0.019  &  0.026 \\
&29284	&  0.019  &  0.026 \\
&29811	&  0.023  &  0.029 \\
&35361	&  0.023  &  0.029 \\
&35399	&  0.019  &  0.026 \\
&35646	&  0.015  &  0.023 \\
&36818	&  0.019  &  0.026 \\
&37098	&  -      &  -     \\
&41843	&  0.017  &  0.024 \\
&41911	&  0.019  &  0.026 \\
&43403	&  0.019  &  0.026 \\
&46649	&  0.021  &  0.027 \\
&47981	&  0.017  &  0.024 \\
&50492	&  0.023  &  0.029 \\
&50514	&  0.021  &  0.027 \\
&52480	&  0.017  &  0.024 \\
&53436	&  -      &  -     \\
&53495	&  -      &  -     \\
&54795	&  0.019  &  0.026 \\
&55181	&  0.021  &  0.027 \\
&56934	&  0.015  &  0.023 \\
&59915	&  0.021  &  0.027 \\
&64232	&  0.017  &  0.024 \\
\hline 
\textbf{S-II} &13311   &  0.019  &  0.026 \\
&13315   &  0.021  &  0.027 \\
&13380   &  0.017  &  0.024 \\
&13390   &  0.021  &  0.027 \\
&13391   &  0.019  &  0.026 \\
&13400   &  0.021  &  0.027 \\
&17449   &  0.017  &  0.024 \\
&17453   &  0.017  &  0.024 \\
&54652   &  0.019  &  0.026 \\
&56934   &  0.015  &  0.023 \\
&57017   &  0.017  &  0.024 \\
&60076   &  0.017  &  0.024 \\
&60131   &  0.019  &  0.026 \\
&60155   &  0.015  &  0.023 \\
&64565   &  0.017  &  0.024 \\
&64636   &  -      &  -     \\
&210288  &  0.015  &  0.023 \\
&210306  &  0.015  &  0.023 \\
\hline 
\textbf{S-III} &24925  &  0.019  &  0.026 \\
&27955  &  0.017  &  0.024 \\
&46647  &  0.015  &  0.023 \\
&54796  &  0.019  &  0.026 \\
\hline \hline
\end{tabular}
\end{table*}

\section{Fields with incomplete colour information}

\begin{figure*}
    \centering\includegraphics[width=7.45cm]{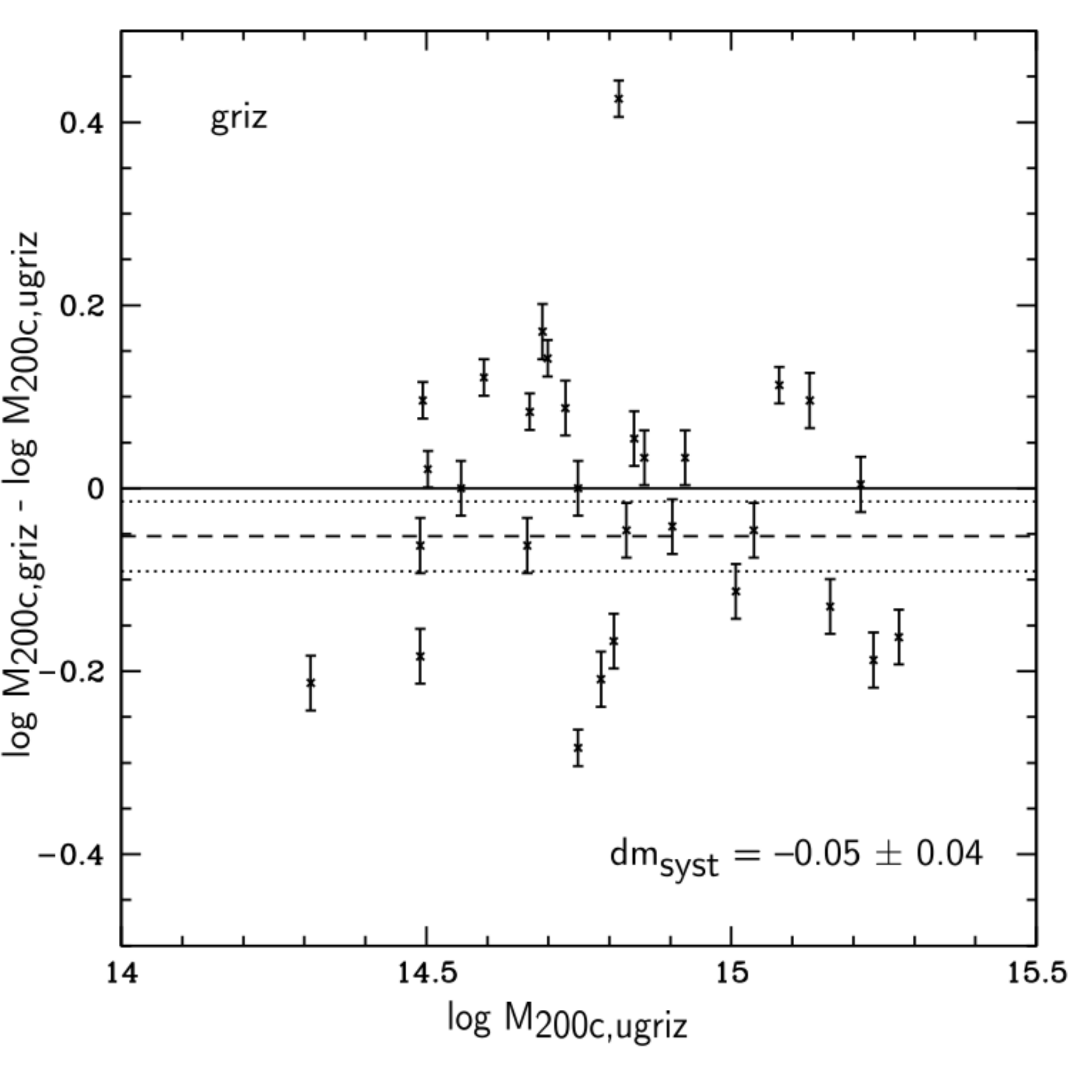}

    \centering\includegraphics[width=7.45cm]{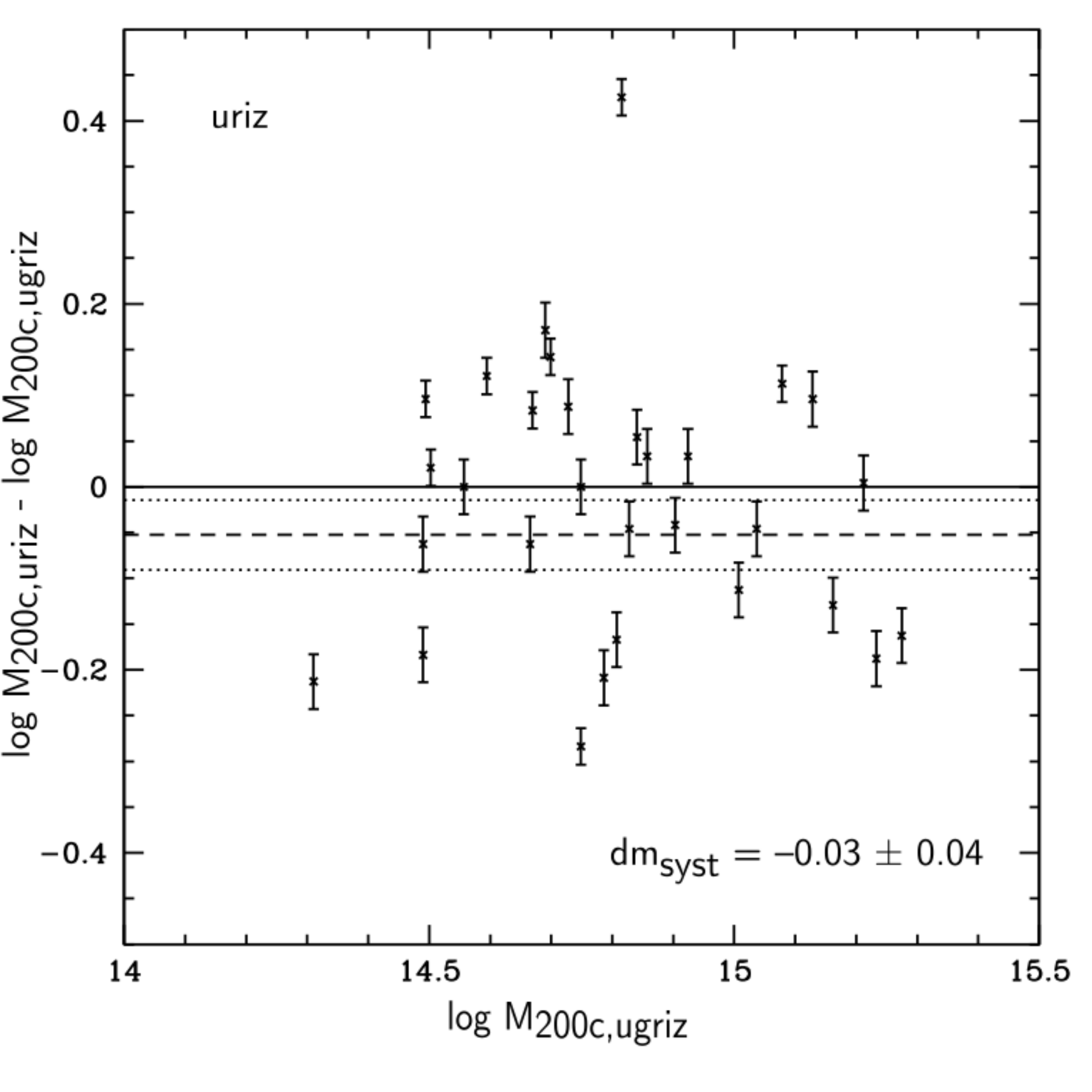}

    \centering\includegraphics[width=7.45cm]{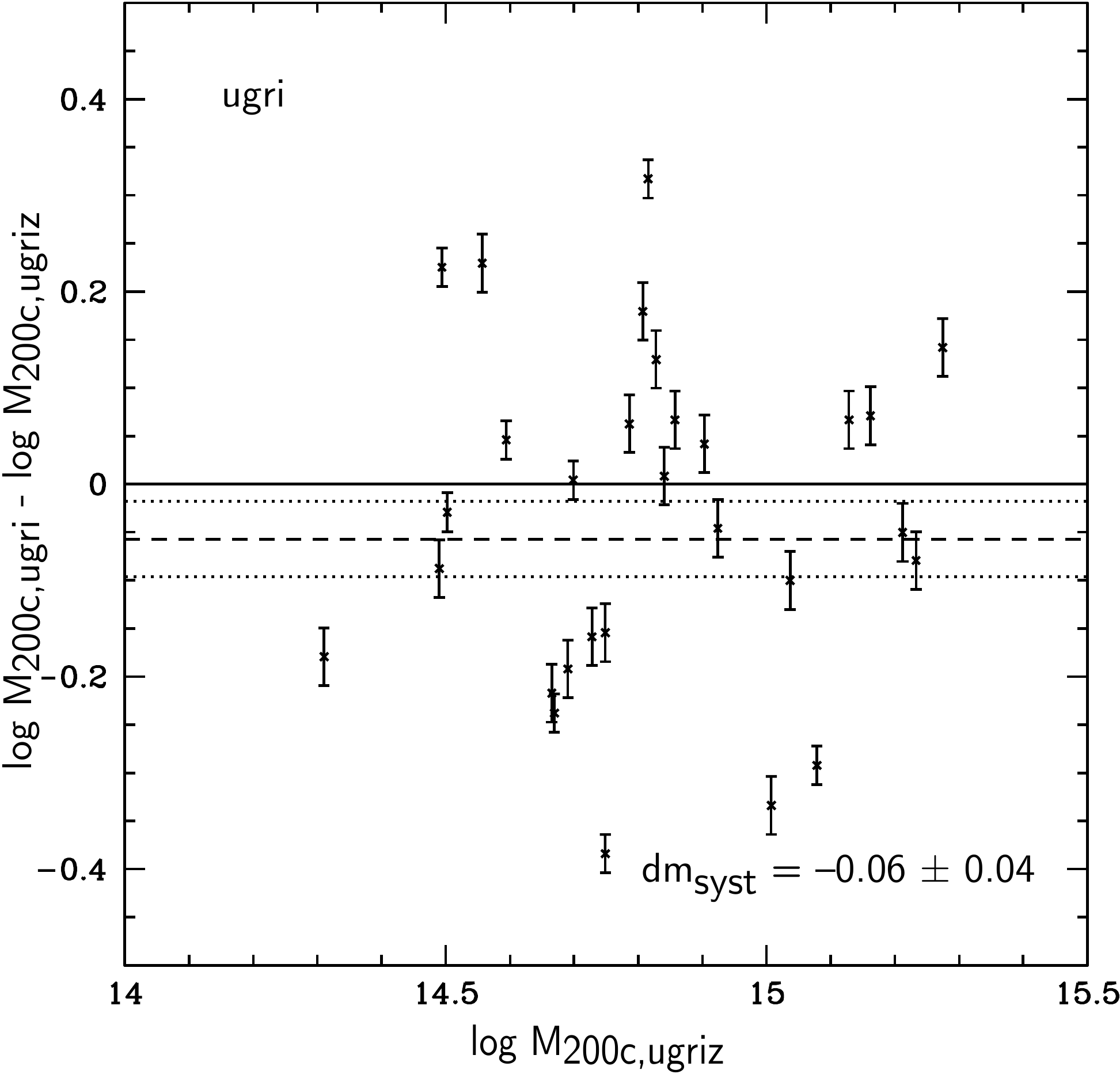}

\caption{Weak lensing mass estimates based on $ugriz$ photometry vs. four-filter photometry in $griz$ (upper left panel), $uriz$ (upper right panel), and $ugri$ (lower panel). The errorbars only show the systematic error in mass measurement since the by far larger statistical error should not significantly depend on the available filter combination. We correct our mass estimates based on $griz$, $uriz$, and $ugri$ photometry and include the uncertainty into our systematic error estimate.}
\label{fig:massbias-griz}

\end{figure*}

The CODEX survey was planned to be covered by five-filter observations ($ugriz$), but in several occasions these observations have remained incomplete yet, resulting in coverage of some clusters only in four or fewer bands. Nonetheless, the colour-magnitude decision tree enables us to use even those pointings with incomplete colour information. However, a reduced accuracy of the measurements cannot be ruled out, which would lead to a scatter weakening the constraints or even inducing a bias on the mass estimates. We therefore have a closer look at those 31 pointings with complete coverage in five bands and recalculate $\beta$ leaving out certain filter information. This step artificially creates full samples with incomplete colour information. In detail we analyze the following filter combinations: $griz$, $ugri$, and $uriz$ (see Fig. \ref{fig:massbias-griz}).

In more detail, first we estimate the geometrical distance ratio $\beta_{\rm{ugriz}}$ (cf. equation~\ref{eqn:beta}) using the complete five-filter $ugriz$ colour information and measure the galaxy cluster mass. This gives us a baseline mass estimate for our galaxy cluster sample. We repeat this mass estimate for each single cluster based on $\beta_{\rm{griz}}$, $\beta_{\rm{ugri}}$ and $\beta_{\rm{uriz}}$ and compare those masses to the five-band mass values. We do this by applying linear regression, weighting each cluster with the corresponding mass uncertainties ($w=1/(\sigma_{m1}^2+\sigma_{m2}^2)$). The uncertainties on the fit parameters are then calculated from jackknife estimates.

Given the large uncertainties in the individual masses, the uncertainties for the fit parameters are quite large as well. However, we do not detect a significant multiplicative bias within the uncertainties.

\subsection{Richness from three-band photometry}
Cluster richness is commonly estimated from four-band $griz$ photometric data. For a small subset of our clusters (3 with $ugri$, 1 with $uriz$), we do not have this full information available. In this section, we investigate whether richnesses estimated from three-band information only ($gri$ or $riz$) show considerable deviations from the fiducial $griz$ case.

\begin{figure*}
     \centering
         \centering
         \includegraphics[width=0.65\textwidth]{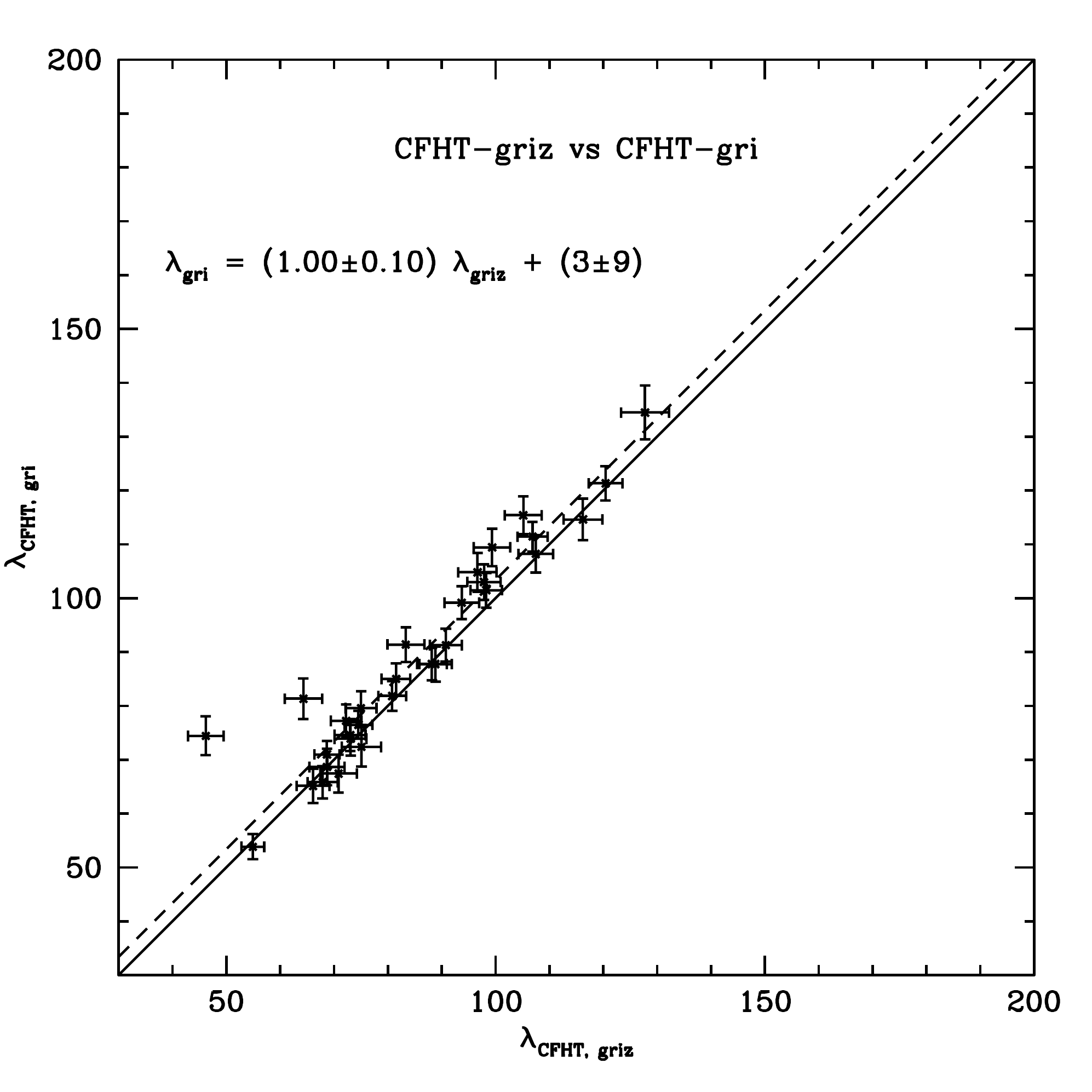}
         \centering
         \includegraphics[width=0.65\textwidth]{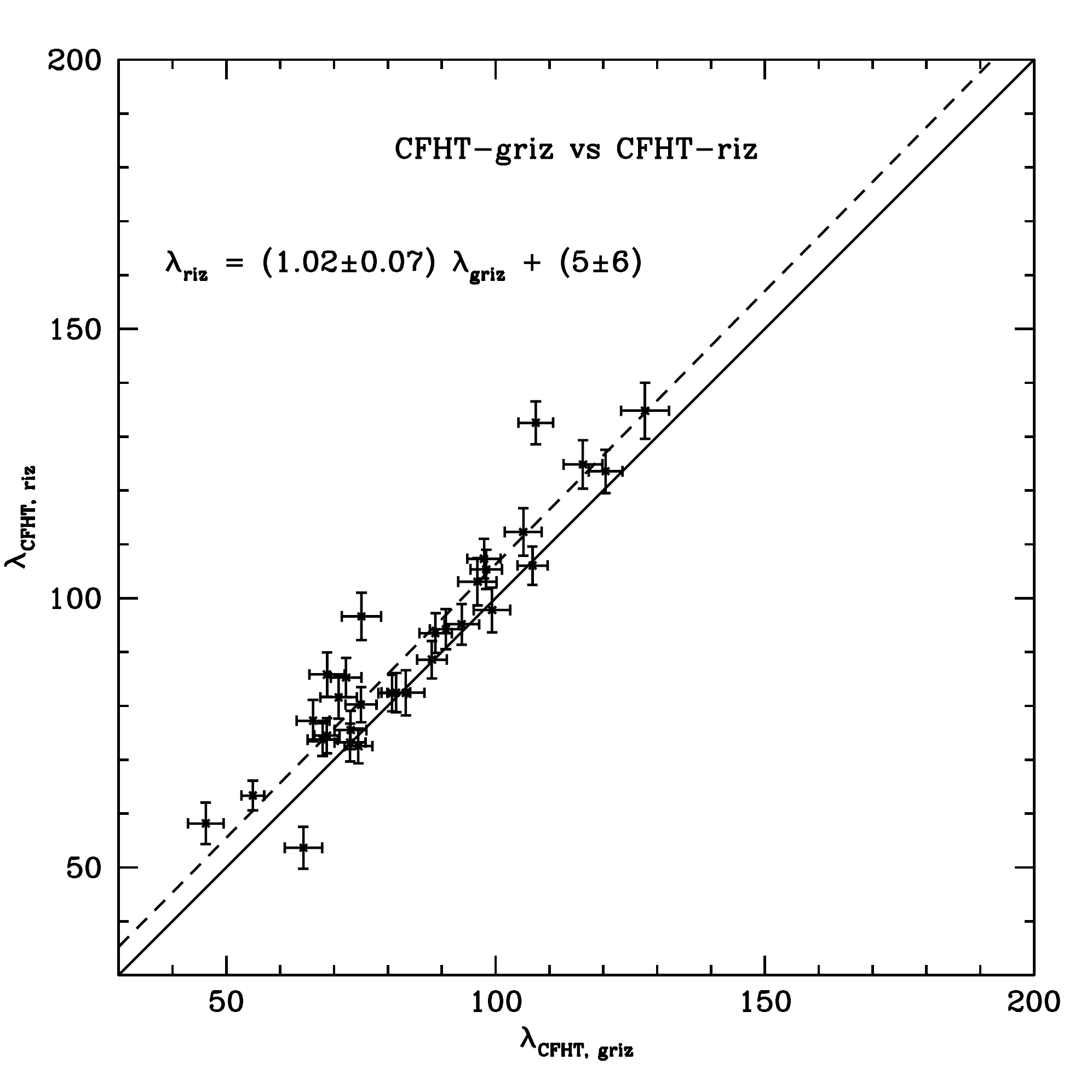}
     \caption{Richness estimated from CFHT imaging, griz vs. gri and griz vs. riz for all clusters covered in griz.}
        \label{fig:richness-richness-sdss}
\end{figure*}

To this end, we run redMaPPer on $gri$ and $riz$ data only, for all clusters with $griz$ observations from our pointed CODEX follow-up data. Comparison of these and the fiducial $griz$ richnesses are shown in (Fig. \ref{fig:richness-richness-sdss}). 
We fit the newly obtained richness estimates based on incomplete colour information with respect to the original values applying a linear regression and find that lack of $g$ or $z$-band does not change the richness estimates significantly.
The best-fit relations of CFHT-griz vs CFHT-gri and CFHT-griz vs CFHT-riz richnesses are approximated by $\lambda_{\mathrm{gri}} = (1.00 \pm 0.10)\lambda_{\mathrm{griz}} + (3 \pm 9)$, and $\lambda_{\mathrm{riz}} = (1.02 \pm 0.07)\lambda_{\mathrm{griz}} + (5 \pm 6)$, respectively. The best-fit relations show the CFHT-griz richness is more consistent with CFHT-riz richness, as a result we use this richness. 

The original plan was to incorporate all 36 clusters from sample S-I into the Bayesian analysis, described in section \ref{sec:application}, but since the calibration is too uncertain for the <5 band clusters, and the sample is too small, we decided to only include the 5 band filter clusters in the final analysis.

\section{Excluded weak lensing samples}

From the initial sample of 407 clusters, and our follow-up observations with different redshift and richness range than the initial sample, we define three lensing samples with distinct selection functions, that are not part of the analysis, due to lacking weak lensing mass information, CFHT richness information, or richness incompleteness. Even though we do not use these clusters to calibrate the richness--mass relation, their weak lensing mass measurements are robust, so we present their mass measurements, along with sky coordinates, X-ray luminosities, spectroscopic and optical redshifts, and SDSS richness.

\begin{itemize}
    \item 
    The definition of the first lensing sample of 36 clusters S-I is given in section \ref{sec:cluster catalog}. The S-I lensing sample is listed in two separate places, in Table \ref{tab:cleanedWL} (cleaned lensing sample of 25 clusters that went into the richness--mass analysis) and Table \ref{tab:primaryWL} (11 excluded clusters from the analysis). 
    \item Our second lensing sample of 18 clusters, hereafter S-II, is selected only by their ROSAT excess. Its position is required to fall inside the CFHTLS footprint, but we do not require a rich optical counterpart in the CFHT observations to be present. Therefore, all clusters in S-II have "-" listed in the CFHT richness and CFHT redshift columns in Appendix, Table \ref{tab:subsamples}.
    For the purpose of feasible lensing analysis, we also require that $z_{\rm{RM,SDSS}}>0.1$. We note that clusters 13390 and 56934 are in both S-I and S-II, but do not have CFHT richness estimates. Unlike S-I and S-III, which are our dedicated observations, S-II shapes are from the public CFHTLenS catalogs \citep{heymans12}, which was before the introduction of self-calibration version of the \emph{lensfit}.
    \item Our third lensing sample of 4 clusters, hereafter S-III, follow the same treatment as S-I, i.e., they are processed with the self-calibrating version of the \emph{lensfit}, as described in section \ref{sec:shape measurement}, following the calibration of \citet{fenechconti17}. Initially, these clusters fulfilled the selection criteria of the primary sample, but were later revised to $z_{\rm{RM,SDSS}} < 0.35$.
    
\end{itemize}

\begin{figure*}
\includegraphics[width=\textwidth]{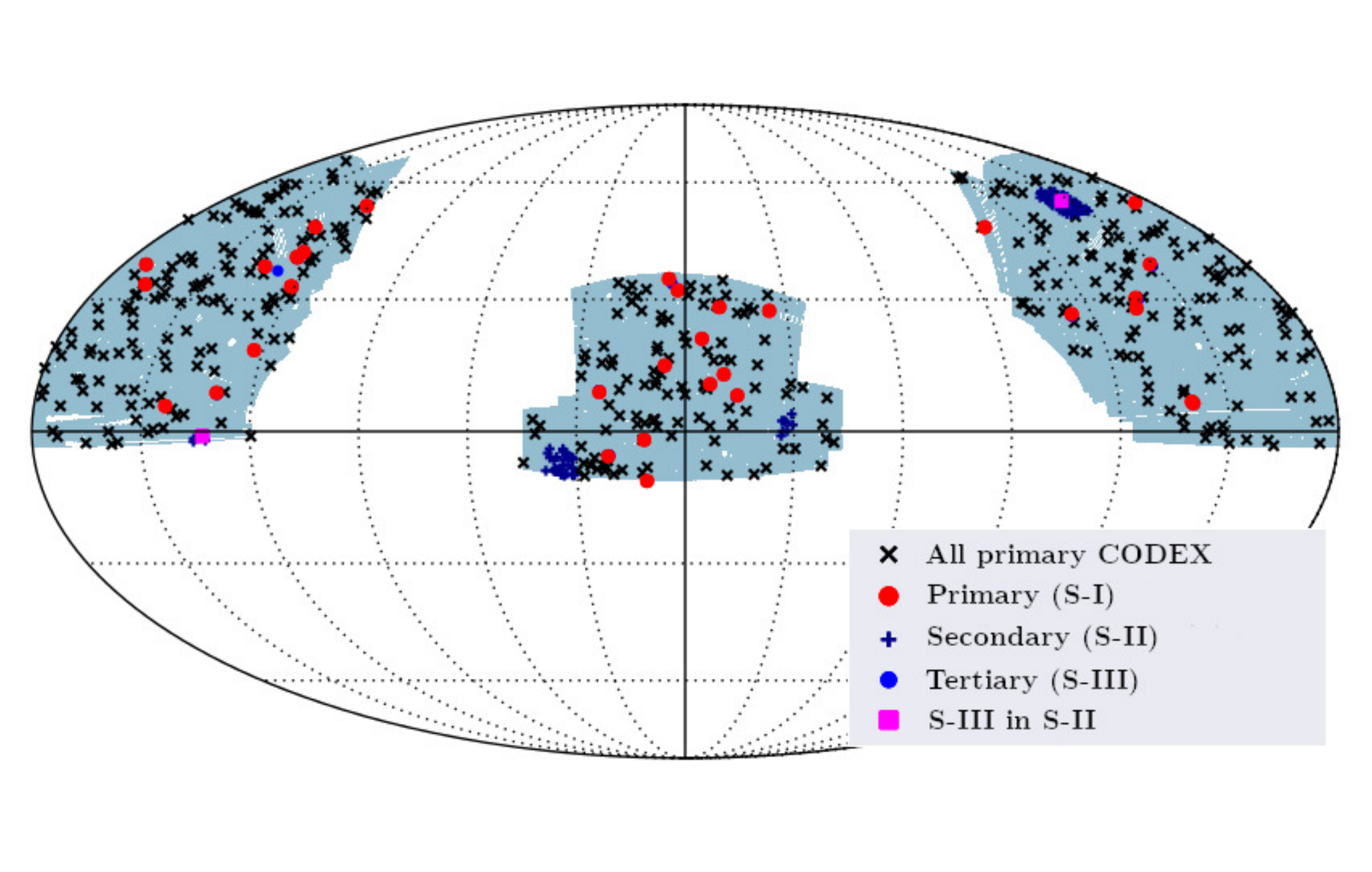}
\caption{Positions of CODEX clusters in the sky. SDSS footprint is shown as grey area. Black crosses correspond to initial CODEX sample of 407 clusters, red points to CODEX lensing sample of 36 clusters (S-I) and dark blue crosses to the secondary sample, (S-II). Additionally, the blue points correspond to tertiary sample (S-III). Most of the tertiary targets (S-III) are secondary objects in primary target pointings and thus overplotted in pink.}
\label{fig:codex_sky}
\end{figure*}

The positions in the sky of clusters in each of these samples is shown in \autoref{fig:codex_sky}.

\subsection{Imaging data for S-II}
In addition to S-I and S-III pointings, for S-II, we make use of the publicly available imaging data from the CFHT Lensing Survey\footnote{http://cfhtlens.org} (CFHTLenS, see \citealt{erben13,heymans12}) to include those CODEX clusters falling into the CFHTLS footprint and download the available reduced imaging data for 87 additional CFHT pointings. A summary of the all used images can be seen in Appendix D. The data reduction steps, detailed in section \ref{sec:imaging data}, for S-II remain the same as in S-I, and S-III.

For the cluster sample S-II falling into the footprint of the CFHTLS we create the corresponding multiband-photometry catalogues in the same way as described in section \ref{sec:photometry}.

For the shape measurement, in the case of the cluster sample S-II we make use of the publicly available \emph{lensfit} shape measurement data of CFHTLenS (see \citealt{miller13}) and download the corresponding catalogues from their website. In contrast to the \emph{lensfit} version used in this work the version used in the CFHTLenS data release was not self-calibrating yet. We therefore include the following correction terms to the measured ellipticities (see Eqns. 17 and 19 in \citealt{heymans12}), a multiplicative one:
\begin{equation}
m(\nu_{\rm SN},r) = \frac{B}{\log(\nu_{\rm SN})} \exp^{-r\, A \, \nu_{\rm SN}} \ ,
\label{eqn:mlensfit}
\end{equation}
with $A=0.057$ and $B=-0.37$, which has to be applied through a weighted ensemble average correction, rather than dividing by $(1 + m)$ on a galaxy-by-galaxy basis, and an additive one:
\begin{equation}
c_2 = \rm Max \left [\frac{{\it F} \log_{10}(\nu_{\rm SN}) - {\it G}}{1 + \left(\frac{r}{r_0}\right)^{\it H}}\,\, ,\,\, 0 \right] \, ,
\label{eqn:clensfit}
\end{equation}
with $F\ =\ 11.910,\ G\ =\ 12.715,\ H\ =\ 2.458,\ r_0\ =\ 0.01'$, which has to be added only to $e_2$.

Since our data likelihood function, in section \ref{sec:application}, assumes that all the clusters in the analysis have both SDSS and CFHT richness estimates, this excludes the entire sample S-II and two clusters (CODEX ID 46647 and 54796) from S-III. For the last two clusters in S-III (CODEX ID 24925 and 27955),with redshifts $z \sim 0.3$, we did not include them in the analysis, as they fell outside of our original redshift region of $0.35<z<0.65$.

\newpage

\begin{table*}
\caption{Lensing sample S-I clusters excluded from the richness--mass analysis.}
\label{tab:primaryWL}
\begin{adjustbox}{angle=90}
\begin{tabular}{c|c|c|c|c|c|c|c|c|c|c|c|c|c}
\hline \hline
CODEX      &  SPIDERS  & R.A.    &  Dec    &  R.A.   &   Dec    & Filters &   z   &    $\rm{z_{RM}}$  &  $\rm{\lambda_{RM}}$    &    $\rm{z_{RM}}$  & $\rm{\lambda_{RM}}$    &  $\rm{\log M_{200, WL}}$ & $\rm{L_X}$ \\
 ID & ID & opt & opt & X-ray & X-ray & CFHT & spec  & SDSS & SDSS & CFHT & CFHT & \ $\rm{M_{\odot}}$ & $\rm[h_{70}^{-2}\ 10^{44}erg/s]$\\
\hline \hline
 12451   & 1\_1235  &  08:04:39  &    53:25:43  &  08:04:38  &    53:25:24 &  ugriz  &    0.584  &  0.566  &  $103 \pm   35$    &  0.583  &   $ 46  \pm  3$ &   $ 14.79 _{ -0.33 } ^{ +0.22 } \pm 0.03 $ &  $  5.9 \pm  2.0 $   \\
 13062   & 1\_1759  &  12:24:51  &    54:19:29  &   12:24:53  &    54:19:45 & griz   &    0.467  &  0.463  &  $ 95 \pm   15$    &  0.467  &   $ 55  \pm  2$ &   $ -     _{       } ^{       }          $  &  $  5.5 \pm  1.7 $  \\
 13390$^a$   & 1\_2042  &  14:14:47  &    54:47:04  &  14:14:49  &    54:46:50 &  ugriz  &    0.618  &  0.623  &  $153 \pm   69$    &  -      &    -            &   $ 15.03 _{ -0.44 } ^{ +0.27 } \pm 0.03 $  &  $ 10.3 \pm  2.2 $  \\
 16566   & 1\_2639  &  08:42:31  &    47:49:19  &  08:42:28  &    47:50:03  &  ugriz  &    0.382  &  0.368  &  $108 \pm    7$    &  0.383  &   $120  \pm  3$ &   $ 14.61 _{ -0.29 } ^{ +0.20 } \pm 0.02 $  &  $  3.1 \pm  1.2 $  \\
 18127   & 1\_3918  &  16:56:53  &    47:48:55  &  16:56:56  &    47:47:27 &  griz   &    0.491  &  0.507  &  $ 88 \pm   24$    &  0.489  &   $ 39  \pm  2$ &   $ 14.93 _{ -1.93 } ^{ +0.57 } \pm 0.05 $  &  $  4.6 \pm  1.1 $  \\
 29811   & 1\_8150  &  11:06:08  &    33:33:40  &  11:06:07  &    33:33:36 &  ugriz  &    0.488  &  0.495  &  $194 \pm   31$    &  -      &     -           &   $ 15.21 _{ -0.18 } ^{ +0.14 } \pm 0.03 $  &  $  5.6 \pm  1.8 $  \\
 35646   & 1\_11555 &  16:23:35  &    26:34:14  &  16:23:42  &    26:33:43  &  griz   &    0.427  &  0.427  &  $108 \pm   10$    &  0.408  &   $ 99  \pm  3$ &   $ 15.19 _{ -2.19 } ^{ +0.31 } \pm 0.04 $  &  $  4.8 \pm  1.1 $   \\
 36818   & 1\_11870 &  22:20:16  &    27:20:03  &  22:20:21  &    27:21:00 &  ugriz  &    0.581  &  0.578  &  $ 80 \pm   64$    &  0.559  &   $ 64  \pm  3$ &   $ 15.04 _{ -0.24 } ^{ +0.18 } \pm 0.03 $  &  $  3.0 \pm  1.5 $  \\
 37098   & 1\_12115 &  23:19:17  &    28:12:01 &  23:19:16  &    28:12:01  &  ugriz  &    0.544  &  0.573  &  $104 \pm   35$    &  0.543  &   $ 75  \pm  4$ &   $ -     _{       } ^{       }          $  &  $ 10.0 \pm  4.4 $  \\
 53436   & 1\_20622 &  14:37:50  &    06:16:41 &  14:37:50  &    06:15:59  &  ugri   &    0.540  &  0.527  &  $ 96 \pm   32$    &  -      &    -            &   $ -     _{       } ^{       }          $  &  $ 16.6 \pm 10.2 $  \\
 53495   & 1\_20670 &  14:39:32  &    06:40:15  &  14:39:21  &    06:38:23 &  ugri   &    0.462  &  0.444  &  $ 90 \pm   23$    &  -      &    -            &   $ -     _{       } ^{       }          $  &  $  6.4 \pm  2.1 $   \\
 56934$^a$   & 1\_22022 &  08:52:17  &   -01:01:36  &  08:52:11  &   -01:01:32  &  ugriz  &    0.459  &  0.476  &  $ 90 \pm   19$    &  -      &    -            &   $ 15.01 _{ -0.28 } ^{ +0.20 } \pm 0.02 $  &  $  4.2 \pm  1.6 $  \\
\hline \hline
\end{tabular}
\end{adjustbox}
\begin{flushleft}
$^a$observed within CFHTLS Wide
\end{flushleft}
\end{table*}

\pagebreak

\begin{table*}
\caption{Weak lensing mass catalogue for the sample S-II (covered by CFHTLS Wide, see \citealt{erben13}), and S-III ($z < 0.35$). }
\label{tab:subsamples}
\begin{adjustbox}{angle=90}
\begin{tabular}{c|c|c|c|c|c|c|c|c|c|c|c|c|c|c}
\hline \hline
Sub- & CODEX      &  SPIDERS  &   R.A.   &   Dec &  R.A.   &   Dec   & Filters &   z   &    $\rm{z_{RM}}$  &  $\rm{\lambda_{RM}}$    &    $\rm{z_{RM}}$  & $\rm{\lambda_{RM}}$    &  $\rm{\log M_{200, WL}}$  & $\rm{L_X}$\\
sample & ID & ID & opt & opt & X  & X & CFHT & spec & SDSS & SDSS & CFHT & CFHT & \ $\rm{M_{\odot}}$ & $\rm[h_{70}^{-2}\ 10^{44}erg/s]$\\
\hline \hline
\textbf{S-II} & 13311    &  1\_1971   &  14:00:40  &    53:26:57  &   14:00:50  &    53:24:34  &  ugriz  &    0.410  &  0.398  &  $ 33 \pm    5$    &  -      &     -         &  $ 14.40      _{ -0.65 }       ^{ +0.32 } \pm 0.03 $ &  $  2.1 \pm 0.7 $ \\
& 13315    &  1\_1974   &  14:19:11  &    53:26:21  & 14:19:20  &    53:25:27  & ugriz  &    0.633  &  0.621  &  $ 41 \pm   73$    &  -      &     -         &  $ 14.95      _{ -0.32 }       ^{ +0.22 } \pm 0.03 $   &  $  7.6 \pm 2.0 $ \\
& 13380    &  1\_2034   &  14:35:28  &    55:07:52  & 14:35:30  &    55:07:49  &  ugriz  &    0.140  &  0.14   &  $ 86 \pm    3$    &  -      &     -         &  $ 14.91      _{ -0.24 }       ^{ +0.18 } \pm 0.02 $   &  $  1.3 \pm 0.1 $\\
& 13390    &  1\_2042   &  14:14:47  &    54:47:04  & 14:14:49  &    54:46:50  & ugriz  &    0.618  &  0.623  &  $153 \pm   69$    &  -      &     -         &  $ 15.03      _{ -0.44 }       ^{ +0.27 } \pm 0.03 $   &  $ 10.3 \pm 2.2 $ \\
& 13391    &  1\_2043   &  14:34:25  &    54:48:51  & 14:34:17  &    54:47:14  & ugriz  &    0.139  &  0.143  &  $ 42 \pm    2$    &  -      &     -         &  $ 14.43      _{ -0.39 }       ^{ +0.24 } \pm 0.03 $   &  $  0.3 \pm 0.1 $\\
& 13400    &  1\_2052   &  14:37:14  &    54:02:14  & 14:37:17  &    54:03:10  &  ugriz  &    0.499  &  0.486  &  $ 40 \pm   13$    &  -      &     -         &  $ 14.52      _{ -0.65 }       ^{ +0.32 } \pm 0.03 $   &  $  2.4 \pm 0.9 $ \\
& 17449    &  1\_3406   &  14:11:21  &    52:12:10  & 14:11:24  &    52:12:30  &  ugriz  &    0.462  &  0.441  &  $ 53 \pm    8$    &  -      &     -         &  $ 14.68      _{ -0.38 }       ^{ +0.25 } \pm 0.02 $   &  $  8.4 \pm 1.6 $\\
& 17453    &  1\_3410   &  14:06:12  &    52:08:13  & 14:06:15  &    52:07:01  &  ugriz  &    0.291  &  0.309  &  $ 36 \pm    3$    &  -      &     -         &  $ 14.49      _{ -0.40 }       ^{ +0.25 } \pm 0.02 $   &  $  0.9 \pm 0.3 $ \\
& 54652    &  1\_21023  &  22:02:22  &    03:52:39  & 22:02:22  &    03:53:02  & ugriz  &    0.491  &  0.534  &  $ 30 \pm   23$    &  -      &     -         &  $ 13.37      _{ -0.37 }       ^{ +1.09 } \pm 0.03 $   &  $  8.2 \pm 2.5 $\\
& 56934    &  1\_22022  &  08:52:17  &   -01:01:36  & 08:52:11  &   -01:01:32  &  ugriz  &    0.459  &  0.476  &  $ 90 \pm   19$    &  -      &     -         &  $ 15.01      _{ -0.28 }       ^{ +0.20 } \pm 0.02 $   &  $  4.2 \pm 1.6 $\\
& 57017    &  1\_22083  &  09:01:31  &   -01:39:17  & 09:01:33  &   -01:38:33  & ugriz  &    0.291  &  0.33   &  $ 96 \pm    7$    &  -      &     -         &  $ 14.91      _{ -0.28 }       ^{ +0.20 } \pm 0.02 $   &  $  1.8 \pm 0.6 $\\
& 60076    &  1\_24090  &  02:14:41  &   -04:34:02  & 02:14:39  &   -04:32:54  & ugriz  &    0.140  &  0.145  &  $ 59 \pm    3$    &  -      &     -         &  $ 14.16      _{ -0.88 }       ^{ +0.35 } \pm 0.02 $   &  $  1.0 \pm 0.2 $\\
& 60131    &  1\_24140  &  02:23:59  &   -08:35:41  & 02:23:54  &   -08:36:15  &  ugriz  &    0.271  &  0.271  &  $ 70 \pm    4$    &  -      &     -         &  $ 14.91      _{ -0.24 }       ^{ +0.18 } \pm 0.03 $   &  $  2.0 \pm 0.6 $\\
& 60155    &  1\_24161  &  02:31:41  &   -04:52:57  & 02:31:46  &   -04:51:38  & ugriz  &    0.185  &  0.188  &  $ 84 \pm    3$    &  -      &     -         &  $ 14.58      _{ -0.43 }       ^{ +0.27 } \pm 0.02 $   &  $  2.3 \pm 0.5 $\\
& 64565    &  1\_24930  &  02:03:29  &   -09:49:00  &  02:03:29  &   -09:49:36  & ugriz  &    0.322  &  0.318  &  $ 80 \pm    4$    &  -      &     -         &  $ 14.99      _{ -0.21 }       ^{ +0.16 } \pm 0.02 $   &  $  3.0 \pm 0.8 $\\
& 64636    &  1\_24941  &  02:23:33  &   -08:57:39  & 02:23:36  &   -08:58:54  &   ugriz  &    0.415  &  0.435  &  $ 45 \pm    7$    &  -      &     -         &  $ -          _{       }       ^{       }          $   &  $  3.9 \pm 1.9 $\\
& 210288   &  2\_1815   &  14:28:38  &    56:51:39  & 14:28:29  &    56:52:52  &  ugriz  &    0.106  &  0.106  &  $ 60 \pm    3$    &  -      &     -         &  $ 14.04      _{ -0.86 }       ^{ +0.35 } \pm 0.02 $   &  $  0.5 \pm 0.1 $\\
& 210306   &  2\_1833   &  14:27:25  &    55:45:00  & 14:27:28  &    55:45:18  &  ugriz  &    0.133  &  0.132  &  $ 59 \pm    3$    &  -      &     -         &  $ 14.39      _{ -0.37 }       ^{ +0.23 } \pm 0.02 $   &  $  0.8 \pm 0.1 $\\
\hline 
\textbf{S-III} &   24925  &  1\_5784   &  08:35:40  &    36:52:26  & 08:35:34  &    36:52:25  & ugriz  &    0.292  &  0.304  &  $ 97 \pm    4$    &  0.288  &   $ 85  \pm  3$ &  $ 14.86      _{ -0.21 }      ^{ +0.16 } \pm 0.03 $   &  $  2.5 \pm 0.6 $\\
& 27955  &  1\_7327   &  00:15:56  &    33:41:43  & 00:15:56  &    33:42:15  & uriz   &    0.303  &  0.31   &  $ 97 \pm    4$    &  0.246  &   $ 79  \pm  4$ &  $ 14.24      _{ -1.24 }      ^{ +1.21 } \pm 0.04 $   &  $  3.0 \pm 0.9 $\\
& 46647  &  1\_17213  &  01:35:45  &    09:10:03  & 01:35:43  &    09:10:08  & ugriz  &    0.261  &  0.26   &  $ 69 \pm    4$    &  -      &   -             &  $ 14.72      _{ -0.28 }      ^{ +0.19 } \pm 0.02 $  &  $  1.5 \pm 0.5 $\\
& 54796  &  1\_21154  &  23:03:25  &    07:56:43  & 23:03:25  &    07:56:47  & ugriz  &    0.159  &  0.168  &  $ 42 \pm    2$    &  -      &   -             &  $ 14.33      _{ -0.43 }      ^{ +0.25 } \pm 0.03 $  &  $  2.4 \pm 0.4 $\\
\hline \hline
\end{tabular}
\end{adjustbox}
\end{table*}

\pagebreak
\onecolumn
\section{Observational properties of the CODEX imaging data}
\label{app:imaging data}
\begin{longtable}{@{}cccccc@{}}
\caption{CODEX cluster sample S-I ($\lambda_{\rm{RM,SDSS}}>60$ and $0.35 < z < 0.65$), targeted fields only}
\label{tab:primary images} \\
\hline \hline
CODEX ID	&	Filter	&	N. of exp.	&	expos. time	&	$m_{\rm{lim}}$	&	Seeing\\
  & & & [s] & [AB mag]  & [$''$]\\
\hline \hline
12451	&	u.MP9301	&	5	&	2601		&	25.20 &	0.83	\\
12451	&	g.MP9401	&	5	&	1051		&	25.12 &	1.08	\\
12451	&	r.MP9601	&	9	&	2492		&	25.01 &	0.94	\\
12451	&	i.MP9702	&	15	&	8403		&	24.89 &	0.61	\\
12451	&	z.MP9801	&	2	&	1081		&	22.56 &	1.04	\\
\hline
13062	&	g.MP9401	&	7	&	1471		&	25.34 &	0.82	\\
13062	&	r.MP9601	&	6	&	1261		&	24.65 &	0.77	\\
13062	&	i.MP9702	&	3	&	1681		&	24.40 &	0.98	\\
13062	&	z.MP9801	&	4	&	2321		&	23.41 &	0.75	\\
\hline
16566	&	u.MP9301	&	8	&	4482	&	24.92 &	1.15	\\
16566	&	g.MP9401	&	3	&	1021	&	24.43 &	0.99	\\
16566	&	r.MP9601	&	6	&	3001	&	24.86 &	1.17	\\
16566	&	i.MP9702	&	8	&	4482	&	24.80 &	0.71	\\
16566	&	z.MP9801	&	8	&	4482	&	23.57 &	0.73	\\
\hline
18127	&	g.MP9401	&	3	&	6301		&	24.71 &	0.82	\\
18127	&	r.MP9601	&	5	&	1291		&	24.48 &	0.74	\\
18127	&	i.MP9702	&	8	&	4482		&	24.60 &	0.54	\\
18127	&	z.MP9801	&	5	&	2701		&	23.34 &	0.78	\\
\hline
24865	&	u.MP9301	&	5	&	2601		&	25.24 &	0.79	\\
24865	&	g.MP9401	&	6	&	1261		&	25.16 &	0.91	\\
24865	&	r.MP9601	&	5	&	1351		&	24.63 &	0.75	\\
24865	&	i.MP9702	&	7	&	3921		&	24.71 &	0.73	\\
24865	&	z.MP9801	&	3	&	1621		&	23.12 &	0.68	\\
\hline
24872	&	u.MP9301	&	5	&	2801	&	25.32 &	1.05	\\
24872	&	g.MP9401	&	3	&	1021	&	25.13 &	0.97	\\
24872	&	r.MP9601	&	3	&	1494	&	24.81 &	0.87	\\
24872	&	i.MP9702	&	8	&	4482	&	24.77 &	0.84	\\
24872	&	z.MP9801	&	4	&	2241	&	23.60 &	0.88	\\
\hline
24877	&	u.MP9301	&	5	&	2801		&	25.32 &	1.05	\\
24877	&	g.MP9401	&	3	&	1021		&	25.13 &	0.97	\\
24877	&	r.MP9601	&	3	&	1494		&	24.81 &	0.87	\\
24877	&	i.MP9702	&	8	&	4482		&	24.77 &	0.84	\\
24877	&	z.MP9801	&	4	&	2241		&	23.60 &	0.88	\\
\hline
24981	&	u.MP9301	&	5	&	2601		&	25.30 &	0.93	\\
24981	&	g.MP9401	&	5	&	1601		&	25.33 &	0.65	\\
24981	&	r.MP9601	&	12	&	7406		&	25.49 &	0.64	\\
24981	&	i.MP9701	&	4	&	961		&	23.80 &	0.89	\\
24981	&	i.MP9702	&	6	&	3361		&	24.59 &	0.60	\\
24981	&	z.MP9801	&	4	&	1441		&	22.91 &	0.81	\\
\hline
25424	&	u.MP9301	&	15	&	7803		&	25.73 &	1.03	\\
25424	&	g.MP9401	&	4	&	841		&	24.99 &	0.76	\\
25424	&	r.MP9601	&	5	&	1351		&	24.51 &	0.65	\\
25424	&	i.MP9702	&	8	&	4482		&	24.73 &	0.65	\\
25424	&	z.MP9801	&	4	&	2161		&	23.27 &	0.46	\\
\hline
25953	&	u.MP9301	&	4	&	2081		&	25.09 &	0.85	\\
25953	&	g.MP9401	&	5	&	1051		&	25.21 &	0.91	\\
25953	&	r.MP9601	&	4	&	991		&	24.58 &	0.84	\\
25953	&	i.MP9702	&	8	&	4482		&	24.92 &	0.72	\\
25953	&	z.MP9801	&	4	&	2161		&	23.62 &	0.83	\\
\hline \hline
\pagebreak
\hline \hline
CODEX ID	&	Filter	&	N. of exp.	&	expos. time	&	$m_{\rm{lim}}$	&	Seeing \\
  & & & [s] & [AB mag]  & [$''$]\\
\hline \hline
27940	&	u.MP9301	&	5	&	2861		&	25.21 &	0.81	\\
27940	&	g.MP9401	&	3	&	1080		&	25.04 &	0.68	\\
27940	&	r.MP9601	&	3	&	1500		&	24.33 &	0.57	\\
27940	&	i.MP9702	&	9	&	5042		&	24.77 &	0.91	\\
27940	&	z.MP9801	&	4	&	2373		&	23.36 &	0.86	\\
\hline
27974	&	u.MP9301	&	4	&	2289		&	25.23 &	0.82	\\
27974	&	g.MP9401	&	3	&	1080		&	25.01 &	0.65	\\
27974	&	r.MP9601	&	3	&	1500		&	24.31 &	0.67	\\
27974	&	i.MP9702	&	8	&	4481		&	24.65 &	0.44	\\
27974	&	z.MP9801	&	4	&	2373		&	23.29 &	0.86	\\
\hline
29283	&	u.MP9301	&	5	&	2601		&	25.22 &	0.90	\\
29283	&	g.MP9401	&	6	&	1261		&	25.23 &	1.20	\\
29283	&	r.MP9601	&	3	&	930		&	24.45 &	0.76	\\
29283	&	i.MP9702	&	8	&	4482		&	24.84 &	0.56	\\
29283	&	z.MP9801	&	4	&	2161		&	23.24 &	0.87	\\
\hline
29284	&	u.MP9301	&	5	&	2601		&	25.22 &	0.90	\\
29284	&	g.MP9401	&	6	&	1261		&	25.23 &	1.20	\\
29284	&	r.MP9601	&	3	&	930		&	24.45 &	0.76	\\
29284	&	i.MP9702	&	8	&	4482		&	24.84 &	0.56	\\
29284	&	z.MP9801	&	4	&	2161		&	23.24 &	0.87	\\
\hline
29811	&	u.MP9301	&	4	&	2081		&	24.85 &	0.84	\\
29811	&	g.MP9401	&	10	&	2102		&	25.45 &	0.93	\\
29811	&	r.MP9601	&	8	&	1981		&	24.86 &	0.78	\\
29811	&	i.MP9702	&	6	&	3361		&	24.51 &	0.44	\\
29811	&	z.MP9801	&	4	&	2161		&	23.26 &	0.51	\\
\hline
35361	&	u.MP9301	&	10	&	5202		&	25.49 &	0.77	\\
35361	&	g.MP9401	&	5	&	1051		&	25.06 &	0.96	\\
35361	&	r.MP9601	&	6	&	1561		&	24.63 &	0.94	\\
35361	&	i.MP9702	&	7	&	3921		&	24.68 &	0.57	\\
35361	&	z.MP9801	&	4	&	2321		&	23.14 &	0.83	\\
\hline
35399	&	u.MP9301	&	7	&	3641		&	24.94 &	0.94	\\
35399	&	g.MP9401	&	3	&	901		&	24.81 &	0.62	\\
35399	&	r.MP9601	&	3	&	1380		&	24.39 &	0.67	\\
35399	&	i.MP9702	&	8	&	4481		&	24.58 &	0.55	\\
35399	&	z.MP9801	&	4	&	2161		&	23.35 &	0.88	\\
\hline
35646	&	g.MP9401	&	5	&	1601		&	25.44 &	0.72	\\
35646	&	r.MP9601	&	12	&	7178		&	25.42 &	0.58	\\
35646	&	i.MP9701	&	5	&	1201		&	24.08 &	0.60	\\
35646	&	z.MP9801	&	7	&	2521		&	23.51 &	0.97	\\
\hline
36818	&	u.MP9301	&	5	&	2861		&	25.29 &	1.07	\\
36818	&	g.MP9401	&	3	&	1020		&	25.03 &	0.82	\\
36818	&	r.MP9601	&	3	&	1501		&	24.48 &	0.88	\\
36818	&	i.MP9702	&	10	&	5602		&	24.73 &	0.45	\\
36818	&	z.MP9801	&	10	&	5884		&	23.69 &	0.63	\\
\hline
37098	&	u.MP9301	&	5	&	2861		&	25.25 &	0.85	\\
37098	&	g.MP9401	&	3	&	1020		&	25.10 &	0.65	\\
37098	&	r.MP9601	&	3	&	1500		&	24.69 &	0.60	\\
37098	&	i.MP9702	&	8	&	4481		&	24.66 &	0.70	\\
37098	&	z.MP9801	&	4	&	2373		&	23.38 &	0.49	\\
\hline
41843	&	u.MP9301	&	5	&	2801		&	25.30 &	0.83	\\
41843	&	g.MP9401	&	3	&	1020		&	25.12 &	0.71	\\
41843	&	r.MP9601	&	6	&	2989		&	24.98 &	0.67	\\
41843	&	i.MP9702	&	8	&	4481		&	24.90 &	0.73	\\
41843	&	z.MP9801	&	8	&	4481		&	23.52 &	0.67	\\
\hline \hline
\pagebreak
\hline \hline
CODEX ID	&	Filter	&	N. of exp.	&	expos. time	&	$m_{\rm{lim}}$	&	Seeing \\
  & & & [s] & [AB mag]  & [$''$]\\
\hline \hline
41911	&	u.MP9301	&	6	&	3361	&	25.37 &	0.86	\\
41911	&	g.MP9401	&	3	&	1020	&	25.10 &	0.91	\\
41911	&	r.MP9601	&	3	&	1494	&	24.49 &	0.99	\\
41911	&	i.MP9702	&	8	&	4481	&	24.74 &	0.65	\\
41911	&	z.MP9801	&	4	&	2241	&	23.33 &	0.64	\\
\hline
43403	&	u.MP9301	&	5	&	2601		&	25.22 &	0.84	\\
43403	&	g.MP9401	&	5	&	1071		&	25.06 &	0.87	\\
43403	&	r.MP9601	&	3	&	630		&	24.25 &	0.87	\\
43403	&	i.MP9702	&	8	&	4481		&	24.70 &	0.54	\\
43403	&	z.MP9801	&	2	&	1080		&	22.47 &	0.41	\\
\hline
46649	&	u.MP9301	&	5	&	2801		&	24.96 &	0.79	\\
46649	&	g.MP9401	&	3	&	1020		&	24.76 &	0.62	\\
46649	&	r.MP9601	&	4	&	1992		&	24.68 &	0.82	\\
46649	&	i.MP9702	&	9	&	5041		&	24.67 &	0.51	\\
46649	&	z.MP9801	&	4	&	2241		&	23.44 &	0.77	\\
\hline
47981	&	u.MP9301	&	5	&	2800		&	25.20 &	1.21	\\
47981	&	g.MP9401	&	3	&	1020		&	25.08 &	1.17	\\
47981	&	r.MP9601	&	4	&	1992		&	24.86 &	0.65	\\
47981	&	i.MP9702	&	9	&	5041		&	24.81 &	0.73	\\
47981	&	z.MP9801	&	4	&	2240		&	23.27 &	0.53	\\
\hline
50492	&	u.MP9301	&	5	&	2801		&	25.16 &	0.56	\\
50492	&	g.MP9401	&	3	&	1080		&	24.91 &	0.56	\\
50492	&	r.MP9601	&	6	&	3001		&	24.87 &	0.42	\\
50492	&	i.MP9702	&	8	&	4481		&	24.58 &	0.70	\\
50492	&	z.MP9801	&	4	&	2241		&	23.11 &	0.58	\\
\hline
50514	&	u.MP9301	&	5	&	2801		&	25.24 &	0.95	\\
50514	&	g.MP9401	&	3	&	1020		&	25.04 &	0.78	\\
50514	&	r.MP9601	&	5	&	2491		&	24.67 &	0.71	\\
50514	&	i.MP9702	&	8	&	4481		&	24.80 &	0.72	\\
50514	&	z.MP9801	&	4	&	2241		&	23.20 &	0.59	\\
\hline
52480	&	u.MP9301	&	5	&	2600		&	25.18 &	0.71	\\
52480	&	g.MP9401	&	3	&	900		&	25.00 &	0.79	\\
52480	&	r.MP9601	&	3	&	1380		&	24.56 &	0.94	\\
52480	&	i.MP9702	&	10	&	5601		&	24.71 &	0.76	\\
52480	&	z.MP9901	&	4	&	2241		&	22.91 &	0.81	\\
\hline
53436	&	u.MP9301	&	7	&	3641		&	25.30 &	0.89	\\
53436	&	g.MP9401	&	3	&	900		&	24.88 &	0.83	\\
53436	&	r.MP9601	&	7	&	3221		&	25.04 &	1.04	\\
53436	&	i.MP9702	&	8	&	4481		&	24.54 &	0.74	\\
\hline
53495	&	u.MP9301	&	7	&	3641		&	25.30 &	0.89	\\
53495	&	g.MP9401	&	3	&	900		&	24.88 &	0.83	\\
53495	&	r.MP9601	&	7	&	3221		&	25.04 &	1.04	\\
53495	&	i.MP9702	&	8	&	4481		&	24.54 &	0.74	\\
\hline
54795	&	u.MP9301	&	5	&	2801		&	25.13 &	0.94	\\
54795	&	g.MP9401	&	3	&	1020		&	24.97 &	1.13	\\
54795	&	r.MP9601	&	3	&	1494		&	24.16 &	0.74	\\
54795	&	i.MP9702	&	8	&	4481		&	24.55 &	0.64	\\
54795	&	z.MP9801	&	4	&	2240		&	23.18 &	0.70	\\
\hline
55181	&	u.MP9301	&	5	&	2801		&	25.02 &	1.18	\\
55181	&	g.MP9401	&	3	&	1080		&	24.94 &	1.19	\\
55181	&	r.MP9601	&	6	&	3001		&	24.75 &	0.74	\\
55181	&	i.MP9702	&	8	&	4481		&	24.56 &	0.62	\\
55181	&	z.MP9801	&	4	&	2240		&	22.99 &	0.67	\\
\hline \hline
\pagebreak
\hline \hline
CODEX ID	&	Filter	&	N. of exp.	&	expos. time	&	$m_{\rm{lim}}$	&	Seeing \\
  & & & [s] & [AB mag]  & [$''$]\\
\hline \hline
59915	&	u.MP9301	&	5	&	2800		&	25.08 &	0.96	\\
59915	&	g.MP9401	&	3	&	1020		&	24.90 &	0.89	\\
59915	&	r.MP9601	&	3	&	1494		&	24.50 &	0.91	\\
59915	&	i.MP9702	&	8	&	4481		&	24.80 &	0.62	\\
59915	&	z.MP9801	&	4	&	2240		&	23.33 &	0.76	\\
\hline
64232	&	u.MP9301	&	5	&	2800		&	25.20 &	0.90	\\
64232	&	g.MP9401	&	3	&	1020		&	25.05 &	0.65	\\
64232	&	r.MP9601	&	3	&	1494		&	24.59 &	0.84	\\
64232	&	i.MP9702	&	6	&	3481		&	24.60 &	0.79	\\
64232	&	z.MP9801	&	4	&	2240		&	23.17 &	0.74	\\
\hline \hline
\end{longtable}
\begin{longtable}{@{}ccccccc@{}}
\caption{CODEX sample S-II covered by CFHTLS Wide (see \citealt{erben13})}
\label{tab:secondary images} \\
\hline \hline
CODEX ID	&	CFHTLS Wide	&	Filter	&	N. of exp.	&	expos. time	&	$m_{\rm{lim}}$ &	Seeing	\\
  & field & & [s] & [AB mag]  & [$''$]\\
\hline \hline
13311	&	W3m3m1	&	u.MP9301	&	5	&	3001	&	25.05 &	0.93	\\
13311	&	W3m3m1	&	g.MP9401	&	5	&	2501	&	25.26 &	0.84	\\
13311	&	W3m3m1	&	r.MP9601	&	4	&	2001	&	24.80 &	0.72	\\
13311	&	W3m3m1	&	i.MP9701	&	7	&	4341	&	24.56 &	0.94	\\
13311	&	W3m3m1	&	z.MP9801	&	6	&	3601	&	23.39 &	0.56	\\
\hline
13315	&	W3m0m1	&	u.MP9301	&	5	&	3001	&	24.99 &	0.78	\\
13315	&	W3m0m1	&	g.MP9401	&	5	&	2501	&	25.59 &	0.76	\\
13315	&	W3m0m1	&	r.MP9601	&	4	&	2001	&	25.00 &	0.61	\\
13315	&	W3m0m1	&	i.MP9701	&	7	&	4306	&	24.63 &	0.54	\\
13315	&	W3m0m1	&	z.MP9801	&	6	&	3601	&	23.39 &	0.62	\\
\hline
13380	&	W3p3p1	&	u.MP9301	&	5	&	3001	&	25.23 &	0.88	\\
13380	&	W3p3p1	&	g.MP9401	&	5	&	2501	&	25.70 &	0.97	\\
13380	&	W3p3p1	&	r.MP9601	&	4	&	2001	&	25.06 &	0.76	\\
13380	&	W3p3p1	&	i.MP9701	&	7	&	4306	&	24.73 &	0.83	\\
13380	&	W3p3p1	&	z.MP9801	&	6	&	3601	&	23.07 &	0.68	\\
\hline
13390	&	W3m0m0	&	u.MP9301	&	5	&	3001	&	25.02 &	0.97	\\
13390	&	W3m0m0	&	g.MP9401	&	5	&	2501	&	25.53 &	0.94	\\
13390	&	W3m0m0	&	r.MP9601	&	4	&	2001	&	24.77 &	0.87	\\
13390	&	W3m0m0	&	i.MP9701	&	7	&	4341	&	24.41 &	0.94	\\
13390	&	W3m0m0	&	z.MP9801	&	5	&	3001	&	23.12 &	0.76	\\
\hline
13390	&	W3m1m0	&	u.MP9301	&	5	&	3001	&	24.91 &	0.69	\\
13390	&	W3m1m0	&	g.MP9401	&	6	&	3001	&	25.66 &	0.99	\\
13390	&	W3m1m0	&	r.MP9601	&	4	&	2001	&	25.09 &	0.66	\\
13390	&	W3m1m0	&	i.MP9701	&	7	&	4306	&	24.24 &	0.53	\\
13390	&	W3m1m0	&	z.MP9801	&	4	&	2401	&	23.01 &	0.71	\\
\hline
13391	&	W3p3m0	&	u.MP9301	&	5	&	3001	&	25.26 &	0.99	\\
13391	&	W3p3m0	&	g.MP9401	&	5	&	2501	&	25.67 &	0.97	\\
13391	&	W3p3m0	&	r.MP9601	&	4	&	2001	&	25.06 &	0.76	\\
13391	&	W3p3m0	&	i.MP9701	&	7	&	4306	&	24.74 &	0.71	\\
13391	&	W3p3m0	&	z.MP9801	&	6	&	3601	&	23.61 &	0.73	\\
\hline
13400	&	W3p3m0	&	u.MP9301	&	5	&	3001	&	25.26 &	0.99	\\
13400	&	W3p3m0	&	g.MP9401	&	5	&	2501	&	25.67 &	0.97	\\
13400	&	W3p3m0	&	r.MP9601	&	4	&	2001	&	25.06 &	0.76	\\
13400	&	W3p3m0	&	i.MP9701	&	7	&	4306	&	24.74 &	0.71	\\
13400	&	W3p3m0	&	z.MP9801	&	6	&	3601	&	23.61 &	0.73	\\
\hline
13400	&	W3p3m1	&	u.MP9301	&	5	&	3001	&	25.24 &	0.89	\\
13400	&	W3p3m1	&	g.MP9401	&	6	&	3001	&	25.71 &	0.89	\\
13400	&	W3p3m1	&	r.MP9601	&	4	&	2001	&	25.05 &	0.79	\\
13400	&	W3p3m1	&	i.MP9701	&	6	&	3691	&	24.87 &	0.85	\\
13400	&	W3p3m1	&	i.MP9702	&	7	&	4306	&	24.71 &	0.68	\\
13400	&	W3p3m1	&	z.MP9801	&	6	&	3601	&	23.64 &	0.64	\\
\hline \hline
\pagebreak
\hline \hline
CODEX ID	&	CFHTLS Wide	&	Filter	&	N. of exp.	&	expos. time	&	$m_{\rm{lim}}$ &	Seeing	\\
  & field & & [s] & [AB mag]  & [$''$]\\
\hline \hline
17449	&	W3m1m2	&	u.MP9301	&	5	&	3001	&	24.48 &	0.86	\\
17449	&	W3m1m2	&	g.MP9401	&	4	&	2001	&	25.40 &	0.88	\\
17449	&	W3m1m2	&	r.MP9601	&	5	&	2501	&	24.93 &	0.65	\\
17449	&	W3m1m2	&	i.MP9701	&	7	&	4306	&	24.34 &	0.65	\\
17449	&	W3m1m2	&	z.MP9801	&	6	&	3601	&	23.45 &	0.67	\\
\hline
17449	&	W3m1m3	&	u.MP9301	&	5	&	3001	&	24.61 &	0.75	\\
17449	&	W3m1m3	&	g.MP9401	&	5	&	2501	&	25.64 &	0.86	\\
17449	&	W3m1m3	&	r.MP9601	&	4	&	2001	&	24.84 &	0.70	\\
17449	&	W3m1m3	&	i.MP9701	&	7	&	4307	&	24.41 &	0.66	\\
17449	&	W3m1m3	&	z.MP9801	&	6	&	3601	&	23.41 &	0.59	\\
\hline
17453	&	W3m2m2	&	u.MP9301	&	5	&	3001	&	25.11 &	0.77	\\
17453	&	W3m2m2	&	g.MP9401	&	5	&	2501	&	25.55 &	0.89	\\
17453	&	W3m2m2	&	r.MP9601	&	4	&	2001	&	24.97 &	0.62	\\
17453	&	W3m2m2	&	i.MP9701	&	7	&	4306	&	24.46 &	0.65	\\
17453	&	W3m2m2	&	z.MP9801	&	6	&	3601	&	23.43 &	0.64	\\
\hline
17453	&	W3m2m3	&	u.MP9301	&	9	&	5402	&	25.57 &	0.86	\\
17453	&	W3m2m3	&	g.MP9401	&	6	&	2501	&	25.62 &	0.89	\\
17453	&	W3m2m3	&	r.MP9601	&	5	&	2001	&	24.97 &	0.80	\\
17453	&	W3m2m3	&	i.MP9701	&	7	&	4306	&	24.40 &	0.73	\\
17453	&	W3m2m3	&	z.MP9801	&	6	&	3601	&	23.22 &	0.82	\\
\hline
54652	&	W4m3p3	&	u.MP9301	&	5	&	3000	&	25.34 &	0.90	\\
54652	&	W4m3p3	&	g.MP9401	&	6	&	3000	&	25.70 &	0.71	\\
54652	&	W4m3p3	&	r.MP9601	&	4	&	2000	&	24.72 &	0.76	\\
54652	&	W4m3p3	&	i.MP9702	&	7	&	4306	&	24.70 &	0.57	\\
54652	&	W4m3p3	&	z.MP9801	&	6	&	3600	&	23.32 &	0.62	\\
\hline
56934	&	W2m0p3	&	u.MP9301	&	5	&	3001	&	25.25 &	0.82	\\
56934	&	W2m0p3	&	g.MP9401	&	5	&	2500	&	25.26 &	0.71	\\
56934	&	W2m0p3	&	r.MP9601	&	6	&	3000	&	25.04 &	0.70	\\
56934	&	W2m0p3	&	i.MP9701	&	7	&	4305	&	24.36 &	0.51	\\
56934	&	W2m0p3	&	z.MP9801	&	6	&	3600	&	23.27 &	0.78	\\
\hline
56934	&	W2m1p3	&	u.MP9301	&	5	&	3000	&	25.38 &	1.07	\\
56934	&	W2m1p3	&	g.MP9401	&	5	&	2500	&	25.67 &	0.75	\\
56934	&	W2m1p3	&	r.MP9601	&	6	&	3001	&	25.16 &	0.66	\\
56934	&	W2m1p3	&	i.MP9701	&	5	&	3075	&	24.48 &	0.63	\\
56934	&	W2m1p3	&	z.MP9801	&	6	&	3600	&	23.46 &	0.69	\\
\hline
57017	&	W2p2p3	&	u.MP9301	&	5	&	3001	&	25.19 &	0.81	\\
57017	&	W2p2p3	&	g.MP9401	&	5	&	2500	&	25.72 &	0.93	\\
57017	&	W2p2p3	&	r.MP9601	&	5	&	2500	&	25.04 &	0.82	\\
57017	&	W2p2p3	&	i.MP9701	&	7	&	4305	&	24.62 &	0.78	\\
57017	&	W2p2p3	&	z.MP9801	&	7	&	4200	&	23.58 &	0.83	\\
\hline
60076	&	W1m1p3	&	u.MP9301	&	5	&	3001	&	25.06 &	0.85	\\
60076	&	W1m1p3	&	g.MP9401	&	5	&	2501	&	25.42 &	0.94	\\
60076	&	W1m1p3	&	r.MP9601	&	4	&	2000	&	24.89 &	0.83	\\
60076	&	W1m1p3	&	i.MP9701	&	7	&	4306	&	24.64 &	0.76	\\
60076	&	W1m1p3	&	z.MP9801	&	10	&	6001	&	23.56 &	0.72	\\
\hline
60131	&	W1p1m2	&	u.MP9301	&	5	&	3000	&	25.38 &	1.03	\\
60131	&	W1p1m2	&	g.MP9401	&	5	&	2500	&	25.60 &	0.76	\\
60131	&	W1p1m2	&	r.MP9601	&	4	&	2000	&	24.86 &	0.69	\\
60131	&	W1p1m2	&	i.MP9701	&	8	&	4921	&	24.86 &	0.70	\\
60131	&	W1p1m2	&	z.MP9801	&	6	&	3600	&	23.50 &	0.72	\\
\hline
60131	&	W1p2m2	&	u.MP9301	&	6	&	3601	&	25.44 &	1.04	\\
60131	&	W1p2m2	&	g.MP9401	&	5	&	2500	&	25.61 &	0.73	\\
60131	&	W1p2m2	&	r.MP9601	&	4	&	2000	&	24.79 &	0.78	\\
60131	&	W1p2m2	&	i.MP9701	&	7	&	4306	&	24.76 &	0.64	\\
60131	&	W1p2m2	&	z.MP9801	&	7	&	4200	&	23.36 &	0.89	\\
\hline \hline
\pagebreak
\hline \hline
CODEX ID	&	CFHTLS Wide	&	Filter	&	N. of exp.	&	expos. time	&	$m_{\rm{lim}}$ &	Seeing	\\
  & field & & [s] & [AB mag]  & [$''$]\\
\hline \hline
60155	&	W1p4p2	&	u.MP9301	&	5	&	3001	&	25.23 &	0.76	\\
60155	&	W1p4p2	&	g.MP9401	&	5	&	2500	&	25.52 &	0.86	\\
60155	&	W1p4p2	&	r.MP9601	&	4	&	2000	&	24.76 &	0.60	\\
60155	&	W1p4p2	&	i.MP9701	&	7	&	4341	&	24.41 &	0.87	\\
60155	&	W1p4p2	&	i.MP9702	&	10	&	6150	&	24.81 &	0.63	\\
60155	&	W1p4p2	&	z.MP9801	&	6	&	3601	&	23.43 &	0.55	\\
\hline
64565	&	W1m4m3	&	u.MP9301	&	5	&	3000	&	25.22 &	0.84	\\
64565	&	W1m4m3	&	g.MP9401	&	7	&	3500	&	25.68 &	0.84	\\
64565	&	W1m4m3	&	r.MP9601	&	4	&	2000	&	24.89 &	0.94	\\
64565	&	W1m4m3	&	i.MP9701	&	7	&	4306	&	24.46 &	0.59	\\
64565	&	W1m4m3	&	z.MP9801	&	6	&	3600	&	23.55 &	0.82	\\
\hline
64636	&	W1p1m2	&	u.MP9301	&	5	&	3000	&	25.38 &	1.03	\\
64636	&	W1p1m2	&	g.MP9401	&	5	&	2500	&	25.60 &	0.76	\\
64636	&	W1p1m2	&	r.MP9601	&	4	&	2000	&	24.86 &	0.69	\\
64636	&	W1p1m2	&	i.MP9701	&	8	&	4921	&	24.86 &	0.70	\\
64636	&	W1p1m2	&	z.MP9801	&	6	&	3600	&	23.50 &	0.72	\\
\hline
64636	&	W1p2m2	&	u.MP9301	&	6	&	3601	&	25.44 &	1.04	\\
64636	&	W1p2m2	&	g.MP9401	&	5	&	2500	&	25.61 &	0.73	\\
64636	&	W1p2m2	&	r.MP9601	&	4	&	2000	&	24.79 &	0.78	\\
64636	&	W1p2m2	&	i.MP9701	&	7	&	4306	&	24.76 &	0.64	\\
64636	&	W1p2m2	&	z.MP9801	&	7	&	4200	&	23.36 &	0.89	\\
\hline
210288	&	W3p2p2	&	u.MP9301	&	5	&	3001	&	25.20 &	0.68	\\
210288	&	W3p2p2	&	g.MP9401	&	5	&	2501	&	25.53 &	0.81	\\
210288	&	W3p2p2	&	r.MP9601	&	4	&	2001	&	24.88 &	0.79	\\
210288	&	W3p2p2	&	i.MP9701	&	7	&	4307	&	24.51 &	0.54	\\
210288	&	W3p2p2	&	z.MP9801	&	6	&	3601	&	22.95 &	0.53	\\
\hline
210288	&	W3p2p3	&	u.MP9301	&	5	&	3001	&	25.25 &	0.99	\\
210288	&	W3p2p3	&	g.MP9401	&	5	&	2501	&	25.49 &	0.74	\\
210288	&	W3p2p3	&	r.MP9601	&	4	&	2001	&	24.93 &	0.67	\\
210288	&	W3p2p3	&	i.MP9701	&	7	&	4306	&	24.56 &	0.69	\\
210288	&	W3p2p3	&	z.MP9801	&	6	&	3601	&	23.82 &	0.63	\\
\hline
210306	&	W3p1p1	&	u.MP9301	&	5	&	3001	&	25.28 &	0.93	\\
210306	&	W3p1p1	&	g.MP9401	&	5	&	2501	&	25.66 &	0.79	\\
210306	&	W3p1p1	&	r.MP9601	&	4	&	2001	&	24.94 &	0.84	\\
210306	&	W3p1p1	&	i.MP9701	&	7	&	4306	&	24.64 &	0.71	\\
210306	&	W3p1p1	&	z.MP9801	&	6	&	3601	&	23.63 &	0.72	\\
\hline
210306	&	W3p2p1	&	u.MP9301	&	5	&	3001	&	25.19 &	0.78	\\
210306	&	W3p2p1	&	g.MP9401	&	6	&	3001	&	25.69 &	0.84	\\
210306	&	W3p2p1	&	r.MP9601	&	4	&	2001	&	24.94 &	0.64	\\
210306	&	W3p2p1	&	i.MP9701	&	9	&	5537	&	24.56 &	0.70	\\
210306	&	W3p2p1	&	z.MP9801	&	7	&	4201	&	23.13 &	0.57	\\
\hline \hline
\end{longtable}
\pagebreak
\begin{longtable}{@{}cccccc@{}}
\caption{CODEX sample S-III ($z \lesssim 0.4$)}
\label{tab:tertiary images} \\
\hline \hline
CODEX ID	&	Filter	&	N. of exp.	&	expos. time	&	$m_{\rm{lim}}$	&	Seeing\\
  & & & [s] & [AB mag]  & [$''$]\\
\hline \hline
24925	&	u.MP9301	&	5	&	2801	&	25.02 &	1.18	\\
24925	&	g.MP9401	&	2	&	680 	&	24.72 &	0.93	\\
24925	&	r.MP9601	&	3	&	1500	&	24.39 &	0.87	\\
24925	&	i.MP9702	&	9	&	5042	&	24.89 &	0.73	\\
24925	&	z.MP9801	&	4	&	2241	&	22.96 &	0.76	\\
\hline
27955	&	u.MP9301	&	7	&	3921		&	25.42 &	0.94	\\
27955	&	r.MP9601	&	4	&	1993		&	24.51 &	0.61	\\
27955	&	i.MP9702	&	8	&	4481		&	24.62 &	0.50	\\
27955	&	z.MP9801	&	4	&	2241		&	23.13 &	0.49	\\
\hline
46647	&	u.MP9301	&	5	&	2801	&	24.96 &	0.79	\\
46647	&	g.MP9401	&	3	&	1020	&	24.76 &	0.62	\\
46647	&	r.MP9601	&	4	&	1992	&	24.68 &	0.82	\\
46647	&	i.MP9702	&	9	&	5041	&	24.67 &	0.51	\\
46647	&	z.MP9801	&	4	&	2241	&	23.44 &	0.77	\\
\hline
54796	&	u.MP9301	&	5	&	2801	&	25.13 &	0.94	\\
54796	&	g.MP9401	&	3	&	1020	&	24.97 &	1.13	\\
54796	&	r.MP9601	&	3	&	1494	&	24.16 &	0.74	\\
54796	&	i.MP9702	&	8	&	4481	&	24.55 &	0.64	\\
54796	&	z.MP9801	&	4	&	2240	&	23.18 &	0.70	\\
\hline \hline
\end{longtable}
\pagebreak
\begin{longtable}{@{}cccccc@{}}
\caption{Additional CODEX galaxy clusters observed in three bands} \\
\hline \hline
CODEX ID	&	Filter	&	N. of exp.	&	expos. time	&	$m_{\rm{lim}}$ &	Seeing	\\
  & & & [s] & [AB mag]  & [$''$]\\
\hline \hline
16463	&	u.MP9301	&	7	&	3641	&	25.40 &	1.09	\\
16463	&	i.MP9702	&	8	&	4482	&	24.74 &	0.57	\\
16463	&	z.MP9801	&	6	&	3241	&	23.47 &	0.77	\\
\hline
16470	&	u.MP9301	&	7	&	3641	&	25.40 &	1.09	\\
16470	&	i.MP9702	&	8	&	4482	&	24.74 &	0.57	\\
16470	&	z.MP9801	&	6	&	3241	&	23.47 &	0.77	\\
\hline
20621	&	g.MP9401	&	4	&	841	&	24.82 &	1.14	\\
20621	&	r.MP9601	&	3	&	631	&	24.16 &	1.06	\\
20621	&	i.MP9702	&	11	&	6162	&	24.84 &	0.61	\\
\hline
20622	&	g.MP9401	&	4	&	841	&	24.82 &	1.14	\\
20622	&	r.MP9601	&	3	&	631	&	24.16 &	1.06	\\
20622	&	i.MP9702	&	11	&	6162	&	24.84 &	0.61	\\
\hline
24747	&	u.MP9301	&	5	&	2601	&	25.20 &	0.73	\\
24747	&	i.MP9702	&	7	&	3921	&	24.64 &	0.61	\\
24747	&	z.MP9801	&	4	&	2241	&	22.81 &	0.90	\\
\hline
25094	&	g.MP9401	&	3	&	630	&	24.81 &	0.74	\\
25094	&	r.MP9601	&	3	&	630	&	24.29 &	0.75	\\
25094	&	i.MP9702	&	1	&	560	&	23.41 &	0.51	\\
\hline
25252	&	g.MP9401	&	3	&	630	&	24.93 &	1.23	\\
25252	&	r.MP9601	&	3	&	630	&	24.34 &	1.07	\\
25252	&	i.MP9702	&	8	&	4482	&	24.74 &	0.60	\\
\hline
29249	&	u.MP9301	&	5	&	2601	&	25.20 &	0.73	\\
29249	&	i.MP9702	&	7	&	3921	&	24.64 &	0.61	\\
29249	&	z.MP9801	&	4	&	2241	&	22.81 &	0.90	\\
\hline
37287	&	u.MP9301	&	3	&	1681	&	24.81 &	1.13	\\
37287	&	i.MP9702	&	8	&	4481	&	24.55 &	0.64	\\
37287	&	z.MP9801	&	4	&	2241	&	22.89 &	0.50	\\
\hline
39323	&	g.MP9401	&	3	&	630	&	24.79 &	0.67	\\
39323	&	r.MP9601	&	3	&	630	&	24.12 &	0.56	\\
39323	&	i.MP9702	&	8	&	4481	&	24.51 &	0.56	\\
\hline
39326	&	g.MP9401	&	3	&	630	&	24.79 &	0.67	\\
39326	&	r.MP9601	&	3	&	630	&	24.12 &	0.56	\\
39326	&	i.MP9702	&	8	&	4481	&	24.51 &	0.56	\\
\hline
39879	&	g.MP9401	&	3	&	630	&	24.86 &	0.73	\\
39879	&	r.MP9601	&	3	&	630	&	24.33 &	0.61	\\
39879	&	i.MP9702	&	8	&	4481	&	24.55 &	0.67	\\
\hline
43881	&	g.MP9401	&	5	&	736	&	24.79 &	0.66	\\
43881	&	r.MP9601	&	4	&	921	&	24.35 &	1.21	\\
43881	&	i.MP9702	&	6	&	715	&	23.90 &	0.84	\\
\hline
46328	&	g.MP9401	&	3	&	1020	&	25.05 &	0.96	\\
46328	&	r.MP9601	&	3	&	1500	&	24.66 &	0.62	\\
46328	&	i.MP9702	&	7	&	3921	&	24.82 &	0.66	\\
\hline
50502	&	g.MP9401	&	3	&	1020	&	25.10 &	0.91	\\
50502	&	r.MP9601	&	3	&	1500	&	24.80 &	0.56	\\
50502	&	i.MP9702	&	8	&	4481	&	24.77 &	0.77	\\
\hline
52484	&	g.MP9401	&	3	&	630	&	24.83 &	0.99	\\
52484	&	r.MP9601	&	5	&	1050	&	24.53 &	0.93	\\
52484	&	i.MP9702	&	8	&	4481	&	24.69 &	0.47	\\
\hline
57892	&	g.MP9401	&	1	&	260	&	23.59 &	1.02	\\
57892	&	r.MP9601	&	5	&	1500	&	24.49 &	0.99	\\
57892	&	i.MP9702	&	1	&	560	&	23.75 &	0.90	\\
\hline
58014	&	g.MP9401	&	5	&	1050	&	25.05 &	0.81	\\
58014	&	r.MP9601	&	3	&	630	&	24.28 &	0.73	\\
58014	&	i.MP9702	&	10	&	5601	&	24.83 &	0.85	\\
\hline \hline
\pagebreak
\hline \hline
CODEX ID	&	Filter	&	N. of exp.	&	expos. time	&	$m_{\rm{lim}}$ &	Seeing	\\
  & & & [s] & [AB mag]  & [$''$]\\
\hline \hline
58114	&	g.MP9401	&	3	&	630	&	24.82 &	0.92	\\
58114	&	r.MP9601	&	3	&	630	&	24.28 &	0.81	\\
58114	&	i.MP9702	&	10	&	5601	&	24.71 &	0.70	\\
\hline
64360	&	g.MP9401	&	3	&	1020	&	25.06 &	0.88	\\
64360	&	r.MP9601	&	3	&	540	&	23.79 &	0.89	\\
64360	&	i.MP9702	&	9	&	5041	&	24.80 &	0.69	\\
\hline
219599	&	g.MP9401	&	3	&	630	&	24.81 &	0.74	\\
219599	&	r.MP9601	&	3	&	630	&	24.29 &	0.75	\\
219599	&	i.MP9702	&	1	&	560	&	23.41 &	0.51	\\
\hline \hline
\end{longtable}
%
%
%
\bsp	
\label{lastpage}
\end{document}